# Estimating Causal Mediation Effects under Correlated Errors

Yi Zhao
*Brown University, Providence RI, USA.*
Xi Luo
*Brown University, Providence RI, USA.*
E-mail: xi.rossi.luo@gmail.com

**Summary**. Causal mediation analysis usually requires strong assumptions, such as ignorability of the mediator, which may not hold in many social and scientific studies. Motivated by a multilevel randomized treatment experiment using functional magnetic resonance imaging (fMRI), this paper proposes a multilevel causal mediation framework for data with hierarchically nested structure, and this framework provides valid inference even if structured unmeasured confounding for the mediator and outcome is present. For the first-level data, we propose a linear structural equation model for a continuous mediator and a continuous outcome, both of which may contain correlated additive errors. A likelihood-based approach is proposed to estimate the model coefficients. The analysis of our estimator characterizes the nonidentifiability issue due to the correlation parameter. To address the identifiability issue and model the variability in multilevel data, we propose to incorporate multiple first-level mediation models across different levels in a unified multilevel mediation framework. All the model coefficients are estimated simultaneously by our optimization algorithms, with innovation to estimate the unknown correlation parameter from data, instead of performing sensitivity analysis. Our asymptotic analysis shows that the correlation parameter is identifiable, and our estimates for the mediation effects are consistent with the parametric convergence rates. Using extensive simulated data and a real fMRI dataset, we demonstrate the improvement of our approaches over existing methods.

*Keywords*: Confounding; Multilevel model; Optimization; Structural equation model

## 1. Introduction

In diverse fields of empirical research, including many in the biological and social science, scientists are often interested in identifying causal mechanisms through which a treatment affects an outcome. Mediation analysis is widely used to quantify the extent of the treatment effect that is mediated by a third variable (called mediator) in the causal pathways. Structural equation modeling (SEM) methods for mediation analysis, such as the Baron-Kenny method (Baron and Kenny, 1986) and its extensions (MacKinnon et al., 2007), can be implemented for complex data. However, the resulting SEM coefficients may not have causal interpretations unless certain causal assumptions are met. A critical and hardly testable assumption is the sequential ignorability assumption, which implies that there are no unmeasured confounders between the mediator and the outcome. Imai et al. (2010) considers this assumption to be "too strong for the typical situations", which is also unlikely to hold in our motivating functional magnetic resonance imaging (fMRI) experiment to be introduced momentarily. In this paper, we propose a novel multilevel SEM framework that estimates from data the effect caused by the mediator-outcome unmeasured confounding. Our framework will then adjust the estimated confounding to yield unbiased and consistent estimates of the causal effects.

Our motivating example for unmeasured confounding and multilevel data comes from an fMRI experiment. A group of healthy participants are instructed to perform motor tasks, responding to a sequence of randomized stimuli. The experiment is organized into three levels, participants, sessions, and trials. In each trial, randomized binary stimuli or instructions, pressing a button or withholding from pressing, are presented to the participants while undergoing fMRI scanning. Scientists are interested in how the stimulus information is processed in the human brain across the population. In particular, based on the prior findings (Duann et al., 2009; Obeso et al., 2013), an important scientific question is to quantify the mediating role of a brain region called the presupplementary motor area (preSMA), when the final motor responses are finally programmed by another brain region called the primary motor cortex (PMC). To address this question, we study both brain activations for each trial, reconstructed from the raw fMRI data. In this example, each trial is a treatment unit, the treatment is the randomized stimulus, the mediator is the preSMA activation, and the outcome is the PMC activation. The scientific goal is to quantify the causal effects through the stimulus–preSMA–PMC pathway and the stimulus–PMC pathway respectively. However, it has been well established that brain activity is a superposition of significant task-unrelated (or spontaneous) activity and potentially weaker task-related activity (Fox et al., 2006), though only the superposition can be measured by fMRI. Since the task-unrelated activity influences both the mediator and outcome regions, it constitutes an unmeasured confounding factor.

The assumptions involved to perform causal mediation analysis have been widely studied in the statistical literature, usually for one-level data, see for example Holland (1988); Robins and Greenland (1992); Angrist et al. (1996); Ten Have et al. (2007); Jo (2008); Albert (2008); Gallop et al. (2009); VanderWeele (2009); Imai et al. (2010); Daniels et al. (2012). However, these approaches assume roughly that there are no unmeasured confounders present in the model, see various variants discussed in a review (Imai et al., 2010). This assumption clearly is unlikely to hold for most settings. Under our experiment, this assumption is violated as the task-unrelated activity poses significant influences on both the mediator and outcome regions. To address this complication, one may adopt the sensitivity analysis (Imai et al., 2010) or apply the instrumental variable (IV) approach under additional structural assumptions (Ten Have et al., 2007; Small, 2011). Lindquist (2012) applied the IV approach for an fMRI dataset when the outcome variable is outside the brain, where the structural assumptions are more likely to hold.

Recently, mediation analysis has been extended to hierarchically organized experiments, probably due to the increasing popularity of such experiments. These approaches are usually developed for two-level data in practice settings. The first-level data are usually modeled by the Baron-Kenny method, and various second-level models are introduced for the estimated parameters, see Krull and MacKinnon (1999) and Kenny et al. (2003) for example. It has not been rigorously studied if the resulting parameters have causal interpretations, though they are expected to suffer from similar limitations of the Baron-Kenny method, such as assuming sequential ignorability.

In this paper, we propose an optimization-based multilevel mediation framework, which we call Correlated-error Mediation Analysis (CMA). We make the following specific contributions. First, we introduce a modeling framework which allows modeling the mediator and outcome variables jointly with correlated errors to model unmeasured confounding. This formulation relaxes the mediator ignorability assumption. Our framework also introduces an integrated approach for modeling two-level or three-level data as we will study in detail in this paper. Second, we study the causal assumptions associated with our framework. We prove that the parameters associated with unmeasured confounding are identifiable and estimable from multilevel data, unlike those untestable assumptions for single level data usually assumed in the literature. Asymptotic analyses also show that our estimates converge to the causal parameters with the parametric rates. Third, we develop efficient computational algorithms to compute for a large number of parameters, for example thousands in our fMRI dataset.

This paper is organized as follows. We first introduce our multilevel SEM for mediation analysis with correlated errors in Section 2. In Section 3, we propose likelihood-based methods to estimate causal coefficients, study the identifiability of model parameters and the asymptotic properties of the estimators. We compare these methods and demonstrate the improvement using extensive simulations in Section 4, and compare the analysis results of different methods in the real fMRI data application in Section 5. We conclude with a discussion in Section 6.

## 2. Model

For simplicity, we will refer to the three levels by trial, session, and participant. Let $Z_{ikj}$ be the treatment assignment for the $j$th trial of the $k$th session of the $i$th participant, where $i = 1, \ldots, N$, $k = 1, \ldots, K_i$, and $j = 1, \ldots, n_{ik}$. Similarly, define $M_{ikj}$ and $R_{ikj}$ for the observed mediator and outcome values from the same multilevel unit. In our experiment, $Z_{ikj}$ is a binary stimulus assignment; $M_{ikj}$ and $R_{ikj}$ are the preSMA and PMC activations respectively for the same trial. It is straightforward to adapt our proposed likelihood framework for varying $K_i$, and thus we will focus on $K_i = K$, as in our experiment, to fix the idea. Unless noted otherwise, we will also use this three-level notation for two-level data, by setting $K = 1$.

Our multilevel mediation model contains two components. At the first level, we propose the following matrix-format mediation model, for every $i$ and $k$,

$$\begin{pmatrix}\mathbf{M}_{ik} & \mathbf{R}_{ik}\end{pmatrix} = \begin{pmatrix}\mathbf{Z}_{ik} & \mathbf{M}_{ik}\end{pmatrix} \begin{pmatrix} A_{ik} & C_{ik} \\ 0 & B_{ik} \end{pmatrix} + \begin{pmatrix}\mathbf{E}_{1_{ik}} & \mathbf{E}_{2_{ik}}\end{pmatrix}, \tag{1}$$

and

$$\mathrm{vec}\left[\begin{pmatrix}\mathbf{E}_{1_{ik}} & \mathbf{E}_{2_{ik}}\end{pmatrix}\right] \sim \mathcal{N}\left(\mathbf{0}, \begin{pmatrix} \sigma^2_{1_{ik}} & \delta_{ik}\sigma_{1_{ik}}\sigma_{2_{ik}} \\ \delta_{ik}\sigma_{1_{ik}}\sigma_{2_{ik}} & \sigma^2_{2_{ik}} \end{pmatrix} \otimes \mathbf{I}_{n_{ik}}\right), \tag{2}$$

where vec$[\cdot]$ is the vectorization operator by stacking columns of a matrix, $\otimes$ is the Kronecker product operator, and $\mathbf{I}_{n_{ik}}$ is the $n_{ik}$-dimensional identity matrix. $A_{ik}$, $B_{ik}$, and $C_{ik}$ are the SEM coefficients in session $k$ of participant $i$. Without loss of generality, we assume that $\mathbf{R}_{ik}$ and $\mathbf{M}_{ik}$ are centered around 0, and thus no intercepts are included in the model. This can be achieved via subtracting the sample means for each variable, and one can also remove the effects of other covariates using regression (Rosenbaum et al., 2002). Standard parametric mediation models (Baron and Kenny, 1986; Imai et al., 2010) without unmeasured confounding are special cases of our model (1)-(2). They are equivalent to assuming $\delta_{ik} = 0$ for every $i$ and $k$, see Section 3.1 for a more detailed comparison. We will instead introduce methods to estimate $\delta_{ik}$ in Sections 3.2 and 3.3.

For higher levels, we propose to pool information across the first-level coefficients in (1) using the following model

$$\mathbf{b}_{ik} = \begin{pmatrix} A_{ik} \\ B_{ik} \\ C_{ik} \end{pmatrix} = \begin{pmatrix} A \\ B \\ C \end{pmatrix} + \begin{pmatrix} \alpha_i \\ \beta_i \\ \gamma_i \end{pmatrix} + \begin{pmatrix} \epsilon^A_{ik} \\ \epsilon^B_{ik} \\ \epsilon^C_{ik} \end{pmatrix} = \mathbf{b} + \mathbf{u}_i + \boldsymbol{\eta}_{ik}. \tag{3}$$

For three-level data, model (3) in essence is a mixed effects model, where $A$, $B$ and $C$ are the fixed effects; $\alpha_i$, $\beta_i$, and $\gamma_i$ are the random effects of $A_{ik}$, $B_{ik}$ and $C_{ik}$, respectively. Similar to many mixed effects models (see Penny et al. (2003) for a review on mixed effects models in fMRI analysis), we assume the random effect $\mathbf{u}_i$ follows a trivariate normal distribution with mean zero and covariance matrix $\boldsymbol{\Psi}$; and $\epsilon^A_{ik}$, $\epsilon^B_{ik}$ and $\epsilon^C_{ik}$ are the random errors in session $k$ of participant $i$, which are identically distributed from a trivariate normal distribution with mean zero and covariance matrix $\boldsymbol{\Lambda}$. $\mathbf{u}_i$ and $\boldsymbol{\eta}_{ik}$, for all $i$ and $k$, are mutually independent. For simplicity, this paper will focus on a diagonal covariance $\boldsymbol{\Psi} = \mathrm{diag}\{\psi^2_\alpha, \psi^2_\beta, \psi^2_\gamma\}$ and $\boldsymbol{\Lambda} = \mathrm{diag}\{\lambda^2_\alpha, \lambda^2_\beta, \lambda^2_\gamma\}$ in our numerical studies for fast and stable covariance estimation. Though our method can be extended to non-diagonal $\boldsymbol{\Psi}$ and $\boldsymbol{\Lambda}$, the computational complexity would increase accordingly. For two-level data or $K = 1$, one cannot estimate the random effects $\mathbf{u}_i$. Thus, we set $\mathbf{u}_i = \mathbf{0}$ and $\boldsymbol{\Psi} = \mathbf{0}$ in model (3) without changing all other modeling assumptions. This case essentially reduces to an ANOVA model for the first level SEM coefficients.

Fig. 1 depicts a conceptual diagram of our multilevel model and the relationship between the parameters across multiple levels. We now describe in detail each modeling component and their causal assumptions.

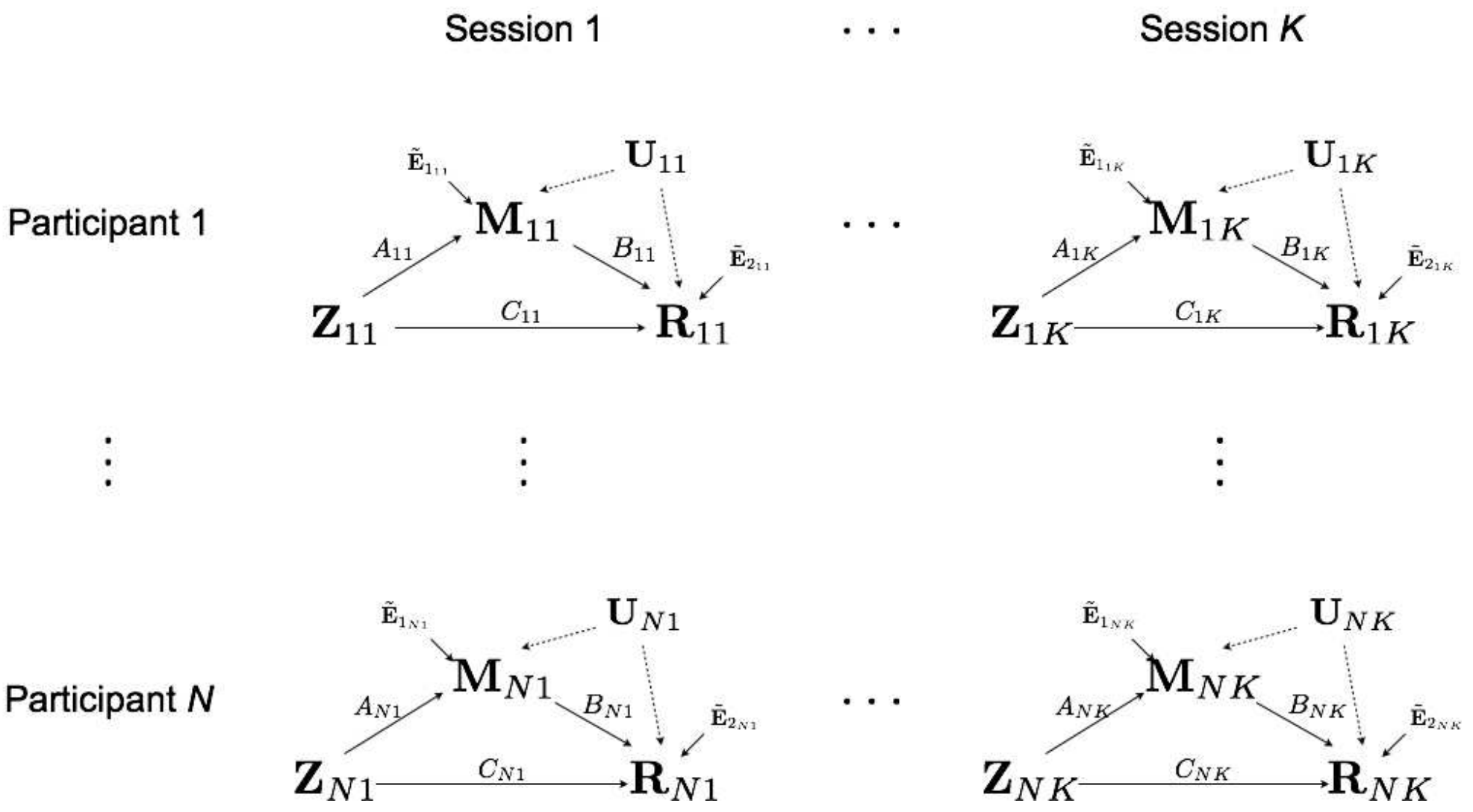

**Fig. 1.** Conceptual causal diagram of our multilevel model. $\mathbf{Z}_{ik}$, $\mathbf{M}_{ik}$ and $\mathbf{R}_{ik}$ are the randomized treatment, mediator and outcome vectors, respectively, in session $k$ of participant $i$, for $i = 1, \ldots, N$ and $k = 1, \ldots, K$. $\mathbf{U}_{ik}$'s are the unmeasured confounders, and $\tilde{\mathbf{E}}_{1_{ik}}$ and $\tilde{\mathbf{E}}_{2_{ik}}$ are the independent model errors. $A_{ik}$, $B_{ik}$ and $C_{ik}$'s are the SEM coefficients.

### *2.1. First-level Model*

For simplicity of notations, when we discuss our first-level model in this section, we denote the vector $\mathbf{Z} = (Z_{ikj}, j = 1, \ldots, n_{ik})$ by omitting the subscripts $i$ and $k$, and similarly omit the subscripts in $\mathbf{M}$, $\mathbf{R}$, $n$ and so on. Using this simplified notation, we write our model as

$$\begin{pmatrix}\mathbf{M} & \mathbf{R}\end{pmatrix} = \begin{pmatrix}\mathbf{Z} & \mathbf{M}\end{pmatrix}\begin{pmatrix}A & C \\ 0 & B\end{pmatrix} + \begin{pmatrix}\mathbf{E}_1 & \mathbf{E}_2\end{pmatrix}, \tag{4}$$

where the errors $\mathbf{E}_1$ and $\mathbf{E}_2$ have correlation $\delta$ as in model (2).

Our model (4) contains the same equations as the common parametric mediation model (Baron and Kenny, 1986; Imai et al., 2010)

$$\mathbf{M} = \mathbf{Z}A + \mathbf{E}_1, \tag{5}$$

$$\mathbf{R} = \mathbf{Z}C + \mathbf{M}B + \mathbf{E}_2. \tag{6}$$

Baron and Kenny (1986) also considered a third but redundant equation

$$\mathbf{R} = \mathbf{Z}C' + \mathbf{E}', \tag{7}$$

where $C'$ is the coefficient of interest and $\mathbf{E}'$ is the noise term. Under models (5)-(7), the average total effect ($C'$) can be decomposed into the indirect effect ($AB$ or $C' - C$) and the direct effect ($C$), see Imai et al. (2010) for a recent review.

The standard mediation models assume independent $\mathbf{E}_1$ and $\mathbf{E}_2$, and thus (5) and (6) are fitted separately. This corresponds to setting $\delta = 0$ in our model (4).

Because it is expected that $\delta \neq 0$ in many practice settings, these standard analyses are usually followed by sensitivity analysis, where the users change the $\delta$ values (sometimes within a hypothesized range) to check its influence on the direct and indirect effect estimates (Imai et al., 2010).

We do not impose this modeling assumption on $\delta$, but rather treat it as a modeling parameter to be estimated from data, see Section 3. Due to the presence of unmeasured confounding in many studies, as in our fMRI experiment described before, it is expected that $\delta \neq 0$. A hypothetical example of this observation is described as follows. Suppose

$$\begin{aligned} \mathbf{M} &= \mathbf{Z}A + g_1\mathbf{U} + \tilde{\mathbf{E}}_1, \\ \mathbf{R} &= \mathbf{Z}C + \mathbf{M}B + g_2\mathbf{U} + \tilde{\mathbf{E}}_2, \end{aligned}$$

where $\mathbf{U}$, $\tilde{\mathbf{E}}_1$, and $\tilde{\mathbf{E}}_2$ are mutually independent normal variables, and $g_1$ and $g_2$ are fixed and unknown scalars. $\mathbf{U}$ here represents the overall effect from all unmeasured confounding factors for both $\mathbf{M}$ and $\mathbf{R}$. Under this model, it is easy to see that the errors in (5) and (6) are correlated when $g_1 g_2 \neq 0$. Later, our approach will remove the influence of $\mathbf{U}$ by accounting the estimation bias due to $\delta \neq 0$.

### 2.1.1. *Assumptions and causal interpretation*

We use Rubin's potential outcome framework (Rubin, 2005) to assess the causal interpretation of our model (4). As Rubin conjectured but not analyzed in Section 7 of his paper, when $Z$ is randomized, a valid approach for mediation analysis is to consider "a bivariate outcome variable" $(M, R)$ in order to infer the direct and indirect effects. We analyze and extend this conjecture in this paper. We here briefly analyze the causal interpretation of our SEM coefficients using potential outcomes

$$\begin{pmatrix}\mathbf{M}(\mathbf{z}) & \mathbf{R}(\mathbf{z},\mathbf{m})\end{pmatrix} = \begin{pmatrix}\mathbf{z} & \mathbf{M}(\mathbf{z})\end{pmatrix}\begin{pmatrix}A & C \\ 0 & B\end{pmatrix} + \begin{pmatrix}\mathbf{E}_1(\mathbf{z}) & \mathbf{E}_2(\mathbf{z})\end{pmatrix}, \tag{8}$$

where all the potential outcomes should be understood as vectors of length $n$. We impose the following assumptions:

- (A1) stable unit treatment value assumption (SUTVA);
- (A2) model (4) correctly specified;
- (A3) the observed bivariate outcome is one realization of the potential outcomes with the observed treatment assignment vector $\mathbf{Z} = \mathbf{z}$;
- (A4) randomized treatment $\mathbf{Z}$ with $0 < \mathbb{P}(\mathbf{Z} = \mathbf{z}) < 1$ for every $\mathbf{z}$, i.e., $\mathbf{Z} \perp \{\mathbf{R}(\mathbf{z}', \mathbf{m}), \mathbf{M}(\mathbf{z})\}$ for all $\mathbf{z}$ and $\mathbf{z}'$; similarly $\mathbf{Z} \perp \{\mathbf{E}_1(\mathbf{z}), \mathbf{E}_2(\mathbf{z}')\}$.

The assumptions (A1)-(A3) are standard regularity assumptions in causal inference, see for example Rubin (1978), Holland (1988) and Imai et al. (2010). Petersen et al. (2006) considered a weaker version of the first part of assumption (A4), and these two versions are equivalent in randomized experiments. For randomized trials, assumption (A4) is automatically satisfied. Assumptions (A1), (A2) and (A4) are imposed to ensure that our approach estimates the causal coefficients consistently.

We do not impose the assumption of ignorability of the mediator, which is a standard assumption in causal mediation analysis, see a review Imai et al. (2010). A version of this assumption is written as, in Imai et al. (2010),

$$R_i(z_i', m_i) \perp M_i(z_i) \mid Z_i = z_i, \tag{9}$$

for all $z_i'$ and $z_i$, $i = 1, \ldots, n$. Under the finest fully randomized causally interpreted structured tree graph (FRCISTG) model, Robins (2003) assumed the following

$$R_i(z_i, m_i) \perp M_i(z_i) \mid Z_i = z_i, \tag{10}$$

together with the first part of assumption (A4), in order to identify causal effects. As discussed in Imai et al. (2010), assumption (10) allows for conditioning on observed post-treatment confounders, but requires an additional no-interaction assumption; while assumption (9) does not depend on post-treatment confounders and the no-interaction assumption is not required. Neither of these two versions of the ignorability assumption holds in (8) when there exists unmeasured confounding or when the errors $\mathbf{E}_1$ and $\mathbf{E}_2$ are correlated.

Momentarily we will introduce our estimator for the coefficients as a function of the correlation $\delta$, the observed bivariate outcome $(\mathbf{R}(\mathbf{Z}), \mathbf{M}(\mathbf{Z}))$ and the randomized treatment assignment $\mathbf{Z}$. Here, based on potential outcomes, we use a generic function $\hat{f}_\delta((\mathbf{R}(\mathbf{z}), \mathbf{M}(\mathbf{z})), \mathbf{z})$ to denote our estimator for a coefficient, say $C$ (or $B$). The exact formula of $\hat{f}_\delta$ will be introduced in Theorem 1 in Section 3.1. Using this generic notation, we prove that our SEM coefficients have causal interpretations because

$$\begin{aligned} \mathbb{E}\left[\hat{f}_\delta((\mathbf{R}(\mathbf{z}), \mathbf{M}(\mathbf{z})), \mathbf{z})\right] &= \mathbb{E}\left[\hat{f}_\delta((\mathbf{R}(\mathbf{z}), \mathbf{M}(\mathbf{z})), \mathbf{z}) \mid \mathbf{Z} = \mathbf{z}\right] \\ &= \mathbb{E}\left[\hat{f}_\delta((\mathbf{R}(\mathbf{Z}), \mathbf{M}(\mathbf{Z})), \mathbf{Z}) \mid \mathbf{Z} = \mathbf{z}\right], \end{aligned}$$

where the first line above uses (A4) and the second line uses (A3). Under (A1), (A2) and the second part of (A4), the last line expectation is consistently estimated using our proposed estimator, as we will describe in Section 3.

## 2.2. Higher-Level models

As motivated by the fMRI experiment, our primary interest is to infer the *population* parameters $A$, $B$ and $C$ in model (3), for either two-level or three-level data. As we will prove in Section 3.1, these parameters are not causally identifiable if one allows $\delta_{ik}$ to vary across $i$ and $k$. In order to make causal interpretation of our estimates, we will need to make the following assumption for the two-level model

- (A5) $\delta_{ik}$ is constant across participants, i.e., $\delta_{ik} = \delta$ for all $i$ and $k = 1$;

or the following for the three-level model

- (A5′) $\delta_{ik}$ is constant across participants and sessions, i.e., $\delta_{ik} = \delta$, for all $i$ and $k$.

Intuitively, either assumption allows us to pool data across levels to improve the estimation of $\delta_{ik}$. As we will prove in Section 3.1.1, $\delta_{ik}$, if allowed to vary with $i$ and $k$, is not identifiable in the likelihood sense. Therefore, these assumptions are minimal for the purpose of model identifiability.

Both assumptions (A5) and (A5′) are weaker than a multilevel mediation model proposed by Kenny et al. (2003). They assumed $\delta_{ik} = \delta = 0$ for all $i$ and $k$ in order to fit (5) and (6) separately for the first-level data, and then adopted the same model formulation for the higher-level data as ours. Their approach thus will suffer from the estimation bias due to nonzero $\delta$ or unmeasured confounding, and thus the resulting estimates may not be interpreted as causal.

Our model framework can also address the following relaxed assumption of (A5′):

- (A5″) $\delta_{ik}$ is constant across participants in session $k$, i.e., $\delta_{ik} = \delta_k$, for all $i = 1, \ldots, N$ and for $\forall\ k \in \{1, \ldots, K\}$.

This assumption essentially allows $\delta$ to vary across sessions, instead of being fixed in (A5′). It is easy to see that this assumption is equivalent to having $K$ versions of assumption (A5). To model three-level data under assumption (A5″), one only needs to fit $K$ two-level models of ours, each for the subset of data when $k = 1, \ldots, K$. Using these $K$ estimates for $\delta_k$, one can also check empirically if (A5′) holds, as we will illustrate the validity of (A5′) for our dataset in Section F.5 of the supplementary materials. Since assumption (A5″) will introduce a minor modification methodologically, we will focus on assumption (A5) and (A5′) in this paper.

Under either (A5) or (A5′), we will prove in Sections 3.2 and 3.3 respectively that $\delta_{ik}$ is identifiable and estimated consistently using our methods when the sample size goes to infinity. As shown in the previous section, our estimates for the first-level parameters $(A_{ik}, B_{ik}, C_{ik})$ are causal asymptotically when $\delta$ is replaced by our consistent estimator. Therefore, our higher-level parameters are also causal, and can be interpreted, to a certain extent, as population averages. Following the term "population inference" (Penny et al., 2003), we will call $AB_d = C' - C$ and $C$ the *population* indirect and direct effects, as they represent the causal effect estimates for a population after accounting for individual variability. Our method also produces the estimate for $AB_p = A \times B$, which is equivalent to $AB_d$ under certain conditions (Kenny et al., 2003).

## 3. Method

We propose a multilevel likelihood framework to estimate all the mediation parameters. Our likelihood criterion takes the following generic form

$$\begin{aligned} \ell &= \sum_{i=1}^{N}\sum_{k=1}^{K} \log \mathbb{P}\left(\mathbf{R}_{ik}, \mathbf{M}_{ik} | \mathbf{Z}_{ik}, \mathbf{b}_{ik}, \delta_{ik}, \sigma_{1_{ik}}, \sigma_{2_{ik}}\right) + \sum_{i=1}^{N} \log \mathbb{P}\left(\mathbf{b}_{i1}, \ldots, \mathbf{b}_{iK} | \mathbf{b}, \boldsymbol{\Lambda}, \boldsymbol{\Psi}\right) \\ &= \ell^{(1)} + \ell^{(2)}, \end{aligned} \tag{11}$$

where $\ell^{(1)}$ is the log-likelihood for the first-level model and $\ell^{(2)}$ is for the higher levels. The specific formulations of $\ell^{(1)}$ and $\ell^{(2)}$ will be introduced in the following sections. In particular, $\ell^{(2)}$ will be the regression likelihood for two-level data when $K = 1$, and the mixed effects likelihood for three-level data.

### *3.1. Method for the first-level model*

Though our integrated method is to maximize $\ell$ (asymptotically), it is worthwhile to discuss the methodological and theoretical issues related to maximizing $\ell^{(1)}$, or the first-level mediation model. As $\ell^{(1)}$ is simply the sum of the likelihood $\ell_{ik}^{(1)}$ of model (1), for every participant $i$ and session $k$, we will focus on $\ell_{ik}^{(1)}$ in this section.

The likelihood $\ell_{ik}^{(1)}$ has six parameters, $(A_{ik}, B_{ik}, C_{ik}, \delta_{ik}, \sigma_{1_{ik}}, \sigma_{2_{ik}})$, for each $i$ and $k$. The exact formulation for $\ell_{ik}^{(1)}$ is given in Section A.1 of the supplementary materials. To characterize the changes in the estimates due to $\delta$ in our model, we will consider first the case when the covariance parameters $(\delta_{ik}, \sigma_{1_{ik}}, \sigma_{2_{ik}})$ are given. The unknown covariance case will be discussed in Section 3.1.1.

THEOREM 1. *Given $(\delta_{ik}, \sigma_{1_{ik}}, \sigma_{2_{ik}})$, the solution that maximizes $\ell_{ik}^{(1)}$ is given by*

$$\begin{aligned}
\hat{A}_{ik} &= (\mathbf{Z}_{ik}^\top \mathbf{Z}_{ik})^{-1} \mathbf{Z}_{ik}^\top \mathbf{M}_{ik}, \\
\hat{C}_{ik} &= (\mathbf{Z}_{ik}^\top \boldsymbol{H}_{\mathbf{M}_{ik}} \mathbf{Z}_{ik})^{-1} \mathbf{Z}_{ik}^\top \boldsymbol{H}_{\mathbf{M}_{ik}} \mathbf{R}_{ik} + \delta_{ik} \frac{\sigma_{2_{ik}}}{\sigma_{1_{ik}}} (\mathbf{Z}_{ik}^\top \mathbf{Z}_{ik})^{-1} \mathbf{Z}_{ik}^\top \mathbf{M}_{ik}, \\
\hat{B}_{ik} &= (\mathbf{M}_{ik}^\top \mathbf{M}_{ik})^{-1} \mathbf{M}_{ik}^\top \left( \mathbf{I}_{n_{ik}} - \mathbf{Z}_{ik} (\mathbf{Z}_{ik}^\top \boldsymbol{H}_{\mathbf{M}_{ik}} \mathbf{Z}_{ik})^{-1} \mathbf{Z}_{ik}^\top \boldsymbol{H}_{\mathbf{M}_{ik}} \right) \mathbf{R}_{ik} - \delta_{ik} \frac{\sigma_{2_{ik}}}{\sigma_{1_{ik}}},
\end{aligned}$$

*where $\mathbf{I}_{n_{ik}}$ is the $n_{ik}$-dimensional identity matrix; $\boldsymbol{H}_{\mathbf{M}_{ik}} = \mathbf{I}_{n_{ik}} - \boldsymbol{P}_{\mathbf{M}_{ik}}$, and $\boldsymbol{P}_{\mathbf{M}_{ik}} = \mathbf{M}_{ik}(\mathbf{M}_{ik}^\top \mathbf{M}_{ik})^{-1} \mathbf{M}_{ik}^\top$ is the projection matrix of $\mathbf{M}_{ik}$.*

This theorem shows how $\delta_{ik}$ and the variance parameters affect the maximum likelihood estimates (MLEs) for $B_{ik}$ and $C_{ik}$ respectively, by two different additive terms related to $\delta_{ik}$. The standard Baron-Kenny estimates (Baron and Kenny, 1986) for $B_{ik}$ and $C_{ik}$ are special cases of ours by setting $\delta_{ik} = 0$ in Theorem 1. The differences between the Baron-Kenny estimates and ours are the biases due to unmeasured confounding, which are corrected in our method. Moreover, the biases of the Baron-Kenny estimates increase when $\delta_{ik}$ moves away from 0, and they are also proportional to the variance ratio $\sigma_{2_{ik}}/\sigma_{1_{ik}}$. The biases are asymptotically independent of the Baron-Kenny estimates (see the proof of Theorem 3 in Section A.6 of the supplementary materials). Finally, with no surprise, our estimates for $A_{ik}$ and the total effect $C'$ are the same as the Baron-Kenny estimates, since $\mathbf{Z}_{ik}$ is randomized.

As our estimator is an MLE, we prove its asymptotic properties in Theorem A.1, see Section A.2 of the supplementary materials. Briefly, our estimator is not only consistent but also achieves the Fisher efficiency, and thus those estimators without accounting for $\delta_{ik} \neq 0$ is asymptotically biased. By the Delta method (Sobel, 1982), we calculate the asymptotic standard errors for the product estimator $\widehat{AB}_{p_{ik}} = \hat{A}_{ik}\hat{B}_{ik}$ and the difference estimator $\widehat{AB}_{d_{ik}} = \hat{C}'_{ik} - \hat{C}_{ik}$ using the asymptotic distributions of $(\hat{A}_{ik}, \hat{B}_{ik}, \hat{C}_{ik})$ and $(\hat{C}_{ik}, \hat{C}'_{ik})$, respectively. Importantly, all these asymptotic standard errors depend on $\delta_{ik}$, see the explicit formulas in Section A.2 of the supplementary materials.

### 3.1.1. *Estimation and identifiability under unknown variances*

It has been known that the parameters $(A_{ik}, B_{ik}, C_{ik}, \delta_{ik}, \sigma_{1_{ik}}, \sigma_{2_{ik}})$ in model (1) are in general not all identifiable without additional assumptions. The non-identifiability issue can be verified using either a rank condition (Hausman, 1983) or comparing the numbers of parameters and equations (Imai et al., 2010). For completeness, we prove the non-identifiability issue in our first-level model from the likelihood perspective.

THEOREM 2. *For every fixed $\delta_{ik} \in (-1, +1)$, $i = 1, \ldots, N$ and $k = 1, \ldots, K$, $\ell_{ik}^{(1)}$ achieves the same maximum (profile) likelihood value $\ell_{ik}^{(1)}(\delta_{ik})$, where the maximum is taken over all the remaining parameters $(A_{ik}, B_{ik}, C_{ik}, \sigma_{1_{ik}}, \sigma_{2_{ik}})$. Moreover, for a given $\delta_{ik}$, the following variance estimates maximize $\ell_{ik}^{(1)}$*

$$\hat{\sigma}_{1_{ik}}^2 = \frac{1}{n_{ik}} \mathbf{M}_{ik}^\top (\mathbf{I}_{n_{ik}} - \boldsymbol{P}_{\mathbf{Z}_{ik}}) \mathbf{M}_{ik}, \quad (12)$$

$$\hat{\sigma}_{2_{ik}}^2 = \frac{1}{n_{ik}(1 - \delta_{ik}^2)} \mathbf{R}_{ik}^\top (\mathbf{I}_{n_{ik}} - \boldsymbol{P}_{\mathbf{M}_{ik}\mathbf{Z}_{ik}} - \boldsymbol{P}_{\mathbf{M}_{ik}}) \mathbf{R}_{ik}, \quad (13)$$

*where $\boldsymbol{P}_{\mathbf{Z}_{ik}} = \mathbf{Z}_{ik}(\mathbf{Z}_{ik}^\top \mathbf{Z}_{ik})^{-1}\mathbf{Z}_{ik}^\top$ and $\boldsymbol{P}_{\mathbf{M}_{ik}\mathbf{Z}_{ik}} = \boldsymbol{H}_{\mathbf{M}_{ik}}\mathbf{Z}_{ik}(\mathbf{Z}_{ik}^\top \boldsymbol{H}_{\mathbf{M}_{ik}}\mathbf{Z}_{ik})^{-1}\mathbf{Z}_{ik}^\top \boldsymbol{H}_{\mathbf{M}_{ik}}$ are projection matrices. The estimates for $A_{ik}$, $B_{ik}$, and $C_{ik}$ are obtained by plugging in the variance estimates above into Theorem 1.*

This theorem shows that the (profile) likelihood function $\ell_{ik}^{(1)}(\delta_{ik})$ achieves the same maximum value, regardless of $\delta_{ik}$. This conclusion holds for both two-level data and three-level data. We illustrate this in Fig. 2a using a simulated two-level dataset (see Section E.2 of the supplementary materials for the simulation setup), where the computed maximum (profile) likelihood value is constant with varying $\delta$. Therefore, one cannot simply maximize $\ell^{(1)}$ to estimate $\delta$. Section B of the supplementary materials also derives the relationship between all these parameters, and presents an example where two generative models with zero or nonzero indirect effects will yield the same data distribution or likelihood.

In comparison, Imai et al. (2010) considered only the first level data setting, and derived the same estimates in (12) and (13) with varying $\delta_{ik}$ as a sensitivity parameter. As shown in Theorem 1, the impact of $\delta_{ik}$ on the estimates can be large if the $\delta_{ik}$ terms dominate. Under such a situation, it is challenging to employ sensitivity analysis because the resulting estimates depend heavily on the (sometimes subjective) choice of $\delta_{ik}$. We illustrate this limitation of sensitivity analysis using our fMRI dataset in Fig. F.9 in the supplementary materials, where totally opposite conclusions can be drawn with only a moderate change in $\delta_{ik}$. Moreover,

**Algorithm 1** An approach to compute all other parameters given $\delta$ and estimate $\delta$ via maximizing the likelihood of model (3) in our two-level mediation model.

**Compute the maximized log-likelihood value of the regression model and coefficient estimates for a given $\delta$:**

1. Estimate $(\mathbf{b}_i, \sigma_{1_i}, \sigma_{2_i})$ for each $i$ using Theorem 1, (12) and (13).
2. Fit model (3) on the estimated $\hat{\mathbf{b}}_i$'s, and estimate $\mathbf{b}$ and $\mathbf{\Lambda}$ via maximum likelihood.
3. Return the maximum log-likelihood value of the regression model.

**When $\delta$ is unknown, apply an optimization algorithm (e.g., Newton's method) to maximize over $\delta$ using the maximum log-likelihood value at Step 3 above.**

the sensitivity analysis approach in Imai et al. (2010) also fails to account for individual variability, which is an important issue in datasets with multiple nested levels (Kenny et al., 2003).

This theorem also shows that it is not easy to avoid either assumption (A5) or (A5′). Suppose we assume $\delta_{ik}$ to be different across $i$ and $k$. For any population effect parameter set $(B, C)$ within a suitable range, one can pick different $\delta_{ik}$ such that for every $i$ and $k$,

$$\hat{B}_{ik} = B \text{ and } \hat{C}_{ik} = C,$$

while they together yield the same likelihood $\ell^{(1)}$ due to Theorem 2. The likelihood term $\ell^{(2)}$ is also the same for these estimates as it depends on $(A, B, C)$ and their first-level counterparts, $A_{ik}, B_{ik}, C_{ik}$, for $i = 1, \ldots, N$ and $k = 1, \ldots, K$. Therefore, there exists multiple parameter estimates such that $\ell$ is the same and the model is not identifiable in the likelihood sense. In contrast, we will prove in the next two sections that the (profile) likelihood will vary with $\delta$ if either (A5) or (A5′) is satisfied.

### *3.2. Method for the two-level model*

For our two-level model, the second term in our likelihood criterion (11) becomes

$$\ell^{(2)} = \sum_{i=1}^{N} \log \mathbb{P}\left(\mathbf{b}_i | \mathbf{b}, \mathbf{\Lambda}\right), \tag{14}$$

where the above is the log-likelihood of regression model (3).

To estimate the parameters, we first propose to maximize this likelihood via a simple two-stage algorithm. In the algorithm, the first step optimizes $(\mathbf{b}_i, \sigma_{1_i}, \sigma_{2_i})$ in the likelihood $\ell^{(1)}$ for a given $\delta_i = \delta$. The second step optimizes $(\mathbf{b}, \mathbf{\Lambda})$ in $\ell^{(2)}$ by plugging in the estimated $\mathbf{b}_i$'s from the first step. Since $\delta$ is usually unknown, we further optimize over $\delta$ to yield the largest likelihood value of the regression model. This idea is summarized in Algorithm 1.

As we prove in the following theorem, this simple algorithm is able to maximizes $\ell$ and the resulting estimate of $\delta$ is consistent asymptotically.

THEOREM 3. *Assume assumptions (A1)-(A5) are satisfied. Let* $\mathbf{Z}_i^\top \mathbf{Z}_i / n_i \to q_i < \infty$ *as* $n = \min_i n_i \to \infty$, *for* $i = 1, \ldots, N$.

*(a) If* $\mathbf{\Lambda}$ *is known, then the two-stage estimator* $\hat{\hat{\delta}}$ *maximizes the profile likelihood of model* (3) *asymptotically, and* $\hat{\hat{\delta}}$ *is* $\sqrt{Nn}$*-consistent.*
*(b) If* $\mathbf{\Lambda}$ *is unknown, then the profile likelihood of model* (3) *has a unique maximizer* $\hat{\delta}$ *asymptotically, and* $\hat{\delta}$ *is* $\sqrt{Nn}$*-consistent, provided that* $1/\varpi = \mathcal{O}_p\left(1/\sqrt{Nn}\right)$ *where* $\varpi = \bar{\kappa}^2/\varrho^2$, $\kappa_i = \sigma_{2_i}/\sigma_{1_i}$, $\bar{\kappa} = (1/N)\sum \kappa_i$, *and* $\varrho^2 = (1/N)\sum(\kappa_i - \bar{\kappa})^2$.

Compared with Theorem 2, this theorem shows that $\delta$ is identifiable under our multilevel model. The intuition is that multilevel observations provide additional information to help avoid the overparameterization issue in the first-level model. The likelihood term $\ell^{(2)}$ has a unique maximizer converging to the true $\delta$, though $\ell^{(1)}$ remains constant.

In practice, the shape of the likelihood $\ell$ with varying $\delta$ depends on the observed data. We propose to check empirically the maximum log-likelihood value of the regression model (3) as a function of $\delta$ using a simulated dataset, see Fig. 2b. For this simulated example, the (profile) likelihood $\ell^{(2)}(\delta)$ is unimodal and achieves its maximum at $\hat{\hat{\delta}} = 0.476$ (the truth is 0.5), while the first-level likelihood in Fig. 2a is flat. These numerical observations are predicted by Theorem 2 and Theorem 3.

### *3.2.1. An alternative coordinate-descent algorithm*

In general, it is challenging to find the global optimum of a generally non-convex function like our $\ell$, especially when $\ell$ contains many parameters. Though Algorithm 1 is simple and consistent for $\delta$, it may not optimize $\ell$ as a whole in finite samples. To address this issue, we propose an alternative algorithm to optimize $\ell$. This algorithm draws on the following properites of $\ell$.

THEOREM 4. *Assume* $\delta_{ik} = \delta$ *is given. The negative of log-likelihood function* (11) *is conditional convex in the parameter sets* $(\sigma_{1_i}^{-1}, \sigma_{2_i}^{-1})$, $(\mathbf{b}_i)$, $\mathbf{b}$, $\mathbf{\Lambda}^{-1}$, *respectively. The conditional optimizer for each parameter set is given in explicit forms in Section A.5 of the supplementary materials.*

Because of this theorem, we propose a block coordinate-descent algorithm to optimize over all the other parameters for a given $\delta$. Each descent step here is computed efficiently using explicit updates, see Section A.5 of the supplementary materials. When $\delta$ is unknown, we then find the $\delta$ that yields the highest maximum profile likelihood value. In practice, we find this approach usually yield estimates very close to the truth in our simulation studies, see the numerical results in Section 4 and Section E of the supplementary materials. This algorithm is summarized in Algorithm 2.

For any data, we again propose to check if $\delta$ can be uniquely determined in any data by visualizing the profile likelihood values (as $\delta$ varies) from Algorithm 2.

**Algorithm 2** An algorithm to compute all other parameters given $\delta$ and estimate $\delta$ in our two-level mediation model using the log-likelihood function (11).

**Compute the profile log-likelihood value and coefficient estimates for a given $\delta$:**

1. Estimate $(\sigma_{1_i}, \sigma_{2_i})$, $\mathbf{b}_i$, $\mathbf{b}$, and $\boldsymbol{\Lambda}$ by maximizing the log-likelihood function (11) over these remaining parameters using block coordinate descent.
2. Return the maximum log-likelihood value.

**When $\delta$ is unknown, apply an optimization algorithm (e.g., Newton's method) to maximize over $\delta$ using the profile log-likelihood value at Step 2 above.**

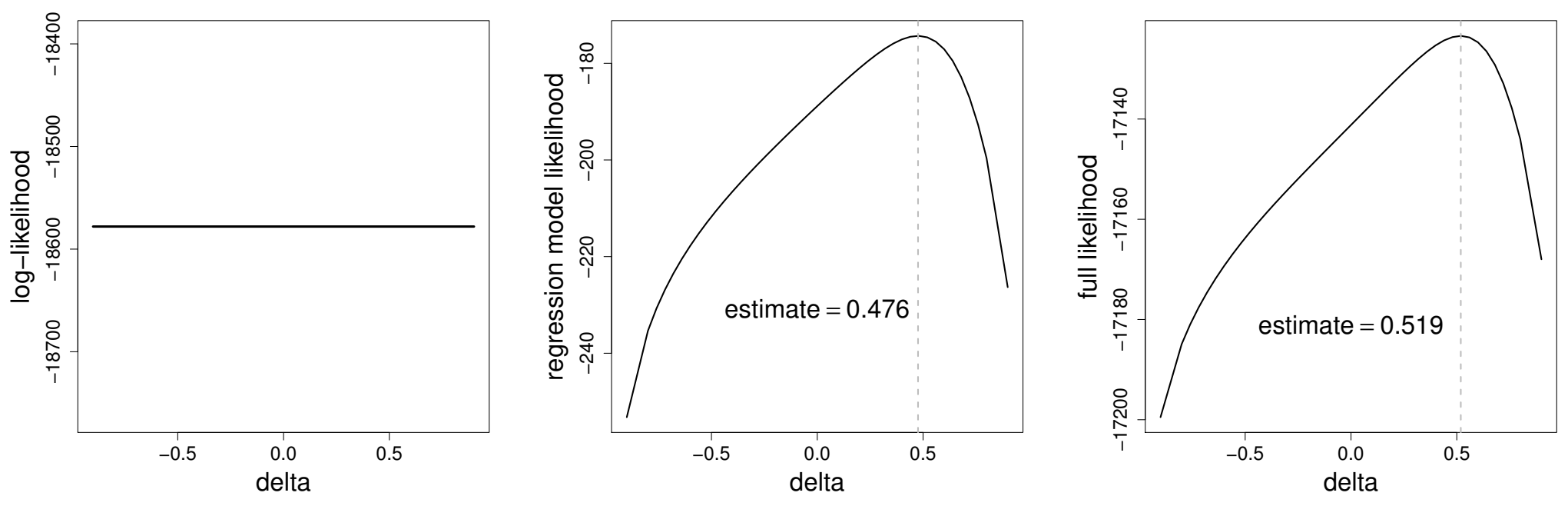

(a) (b) (c)

**Fig. 2.** The log-likelihood functions for (a) the first-level model ($\ell^{(1)}$), (b) the higher-level model ($\ell^{(2)}$), and (c) the two-level model ($\ell$) of a simulated two-level dataset. The true $\delta$ value is 0.5. The dashed lines in (b) and (c) are the estimates from the two-stage algorithm and the block coordinate-descent algorithm, respectively.

In the toy simulation example described before, the numerical value of the log-likelihood $\ell$ is a unimodal function of $\delta$, see Fig. 2c. It yields a slightly better estimate ($\hat{\delta} = 0.519$) with lower bias than Algorithm 1 does in this example.

### *3.3. Method for the three-level model*

For our three-level model, the second term of our likelihood criterion becomes

$$\ell^{(2)} = \sum_{i=1}^{N} \log \mathbb{P}\left(\mathbf{b}_{i1}, \ldots, \mathbf{b}_{iK} | \mathbf{b}, \boldsymbol{\Psi}, \boldsymbol{\Lambda}\right), \tag{15}$$

where the above is now the log-likelihood of the mixed effects model (3).

We first propose a two-stage algorithm for optimizing the likelihood criterion. This algorithm is very similar to Algorithm 1. We first optimize the first-level term $\ell^{(1)}$ with a given $\delta$, and then optimize the mixed effects likelihood $\ell^{(2)}$ using the estimated coefficients from the previous step. The modification to Algorithm 1 is to replace the model and likelihood with the mixed effects model and its likelihood, respectively. Thus the description of this algorithm is omitted here. $\delta$ is again estimated via maximizing the profile likelihood $\ell^{(2)}$ (after also maximizing over $\mathbf{\Psi}$ and $\mathbf{\Lambda}$ if unknown) from the previous two steps. The following two theorems show that this algorithm identifies $\delta$ with asymptotic consistency, when $\mathbf{\Psi}$ and $\mathbf{\Lambda}$ are either known or estimated.

THEOREM 5. *Assume (A1)-(A5′) are satisfied. Let* $\mathbf{Z}_{ik}^\top \mathbf{Z}_{ik}/n_{ik} \to q_{ik} < \infty$ *as* $n = \min_{i,k} n_{ik} \to \infty$, *for* $i = 1, \ldots, N$ *and* $k = 1, \ldots, K$. *Suppose* $\mathbf{\Psi}$ *and* $\mathbf{\Lambda}$ *in model* (3) *are known. Then, for fixed* $K$, *the two-stage estimator* $\hat{\delta}$ *maximizes the profile likelihood of the mixed effects model, and* $\hat{\delta}$ *is* $\sqrt{Nn}$*-consistent.*

THEOREM 6. *Assume (A1)-(A5′) are satisfied. Suppose that* $\mathbf{\Psi}$ *and* $\mathbf{\Lambda}$ *are estimated via profile likelihood using the mixed model. Suppose* $1/\varpi = \mathcal{O}_p\left(1/\sqrt{Nn}\right)$ *where* $\varpi = \bar{\kappa}^2/\varrho^2$, $\kappa_{ik} = \sigma_{2_{ik}}/\sigma_{1_{ik}}$, $\bar{\kappa} = (nN)^{-1}\sum \kappa_{ik}$, *and* $\varrho^2 = (nN)^{-1}\sum(\kappa_{ik} - \bar{\kappa})^2$. *Then, for fixed* $K$, *the two-stage estimator* $\hat{\delta}$ *maximizes the profile likelihood of the mixed effects model, and* $\hat{\delta}$ *is* $\sqrt{Nn}$*-consistent, provided that either one of the following conditions holds:*

*(a) if* $\lambda_\alpha^2 \geq \psi_\alpha^2$, $K \geq 2$;
*(b) if* $\lambda_\alpha^2 < \psi_\alpha^2$,

$$K \geq \frac{\lambda_\gamma^2}{\psi_\gamma^2} \cdot \frac{\lambda_\alpha^2}{\psi_\alpha^2 - \lambda_\alpha^2} + 1.$$

These two theorems show that $\delta$ is identifiable and estimated consistently under certain regularity conditions. When the covariance matrices are unknown, we also require some additional regularity conditions, such as the minimal $K$ condition, in order to estimate the variance components well. Theoretically, the minimal $K$ condition holds automatically if one sets $K \to \infty$ in a typical asymptotic analysis setting. We here consider the fixed $K$ setting because it is less restrictive for practical examples. When $K$ is fixed, the convergence rates only depend on the number of participants and the number of trials within each session. These rates of our SEM modeling are consistent with those of standard linear mixed effects models (Nie, 2007).

In practice, we propose the following approach to check the condition on $K$ for finite samples and real data. Since $\lambda_\alpha^2$ and $\psi_\alpha^2$ depend only on $A_{ik}$'s, independent of $\delta$ as shown by Theorem 1, we estimate $\lambda_\alpha^2$ and $\psi_\alpha^2$ unbiasedly from the data without knowing $\delta$. If these two estimates satisfy condition (a) in Theorem 6, then we only need to verify $K > 2$. If these two estimates satisfy condition (b) instead, we then compute the lower bounds of $K$ in the theorem by varying $\delta$ values in a range, and verify if $K$ is larger than the maximum of the lower bounds.

### *3.3.1. An alternative coordinate-descent algorithm*

Similar to Section 3.2.1, we propose an alternative coordinate-descent algorithm to explore its improvement in finite samples. Analogous to Theorem 4, we prove that the resulting likelihood, named as m-likelihood (marginal-likelihood), is also conditional convex and all the iterative updates are given in explicit forms. Moreover, these updates are given explicitly. As these results are parallel to those in Section 3.2.1, we include them in Section A.5.2 of the supplementary materials.

### *3.3.2. An alternative likelihood formulation*

In mixed effects modeling, alternative likelihood-based criteria have been proposed to improve finite sample performance. Because our framework is flexible enough, and these different criteria can be incorporated as well. Following Lee and Nelder (1996), we also consider the hierarchical-likelihood (h-likelihood) criterion, which replaces $\ell^{(2)}$ with

$$\ell^{(2)} = \sum_{i=1}^{N} \sum_{k=1}^{K} \log \mathbb{P} \left( \mathbf{b}_{ik} | \mathbf{u}_i, \mathbf{b}, \mathbf{\Lambda} \right) + \sum_{i=1}^{N} \log \mathbb{P} \left( \mathbf{u}_i | \mathbf{\Psi} \right). \tag{16}$$

This criterion includes the random effects $\mathbf{u}_i$, $i = 1, \ldots, N$. The introduction of the random effects likelihood term above is sometimes viewed as a computational approach to simplify the mixed effects likelihood, especially when it is not straightforward to integrate over this term explicitly. Lee and Nelder (1996) proved that maximizing hierarchical likelihood is asymptotically equivalent to maximizing standard likelihood, so we do not expect this will yield very different estimates from the previous two algorithms when the sample size is reasonably large. Since the last term in $\ell^{(2)}$ above is sometimes viewed as a penalty term to stabilize the estimates in finite samples, it may also introduce some estimation bias, though this is usually negligible in practice, see Commenges et al. (2009) for a review. Similar as before, we prove that the hierarchical likelihood is conditional convex and we propose a block coordinate descent algorithm to compute the estimates. These results are analogous to those in Section 3.2.1, and included in Section A.5.1 of the supplementary materials.

### *3.4. Inference*

Due to the complexity of our multilevel model, the distributions of our estimated parameters, especially the indirect effect and $\delta$, may deviate from normal in finite samples. We propose to use wild bootstrap (Wu, 1986) to compute the confidence intervals.

## 4. Simulation study

In this section, we compare our estimators with others under the three-level model when $\delta$ is unknown. The simulation results of the first-level and two-level model are presented in Section E of the online supplementary materials. The methods include our two-stage mixed effects algorithm (CMA-ts) from Section 3.3, our coordinate descent algorithm for mixed effects likelihood (CMA-m) from Section 3.3.1, our coordinate descent algorithm for hierarchical likelihood (CMA-h) from Section 3.3.2, our first-level method (CMA-$\delta$) from Section 3.1, the linear mixed effects SEM (KKB) method (Kenny et al., 2003), and the Baron-Kenny (BK) method (Baron and Kenny, 1986). Neither KKB or BK can estimate $\delta$ as they assume that there is no unmeasured confounding (or $\delta = 0$). Because our CMA-$\delta$ method allows $\delta$ as input, we will use the true $\delta$ value as input to assess the oracle performance of this single level method in multilevel data. Since both BK and CMA-$\delta$ are developed for one-level data, we apply them by concatenating the multilevel data from all sessions and all participants. Our CMA methods are implemented using our developed R package macc, KKB using the lme4 package, and BK via standard regression.

We set the total number of participants $N = 50$ and the number of sessions $K = 4$. Under each session and for each participant, the number of trials is a random draw from the Poisson distribution with mean 100. Our main objective is to identify the population direct effect (denoted by $C$) and the population indirect effect (denoted by $C' - C$ or $AB$). From the product definition of the indirect effect, the null hypothesis is $\mathrm{H}_0 : AB = 0$, which includes three scenarios for $A$ and $B$. That is, a) $A = 0$, $B \neq 0$; b) $A \neq 0$, $B = 0$; and c) $A = B = 0$. Since all methods yield unbiased estimate for $A$ (independent of $\delta$), here we only present the scenario of b) $A \neq 0$, $B = 0$ for the null of $AB$. Under the alternatives, we set the population level $A = 0.5$, $B = -1$, and $C = 0.5$. Both $\boldsymbol{\Psi}$ and $\boldsymbol{\Lambda}$ are set to be diagonal, and the variance components are $\psi_\alpha^2 = \psi_\beta^2 = \psi_\gamma^2 = 0.5$ and $\lambda_\alpha^2 = \lambda_\beta^2 = \lambda_\gamma^2 = 0.5$. For each participant in each session, the variances of the errors in the first level mediation model are $\sigma_{1_{ik}} = 1$ and $\sigma_{2_{ik}} = 2$, for $i = 1, \ldots, N$ and $k = 1, \ldots, K$. The correlation between the errors (denoted by $\delta$) is either 0.5 or 0, to simulate the settings with and without unmeasured confounding respectively. The simulation is repeated 200 times.

Table 1 presents the point estimates from all the methods considered. From the table, CMA-ts, CMA-h, and CMA-m have small biases in estimating $\delta$. CMA-h has slightly lower biases than CMA-ts and CMA-m on average. The KKB, BK and CMA-$\delta$ estimates yield large biases in $B$ and $C$, as well as in the indirect effect when the true $\delta$ is nonzero.

To validate our consistency theory and compare the finite sample performance of our algorithms, we expand the simulation of the first case in Table 1. We consider $N = n_{ik} = 50, 200, 500, 1000$ and $K = 4, 10$. Fig. 3 shows that the estimates of $\delta$ by our algorithms (CMA-ts, CMA-h, and CMA-m) approach the true values as the number of trials and the number of participants increases, and this convergence does not depend on the increase of the number of sessions, as predicted by our theory. CMA-h has the lowest bias in finite samples, probably due to the regularization term as discussed in Section 3.3.2. Fig. E.5 in the supplementary materials compares the biases in estimating the direct and indirect effects, and CMA-h also achieves the smallest biases.

To test whether our methods are robust to the magnitude of unmeasured confounding, we simulate from the previous case with varying $\delta \in (-1, 1)$. Fig. 4 shows that our methods yield lower estimation biases for $\delta$, and also lower biases than other competing methods in terms of estimating $AB$. Across different $\delta$ values, CMA-h has the lowest biases among all methods. This is also consistent with the simulation results earlier. The biases in KKB, BK and CMA-$\delta$ increase dramatically as $|\delta|$ approaches to one, while our multilevel methods have numerically negligible biases across all $\delta$ values. KKB has the largest bias, followed by BK and CMA-$\delta$, and the biases can be as large as 200%. The biases in estimating $B$, $C$ and $AB$ of BK and KKB are approximately a linear function of $\delta$, as predicted by our theory, see Section D of the supplementary materials.

**Table 1.** Average point estimates and empirical standard errors (in brackets) of CMA-ts, CMA-h CMA-m, CMA-$\delta$, KKB and BK. $\delta$ is estimated in our CMA methods. It is set to zero in KKB and BK, and set to the true $\delta$ in CMA-$\delta$.

| *Method* | $\delta$ | $C$ | $B$ | $AB_p$ | $AB_d$ |
|---|---|---|---|---|---|
| True value | 0.5 | 0.5 | -1 | -0.5 | -0.5 |
| CMA-ts | 0.476 (0.029) | 0.473 (0.121) | -0.940 (0.129) | -0.455 (0.119) | -0.450 (0.147) |
| CMA-h | 0.502 (0.031) | 0.504 (0.121) | -1.006 (0.136) | -0.488 (0.129) | -0.482 (0.153) |
| CMA-m | 0.541 (0.037) | 0.557 (0.131) | -1.117 (0.155) | -0.541 (0.143) | -0.535 (0.165) |
| CMA-$\delta$ | - | 0.719 (0.163) | -1.436 (0.127) | -0.699 (0.174) | -0.699 (0.174) |
| KKB | - | 0.016 (0.171) | 0.000 (0.105) | 0.001 (0.053) | 0.006 (0.110) |
| BK | - | 0.183 (0.173) | -0.337 (0.122) | -0.163 (0.068) | -0.163 (0.068) |
| True value | 0.5 | 0.5 | 0 | 0 | 0 |
| CMA-ts | 0.474 (0.029) | 0.474 (0.117) | 0.052 (0.132) | 0.025 (0.069) | 0.021 (0.117) |
| CMA-h | 0.499 (0.030) | 0.506 (0.117) | -0.014 (0.134) | -0.007 (0.070) | -0.011 (0.116) |
| CMA-m | 0.538 (0.040) | 0.560 (0.129) | -0.121 (0.170) | -0.062 (0.090) | -0.065 (0.132) |
| CMA-$\delta$ | - | 0.717 (0.160) | -0.442 (0.129) | -0.221 (0.091) | -0.221 (0.091) |
| KKB | - | 0.013 (0.155) | 0.985 (0.112) | 0.485 (0.125) | 0.482 (0.155) |
| BK | - | 0.174 (0.163) | 0.656 (0.124) | 0.322 (0.093) | 0.322 (0.093) |
| True value | 0 | 0.5 | -1 | -0.5 | -0.5 |
| CMA-ts | -0.001 (0.040) | 0.490 (0.125) | -0.988 (0.137) | -0.485 (0.132) | -0.479 (0.158) |
| CMA-h | -0.001 (0.044) | 0.490 (0.126) | -0.988 (0.141) | -0.485 (0.133) | -0.478 (0.158) |
| CMA-m | 0.000 (0.086) | 0.493 (0.151) | -0.991 (0.197) | -0.488 (0.154) | -0.481 (0.176) |
| CMA-$\delta$ | - | 0.498 (0.144) | -0.991 (0.120) | -0.485 (0.126) | -0.485 (0.126) |
| KKB | - | 0.491 (0.122) | -0.991 (0.108) | -0.485 (0.119) | -0.479 (0.151) |
| BK | - | 0.498 (0.144) | -0.991 (0.120) | -0.485 (0.126) | -0.485 (0.126) |

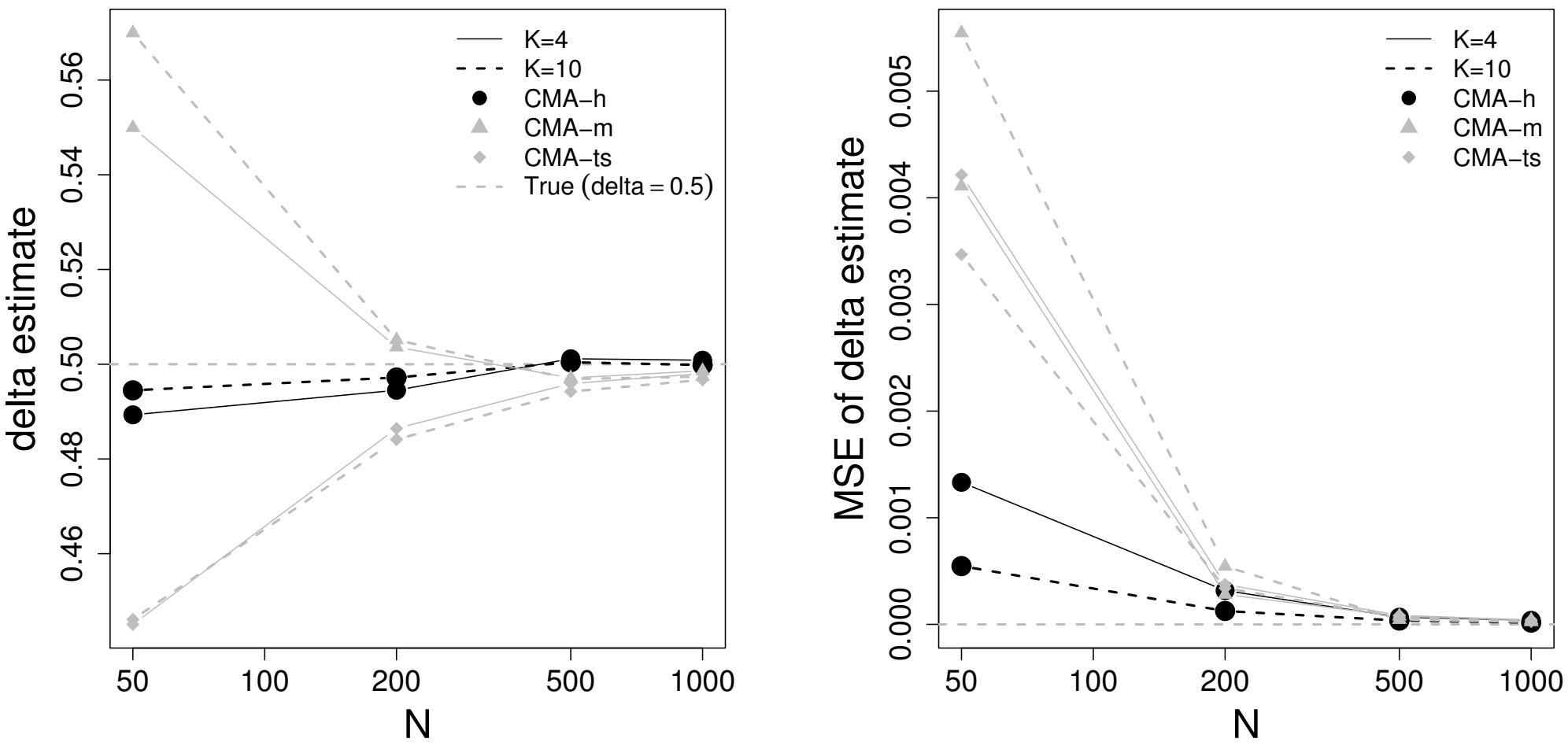

(a) (b)

**Fig. 3.** (a) Average point estimates for $\delta$ and (b) the mean squared errors for $\hat{\delta}$ by CMA-ts, CMA-h and CMA-m. The solid circles are from CMA-h, the solid triangles are from CMA-m, and the solid diamonds are from CMA-ts. The dashed line shows the true value of $\delta$ in (a), and zero in (b).

## 5. Application

We apply our proposed model and methods on an fMRI dataset. In the experiment, $N = 96$ participants consented to fMRI scanning while performing a response conflict task, where the conflict occurs between the GO trial (pressing a button when seeing a "circle" on the screen) and the STOP trial (withholding the press when seeing a "cross"). Each participant $i$, $i = 1, \ldots, N$, was scanned in $K_i = K = 4$ sessions. Each session $k$ is about ten minutes in length with $n_{ik}$ (median 90) randomized STOP/GO trials. For participant $i$ in session $k$, with probability $3/4$, the $j$th trial is a GO trial ($Z_{ikj} = 0$), and with probability $1/4$, it is a STOP trial ($Z_{ikj} = 1$). The experiment paradigm has been described in details in Duann et al.

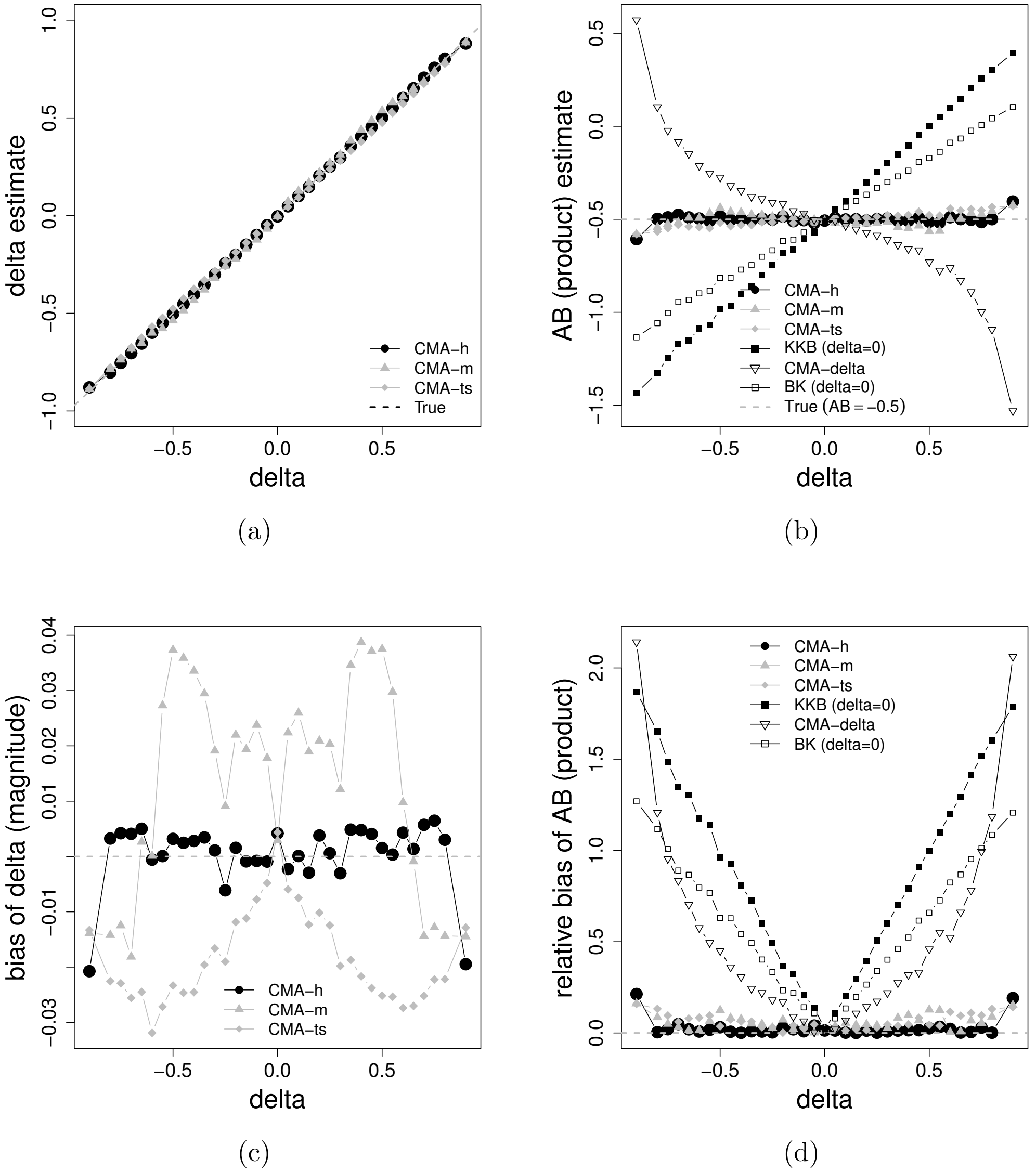

**Fig. 4.** Point estimates of (a) $\delta$ and (b) $AB$, (c) the bias of $\hat{\delta}$, and (d) the relative bias of $\widehat{AB}_p$ with varying $\delta$. The solid circles are from CMA-h, the solid triangles are from CMA-m, the solid diamonds are from CMA-ts, the solid squares are from KKB with $\delta = 0$, the triangles are from CMA-$\delta$ with the true $\delta$ value, and the squares are from BK with $\delta = 0$. The dashed lines are the true parameter values in (a) and (b), and zeros in (c) and (d).

(2009) and Luo et al. (2012). The scientists are interested in understanding how different brain regions and pathways are stimulated in this experiment. Prior modeling efforts have identified various pathways (Duann et al., 2009). In this study, we investigate a brain pathway from the presupplementary motor area (preSMA) to the primary motor cortex (PMC). The latter is a region that has been known to carry out the primary function of movements, and the former is a primary region for motor response prohibition (Duann et al., 2009). The existence of this pathway has been confirmed recently using transcranial magnetic stimulation (Obeso et al., 2013).

When the participants were performing these tasks, fMRI recorded blood-oxygen-level dependent (BOLD) signals from two brain regions, preSMA and PMC. The BOLD measures are sampled every two seconds, resulting 295 data points for each session. BOLD measures are only proxies to neural activities or loosely speaking brain activations as they are complicated by the haemodynamic response (HRF) (Lindquist, 2008). We adopt a widely used single trial analysis approach (Rissman et al., 2004; Atlas et al., 2010; Luo et al., 2012) on the averaged signals from the two brain regions. This approach removes the temporal correlations in the time series by multiplying the "pre-whitening" matrix in conjunction with filtering (Friston et al., 2000), and extracted single trial brain activations $M_{ikj}$ of preSMA and $R_{ikj}$ of PMC for each treatment $Z_{ikj}$ using a canonical HRF model. These steps are implemented in standard neuroimaging software SPM (Friston et al., 1994).

In this application, we are interested in quantifying the causal effect of the stimuli going through preSMA and the effect that does not go through preSMA. Neuroimaging analysis usually concerns about population estimates rather than individual variability. Both our two-level and three-level models are capable of estimating the population effects, and we will here focus on applying our proposed three-level model for the sake of space. Additional data analyses and validating our modeling assumptions are included in Section E of the online supplementary materials.

As described in the introduction, the mediator-outcome confounding factor may come from several sources in fMRI experiments. First, systematic errors, such as head motions, usually influence both brain activities $M$ and $R$, see the discussion in Sobel and Lindquist (2014). Second, brain activity under a task can be reliably modeled by a linear superposition of task-related activity and (spontaneous) task-unrelated activity. The task-unrelated activity is shown to account for a significant fraction of the variation in brain activity (Cole et al., 2014; Fox et al., 2006). Third, other brain regions that are not included in the model may also influence the two brain regions considered here, for example, a third region inferior frontal gyrus may influence both preSMA and PMC (Obeso et al., 2013). Since head motions are robustly estimated in standard neuroimaging processing pipelines, we will treat head motions as a source of confounding that can be calculated, and will use two analyses with and without adjusting for head motions to assess the robustness of our methods.

In the first analysis, we intentionally do not adjust head motions using regression (Rosenbaum et al., 2002), and thus head motions contribute to unmeasured confounding. Fig. 5a presents the estimated $\delta$ using our CMA methods. All CMA methods yield similar estimates for $\delta$, which are all far from zero. The differences between them are consistent with the simulation results for finite samples, where CMA-ts and CMA-m slightly underestimate and overestimate $\delta$, respectively. Table 2 and Table F.4 in the supplementary materials compare the inference results of different one-level and three-level methods. Our three-level CMA methods yield very different results from other methods. For example, CMA-h yields the estimates -0.465 (95% confidence interval: -0.578, -0.325) for $\delta$, $0.293(0.237, 0.350)$ for $AB_p$, $0.273(0.215, 0.331)$ for $AB_d$ (see Table F.4), and $-0.177(-0.247, -0.108)$ for $C$ with 500 wild bootstrap samples. Because both KKB and BK ignored the unmeasured confounding effect, the indirect effect estimates by KKB and BK are about 50% less of ours. The direct effect estimates of CMA-ts, CMA-h and CMA-m are significant at the 5% level, while the KKB and BK estimates are not significant.

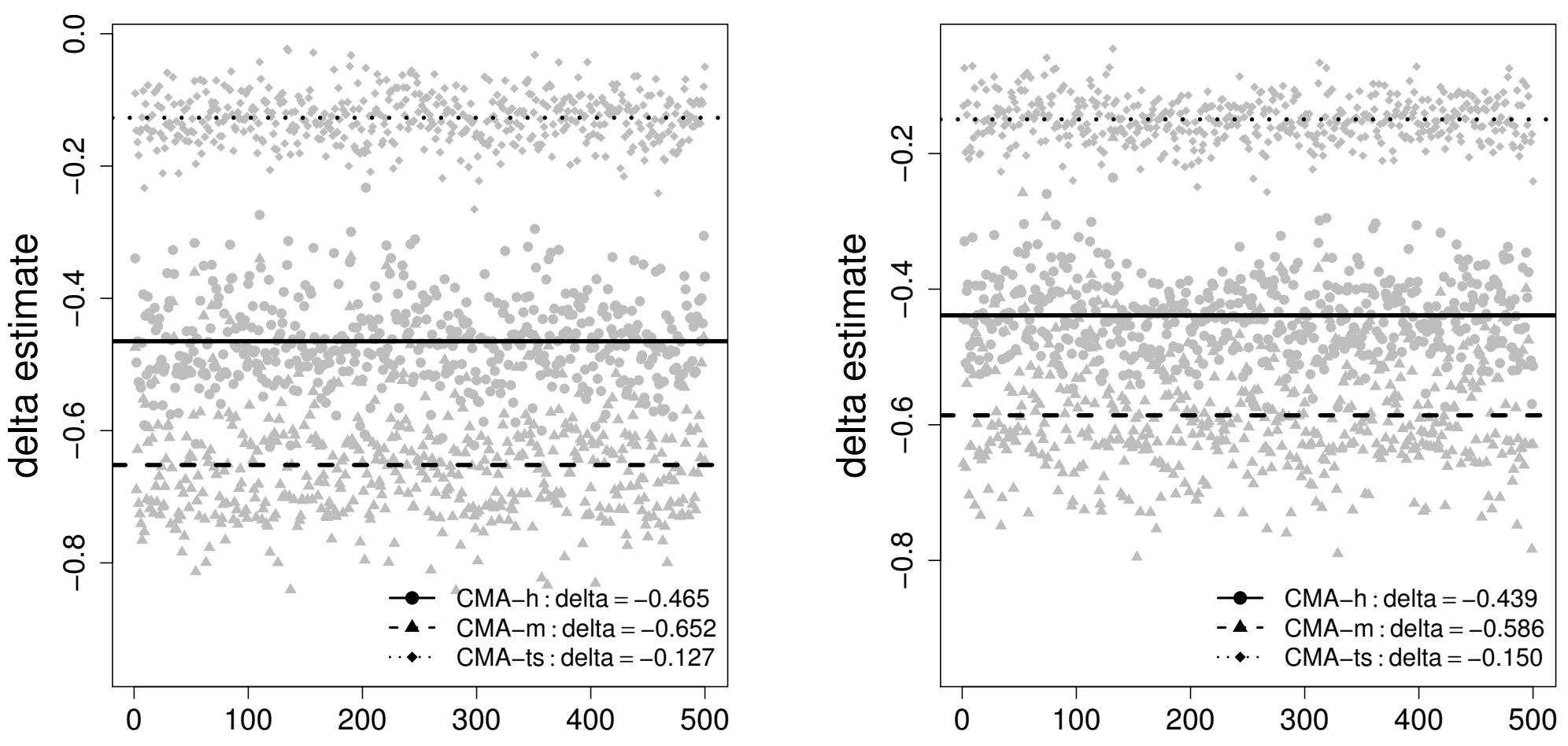

(a) (b)

**Fig. 5.** 500 bootstrapped $\delta$ estimates for the fMRI data (a) without motion correction and (b) with motion correction using our proposed CMA methods. The solid circles are from CMA-h, the solid triangles are from CMA-m, and the solid diamonds are from CMA-ts. The lines are the bootstrap averages.

Our methods also lead to important and interpretable scientific conclusions. The results of our methods show that the STOP stimulus increases the preSMA activity which then further increases the PMC activity via the preSMA-PMC pathway, while the STOP stimulus directly decreases the PMC activity. The indirect effect estimates are about two folds larger than the direct effect, which quantifies the important role of preSMA in motor prohibition. The direct effect estimates are negative (and significant at the 5% level), and this is consistent with the expectation that the participants are instructed to withhold motor movement during the STOP trials. As a comparison, KKB and BK without accounting for unmeasured confounding, not only underestimate the role of preSMA, but also miss the significant

**Table 2.** Average estimates and 95% confidence intervals from our multilevel CMA methods, CMA-$\delta$ method with $\delta$ estimated from CMA-h, KKB and BK on the fMRI dataset with and without motion correction (MC), using 500 bootstrap samples.

| *Data* | *Method* | $\delta$ | $C$ | $AB_p$ |
|---|---|---|---|---|
| Without MC | CMA-ts | -0.127 (-0.201, -0.056) | -0.051 (-0.099, -0.002) | 0.166 (0.142, 0.191) |
| | CMA-h | -0.465 (-0.578, -0.325) | -0.177 (-0.247, -0.108) | 0.293 (0.237, 0.350) |
| | CMA-m | -0.652 (-0.794,-0.456) | -0.288 (-0.412, -0.164) | 0.406 (0.289, 0.523) |
| | CMA-$\delta$ | - | -0.214 (-0.289, -0.138) | 0.295 (0.231, 0.359) |
| | KKB | - | -0.007 (-0.050, 0.036) | 0.123 (0.119, 0.127) |
| | BK | - | -0.029 (-0.071, 0.012) | 0.111 (0.107, 0.115) |
| With MC | CMA-ts | -0.149 (-0.212, -0.04) | -0.074 (-0.122, -0.025) | 0.155 (0.134, 0.176) |
| | CMA-h | -0.439 (-0.532, -0.322) | -0.178 (-0.239, -0.117) | 0.260 (0.215, 0.306) |
| | CMA-m | -0.586 (-0.730, -0.410) | -0.253 (-0.357, -0.150) | 0.336 (0.241, 0.432) |
| | CMA-$\delta$ | - | -0.225 (-0.290, -0.160) | 0.261 (0.210, 0.312) |
| | KKB | - | -0.023 (-0.069, 0.022) | 0.106 (0.102, 0.110) |
| | BK | - | -0.061 (-0.105, 0.017) | 0.097 (0.093, 0.101) |

direct prohibition effect.

In the second analysis, we validate the robustness of our method against varying magnitude of unmeasured confounding. We now adjust head motions in the data processing step, which is expected to reduce the magnitude of unmeasured confounding. Indeed, both CMA-h and CMA-m yield slightly smaller estimates of $\delta$ in magnitude after correcting for head motions, though this change is not significant. Our estimates for other parameters are similar to before, see Table 2 and Table F.4 for the details. This confirms that our method is stable under varying magnitude of confounding. In contrast, the indirect effect estimates by KKB and BK are outside the corresponding confidence intervals after motion correction. This suggests that the KKB and BK estimates are sensitive to whether head motions, a known confounding factor, is adjusted.

## 6. Discussion

In this study, we propose a multilevel mediation modeling framework for data with a hierarchically nested structure. This framework simultaneously address the unmeasured confounding issue and individual variation in causal mediation analysis. For our one-level, two-level, and three-level mediation models, we introduce a few optimization-based methods along with efficient algorithms for computing multiple model parameters, especially a large number of parameters in the multilevel mediation models. We prove that these methods will estimate consistently the magnitude of unmeasured confounding in our multilevel models, which is mathematically impossible for single-level mediation models. We use extensive simulations and a real fMRI dataset to show that the resulting estimates correct for biases effectively in finite samples, and are robust to the magnitude of the unmeasured confounding effect. Though the method is motivated by an fMRI experiment, the methodology can certainly be extended to many other studies when the correlated errors are believed to be present due to confounding.

The proposed framework may be extended to other types of mixed effects models for multilevel data. Given that there are many variants in the literature on specifying the random effects and covariance structures, we will leave to future research the analysis of more complex multilevel SEMs. For example, it is interesting to study if $\delta_{ik}$ can also be treated in a mixed effects model. Moreover, it is also interesting to test whether the finite sample estimation accuracy can be improved by using a global optimization algorithm.

For randomized studies with covariates, one can conduct covariate adjustment on both the mediator and the outcome first and then apply our method, similar to what we have done on the fMRI data with motion correction. Another option is to include the covariates in our first-level model. However, this will make the computation more challenging as the number of parameters will increase. The optimization problem also becomes more complicated when there are different covariates at different levels, see Section C of the supplementary materials for more discussions. Both approaches for covariate adjustment can be applied to observational studies under certain assumptions, see Rosenbaum et al. (2002) and Section C of the supplementary materials.

We also leave to future work on various extensions of our proposed framework, including interactions (MacKinnon et al., 2007; Valeri and VanderWeele, 2013), and functional mediation (Lindquist, 2012). It is also possible to consider mediator and outcome from other distributions, as our framework is based on maximum likelihood. However, it is not straightforward to quantify the unmeasured confounding effect using correlations in other distributions. It is also interesting to study nonlinear mediation models as well.

## 7. Supplementary materials

In the online supplementary, we include additional theorems with their proofs, and additional simulation and real data analysis results.

# SUPPLEMENT TO ESTIMATING CAUSAL MEDIATION EFFECTS UNDER CORRELATED ERRORS

YI ZHAO AND XI LUO

## A Theory and proof

### A.1 Proof of Theorem 1

*Proof.* To keep the following proof uncluttered, we drop the participant index $i$ and session index $k$ hereafter. Under model (4), the conditional distribution of $\mathbf{M}$ and $\mathbf{R}$ is

$$\begin{aligned} \mathbf{M} \mid \mathbf{Z} &\sim \mathcal{N}\left(\mathbf{Z}A, \sigma_1^2 \mathbf{I}_n\right), \\ \mathbf{R} \mid \mathbf{M}, \mathbf{Z} &\sim \mathcal{N}\left(\mathbf{Z}C + \mathbf{M}B + \kappa(\mathbf{M} - \mathbf{Z}A), \sigma_2^2(1-\delta^2)\mathbf{I}_n\right), \end{aligned}$$

where $\kappa = \sigma_2/\sigma_1$. The log-likelihood function is

$$\begin{aligned} \ell &= -\frac{n}{2}\log \sigma_1^2\sigma_2^2(1-\delta^2) - \frac{1}{2\sigma_1^2}(\mathbf{M} - \mathbf{Z}A)^\top(\mathbf{M} - \mathbf{Z}A) \\ &\quad - \frac{1}{2\sigma_2^2(1-\delta^2)}\left((\mathbf{R} - \mathbf{M}B - \mathbf{Z}C) - \kappa(\mathbf{M} - \mathbf{Z}A)\right)^\top \left((\mathbf{R} - \mathbf{M}B - \mathbf{Z}C) - \kappa(\mathbf{M} - \mathbf{Z}A)\right). \end{aligned}$$

To maximize the log-likelihood function, solving the following

$$\begin{aligned} \frac{\partial \ell}{\partial A} &= \frac{1}{\sigma_1^2}\mathbf{Z}^\top(\mathbf{M} - \mathbf{Z}A) - \frac{1}{\sigma_2^2(1-\delta^2)}\kappa\mathbf{Z}^\top\left((\mathbf{R} - \mathbf{M}B - \mathbf{Z}C - \kappa\mathbf{M}) + \kappa\mathbf{Z}A\right) = 0, \\ \frac{\partial \ell}{\partial C} &= \frac{1}{\sigma_2^2(1-\delta^2)}\mathbf{Z}^\top\left((\mathbf{R} - bMB - \kappa(\mathbf{M} - \mathbf{Z}A)) - \mathbf{Z}C\right) = 0, \\ \frac{\partial \ell}{\partial B} &= \frac{1}{\sigma_2^2(1-\delta^2)}\mathbf{M}^\top\left((\mathbf{R} - \mathbf{Z}C - \kappa(\mathbf{M} - \mathbf{Z}A)) - \mathbf{M}B\right) = 0, \end{aligned}$$

yields

$$\begin{aligned} \hat{A} &= (\mathbf{Z}^\top\mathbf{Z})^{-1}\mathbf{Z}^\top\mathbf{M}, \\ \hat{C} &= (\mathbf{Z}^\top(\mathbf{I}_n - \boldsymbol{P}_{\mathbf{M}})\mathbf{Z})^{-1}\mathbf{Z}^\top(\mathbf{I}_n - \boldsymbol{P}_{\mathbf{M}})\mathbf{R} + \delta\frac{\sigma_2}{\sigma_1}(\mathbf{Z}^\top\mathbf{Z})^{-1}\mathbf{Z}^\top\mathbf{M}, \\ \hat{B} &= (\mathbf{M}^\top\mathbf{M})^{-1}\mathbf{M}^\top\left(\mathbf{I}_n - \mathbf{Z}(\mathbf{Z}^\top(\mathbf{I}_n - \boldsymbol{P}_{\mathbf{M}})\mathbf{Z})^{-1}\mathbf{Z}^\top(\mathbf{I}_n - \boldsymbol{P}_{\mathbf{M}})\right)\mathbf{R} - \delta\frac{\sigma_2}{\sigma_1}. \end{aligned}$$

□

### A.2 Asymptotic property of the first-level model estimators introduced in Theorem 1

In this section, we discuss the asymptotic properties of our estimators for the parameters in the first-level model. For convenience, the participant index $i$ and session index $k$ are omitted hereafter.

### A.2.1 Asymptotic property of $(\hat{A}, \hat{C}, \hat{B})$

**Theorem A.1.** *Assume (A1), (A2) and (A4). Suppose* $\mathbf{Z}^\top\mathbf{Z}/n \to q < \infty$ *as* $n \to \infty$*, then the estimators in Theorem 1 converge asymptotically as*

$$\sqrt{n}\left(\begin{pmatrix}\hat{A}\\ \hat{C}\\ \hat{B}\end{pmatrix} - \begin{pmatrix}A\\ C\\ B\end{pmatrix}\right) \xrightarrow{\mathcal{D}} \mathcal{N}\left(\mathbf{0}, \boldsymbol{V}\right),$$

*where* $\boldsymbol{V}$ *is the inverse Fisher's information matrix of* $(A, C, B)$*,*

$$\boldsymbol{V} = \begin{pmatrix} \sigma_1^2/q & \delta\sigma_1\sigma_2/q & 0 \\ \delta\sigma_1\sigma_2/q & \sigma_2^2(qA^2 + \sigma_1^2 - qA^2\delta^2)/q\sigma_1^2 & -A\sigma_2^2(1-\delta^2)/\sigma_1^2 \\ 0 & -A\sigma_2^2(1-\delta^2)/\sigma_1^2 & \sigma_2^2(1-\delta^2)/\sigma_1^2 \end{pmatrix}.$$

*Proof.* The Fisher information matrix of the coefficients $A$, $C$ and $B$ is

$$I(A, C, B) = \frac{1}{\sigma_1^2\sigma_2^2(1-\delta^2)} \begin{pmatrix} \sigma_2^2\mathbf{Z}^\top\mathbf{Z} & -\delta\sigma_1\sigma_2\mathbf{Z}^\top\mathbf{Z} & -\delta\sigma_1\sigma_2 A\mathbf{Z}^\top\mathbf{Z} \\ -\delta\sigma_1\sigma_2\mathbf{Z}^\top\mathbf{Z} & \sigma_1^2\mathbf{Z}^\top\mathbf{Z} & \sigma_1^2 A\mathbf{Z}^\top\mathbf{Z} \\ -\delta\sigma_1\sigma_2 A\mathbf{Z}^\top\mathbf{Z} & \sigma_1^2 A\mathbf{Z}^\top\mathbf{Z} & \sigma_1^2 A^2\mathbf{Z}^\top\mathbf{Z} + n\sigma_1^4 \end{pmatrix}.$$

The inverse of the Fisher information matrix is

$$I^{-1}(A, C, B) = \begin{pmatrix} n\sigma_1^2/\mathbf{Z}^\top\mathbf{Z} & n\delta\sigma_1\sigma_2/\mathbf{Z}^\top\mathbf{Z} & 0 \\ n\delta\sigma_1\sigma_2/\mathbf{Z}^\top\mathbf{Z} & \sigma_2^2(A^2\mathbf{Z}^\top\mathbf{Z} + n\sigma_1^2 - \delta^2A^2\mathbf{Z}^\top\mathbf{Z})/\sigma_1^2\mathbf{Z}^\top\mathbf{Z} & -\sigma_2^2(1-\delta^2)A/\sigma_1^2 \\ 0 & -\sigma_2^2(1-\delta^2)A/\sigma_1^2 & \sigma_2^2(1-\delta^2)/\sigma_1^2 \end{pmatrix}.$$

Since $\hat{A}$, $\hat{C}$ and $\hat{B}$ are MLEs of the coefficient parameters, they are consistent. Under the assumption that $\mathbf{Z}^\top\mathbf{Z}/n \to q$ as $n \to \infty$, the asymptotic joint distribution of $(\hat{A}, \hat{C}, \hat{B})$ is

$$\sqrt{n}\left(\begin{pmatrix}\hat{A}\\ \hat{C}\\ \hat{B}\end{pmatrix} - \begin{pmatrix}A\\ C\\ B\end{pmatrix}\right) \xrightarrow{\mathcal{D}} \mathcal{N}\left(\mathbf{0}, \boldsymbol{V}(A, C, B)\right),$$

where

$$\boldsymbol{V}(A, C, B) = \begin{pmatrix} \sigma_1^2/q & \delta\sigma_1\sigma_2/q & 0 \\ \delta\sigma_1\sigma_2/q & \sigma_2^2(qA^2 + \sigma_1^2 - A^2\delta^2 q)/q\sigma_1^2 & -A\sigma_2^2(1-\delta^2)/\sigma_1^2 \\ 0 & -A\sigma_2^2(1-\delta^2)/\sigma_1^2 & \sigma_2^2(1-\delta^2)/\sigma_1^2 \end{pmatrix}.$$

□

### A.2.2 Asymptotic property of $(\hat{C}, C')$

**Theorem A.2.** *Under the same conditions in Theorem A.1, the estimator* $\hat{C}'$ *and the estimator* $\hat{C}$ *in Theorem 1 converge as*

$$\sqrt{n}\left(\begin{pmatrix}\hat{C}'\\ \hat{C}\end{pmatrix} - \begin{pmatrix}C + AB\\ C\end{pmatrix}\right) \xrightarrow{\mathcal{D}} \mathcal{N}\left(\mathbf{0}, \boldsymbol{V}'\right),$$

*where*

$$\boldsymbol{V}' = \begin{pmatrix} (B^2\sigma_1^2 + 2B\delta\sigma_1\sigma_2 + \sigma_2^2)/q & (\sigma_2^2 + B\delta\sigma_1\sigma_2)/q \\ (\sigma_2^2 + B\delta\sigma_1\sigma_2)/q & \sigma_2^2(qA^2 + \sigma_1^2 - qA^2\delta^2)/q\sigma_1^2 \end{pmatrix}.$$

*Proof.* Similar to the proof of Theorem A.1. □

### A.2.3 Asymptotic property of $\widehat{AB}_p$ and $\widehat{AB}_d$

**Corollary A.1.** *Under the same conditions in Theorem A.1, the two estimators of mediation effect, $\widehat{AB}_p$ and $\widehat{AB}_d$, are asymptotically equivalent. The asymptotic distribution of $\widehat{AB}_d$ (or $\widehat{AB}_p$) is*

$$\sqrt{n}\left(\widehat{AB}_d - AB\right) \xrightarrow{\mathcal{D}} \mathcal{N}\left(0, \frac{\sigma_1^2}{q}B^2 + \frac{\sigma_2^2(1-\delta^2)}{\sigma_1^2}A^2\right). \tag{A.1}$$

*Proof.* This can be proved through multivarite Delta method. □

## A.3 Proof of Theorem 2

*Proof.* Here, for simplicity,the participant index $i$ and session index $k$ are omitted.

Given $\delta$, $\sigma_1$ and $\sigma_2$, $A$, $C$ and $B$ are estimated by Theorem 1, and yield the profile likelihood of $(\delta, \sigma_1, \sigma_2)$ as

$$\ell\left(\sigma_1, \sigma_2, \delta | \mathbf{Z}, \mathbf{M}, \mathbf{R}\right) = -\frac{n}{2}\log \sigma_1^2\sigma_2^2(1-\delta^2) - \frac{1}{2\sigma_1^2}\mathbf{M}^\top(\mathbf{I}_n - \boldsymbol{P}_\mathbf{Z})\mathbf{M} - \frac{1}{2\sigma_1^2(1-\delta^2)}\mathbf{R}^\top(\mathbf{I}_n - \boldsymbol{P}_{\mathbf{MZ}} - \boldsymbol{P}_\mathbf{M})\mathbf{R}.$$

This gives the estimator of $\sigma_1$ and $\sigma_2$ with given $\delta$,

$$\begin{aligned} \hat{\sigma}_1^2 &= \frac{1}{n}\mathbf{M}^\top(\mathbf{I}_n - \boldsymbol{P}_\mathbf{Z})\mathbf{M}, \\ \hat{\sigma}_2^2 &= \frac{1}{n(1-\delta^2)}\mathbf{R}^\top(\mathbf{I}_n - \boldsymbol{P}_{\mathbf{MZ}} - \boldsymbol{P}_\mathbf{M})\mathbf{R}. \end{aligned}$$

Plug in these estimators, the profile log-likelihood of $\delta$ is

$$\ell\left(\delta | \mathbf{Z}, \mathbf{M}, \mathbf{R}\right) = -\frac{n}{2}\log\left(\frac{1}{n^2}\left(\mathbf{M}^\top(\mathbf{I}_n - \boldsymbol{P}_\mathbf{Z})\mathbf{M}\right)\left(\mathbf{R}^\top(\mathbf{I}_n - \boldsymbol{P}_{\mathbf{MZ}} - \boldsymbol{P}_\mathbf{M})\mathbf{R}\right)\right) - n,$$

which is a constant function of $\delta$. □

## A.4 Another way of estimating $(\sigma_{1_{ik}}, \sigma_{2_{ik}})$ with given $\delta_{ik}$

In this section, we discuss another approach of calculating the estimator of $(\sigma_{1_{ik}}, \sigma_{2_{ik}})$ with given $\delta_{ik}$ in the first-level model. Again, here, for convenience, we drop the participant index $i$ and session index $k$. We consider the following moment calculation, since MLEs are also moment estimators under Gaussian distributions. Intuitively, the cause of non-identifiability is that the number of the parameters in our mediation model is one more than the number of moment equations. Plugging (5) into (6), it yields

$$\mathbf{R} = \mathbf{Z}C' + \mathbf{E}_1 B + \mathbf{E}_2 = \mathbf{Z}C' + \mathbf{E}'. \tag{A.2}$$

It has been known that $A$ and $C'$ can be estimated unbiasedly using regression based on (5) and (7). The covariance matrix $\mathbf{\Sigma}_B$ of $(E_1, E')$ can be then estimated using the sample covariance of the residuals as

$$\hat{\mathbf{\Sigma}}_B = \frac{1}{n}\begin{pmatrix} (\mathbf{M} - \mathbf{Z}\hat{A})^\top(\mathbf{M} - \mathbf{Z}\hat{A}) & (\mathbf{M} - \mathbf{Z}\hat{A})^\top(\mathbf{R} - \mathbf{Z}\hat{C}') \\ (\mathbf{R} - \mathbf{Z}\hat{C}')^\top(\mathbf{M} - \mathbf{Z}\hat{A}) & (\mathbf{R} - \mathbf{Z}\hat{C}')^\top(\mathbf{R} - \mathbf{Z}\hat{C}') \end{pmatrix}. \tag{A.3}$$

Since $\hat{A}$ can be unbiasedly estimated, independent of $\delta$, we estimate $\sigma_1^2$ by

$$\hat{\sigma}_1^2 = \hat{\mathbf{\Sigma}}_B\left(1, 1\right), \tag{A.4}$$

where $\hat{\mathbf{\Sigma}}_B\left(i,j\right)$ is the $(i,j)$th entry of $\hat{\mathbf{\Sigma}}_B$. From (A.2), the two $\delta$-related entries in the population covariance $\mathbf{\Sigma}_B$ are

$$\begin{aligned}\mathrm{Cov}\left(\mathbf{E}_1, \mathbf{E}'\right) &= B\sigma_1^2 + \delta\sigma_1\sigma_2, & \text{(A.5)}\\ \mathrm{Var}\left(\mathbf{E}'\right) &= B^2\sigma_1^2 + 2B\delta\sigma_1\sigma_2 + \sigma_2^2. & \text{(A.6)}\end{aligned}$$

Therefore, the three parameters $(B, \sigma_2, \delta)$ cannot be uniquely determined from the two equations above. It is also easy to see that $(B, \sigma_2)$ can be uniquely determined from the two equations once $\delta$ is fixed. Given $\delta$ and $\hat{\mathbf{\Sigma}}_B$, $(B, \sigma_2)$ can be estimated from (A.5) and (A.6) by

$$\hat{B} = \frac{\hat{\mathbf{\Sigma}}_B(1,2)}{\hat{\sigma}_1^2} - \frac{\delta}{\hat{\sigma}_1^2\sqrt{1-\delta^2}}\sqrt{\hat{\sigma}_1^2\hat{\mathbf{\Sigma}}_B(2,2) - \hat{\mathbf{\Sigma}}_B^2(1,2)} \tag{A.7}$$

and

$$\hat{\sigma}_2^2 = \frac{1}{\hat{\sigma}_1^2(1-\delta^2)}\left[\hat{\sigma}_1^2\hat{\mathbf{\Sigma}}_B(2,2) - \hat{\mathbf{\Sigma}}_B^2(1,2)\right]. \tag{A.8}$$

This formulation for estimating $(\sigma_1, \sigma_2)$ will be used in the proof of Theorems 3, 5 and 6.

## A.5 The conditional convexity of the log-likelihood function (11)

Since the two-level likelihood function is a special case of the three-level likelihood function, in this section, we will focus on the three-level model. For the higher-level mixed effects model, we consider two formulations of the likelihood function, i.e., the marginal likelihood (m-likelihood) and the hierarchical likelihood (h-likelihood). Here, we will discuss the conditional convexity and provide the iterative updates of the block coordinate-descent algorithm under these two formulations separately.

### A.5.1 The h-likelihood algorithm

Under assumption (A5′) that $\delta_{ik} = \delta$ (for all $i = 1, \ldots, N$ and $k = 1, \ldots, K$), replacing the higher-level model likelihood in function (11) with the h-likelihood function (16), we have for the three-level model,

$$\begin{aligned}\ell_h &= \sum_{i=1}^{N}\sum_{k=1}^{K}\log\mathbb{P}\left(\mathbf{R}_{ik}, \mathbf{M}_{ik}|\mathbf{Z}_{ik}, \delta, \mathbf{b}_{ik}, \sigma_{1_{ik}}, \sigma_{2_{ik}}\right) + \sum_{i=1}^{N}\sum_{k=1}^{K}\log\mathbb{P}\left(\mathbf{b}_{ik}|\mathbf{u}_i, \mathbf{b}, \mathbf{\Lambda}\right) + \sum_{i=1}^{N}\log\mathbb{P}\left(\mathbf{u}_i|\mathbf{\Psi}\right)\\ &= \ell^{(1)} + \ell_h^{(2)}. \end{aligned} \tag{A.9}$$

**Theorem A.3.** *Assume $\delta$ is given. The negative h-likelihood function above is conditional convex in the parameter sets, $(\sigma_{1_{ik}}^{-1}, \sigma_{2_{ik}}^{-1})$, $(\mathbf{b}_{ik})$, $(\mathbf{u}_i)$, $\mathbf{\Psi}^{-1}$, $\mathbf{\Lambda}^{-1}$, respectively.*

*Proof.* Since $\sigma_{1_{ik}}$ and $\sigma_{2_{ik}}$ are in the $\ell^{(1)}$ function and separable, we only need to prove the convexity of the negative log-likelihood under the first-level model. For the first-level model, by omitting $i$ and $k$,

$$\boldsymbol{Y} = \boldsymbol{X}\mathbf{\Theta} + \boldsymbol{E}, \quad \mathrm{vec}\left[\boldsymbol{E}\right] \sim \mathcal{N}\left(\mathbf{0}, \mathbf{\Sigma}\otimes\mathbf{I}_n\right),$$

where

$$\boldsymbol{Y} = \begin{pmatrix}\mathbf{M} & \mathbf{R}\end{pmatrix}, \quad \boldsymbol{X} = \begin{pmatrix}\mathbf{Z} & \mathbf{M}\end{pmatrix}, \quad \mathbf{\Theta} = \begin{pmatrix}A & C\\ 0 & B\end{pmatrix}, \quad \boldsymbol{E} = \begin{pmatrix}\mathbf{E}_1 & \mathbf{E}_2\end{pmatrix}.$$

The covariance matrix $\mathbf{\Sigma}$ can be written as

$$\mathbf{\Sigma} = \begin{pmatrix}\sigma_1^2 & \delta\sigma_1\sigma_2\\ \delta\sigma_1\sigma_2 & \sigma_2^2\end{pmatrix} = \boldsymbol{D}\mathbf{\Omega}\boldsymbol{D}, \quad \boldsymbol{D} = \begin{pmatrix}\sigma_1 & 0\\ 0 & \sigma_2\end{pmatrix}, \quad \mathbf{\Omega} = \begin{pmatrix}1 & \delta\\ \delta & 1\end{pmatrix}.$$

The log-likelihood function of the first-level model is

$$\begin{aligned}\ell^{(1)} &= -n\log\det(\boldsymbol{\Sigma}) - \mathrm{tr}\left[(\boldsymbol{Y}-\boldsymbol{X\Theta})\boldsymbol{\Sigma}^{-1}(\boldsymbol{Y}-\boldsymbol{X\Theta})^\top\right] \\ &= n\log\det(\boldsymbol{D}^{-1}\boldsymbol{\Omega}^{-1}\boldsymbol{D}^{-1}) - \mathrm{tr}\left[\boldsymbol{D}^{-1}\boldsymbol{\Omega}^{-1}\boldsymbol{D}^{-1}(\boldsymbol{Y}-\boldsymbol{X\Theta})^\top(\boldsymbol{Y}-\boldsymbol{X\Theta})\right].\end{aligned}$$

Let $\boldsymbol{S} = (\boldsymbol{Y}-\boldsymbol{X\Theta})^\top(\boldsymbol{Y}-\boldsymbol{X\Theta})$, then

$$\ell^{(1)} = 2n\log\sigma_1^{-1}+2n\log\sigma_2^{-1}+n\log(\Omega_{11}^{-1}\Omega_{22}^{-1}-\Omega_{12}^{-2})-\left(\sigma_1^{-2}\Omega_{11}^{-1}S_{11}+2\sigma_1^{-1}\sigma_2^{-1}\Omega_{12}^{-1}S_{12}+\sigma_2^{-2}\Omega_{22}^{-1}S_{22}\right),$$

where $\Omega_{ij}^{-1}$ is the $(i,j)$ element of $\boldsymbol{\Omega}^{-1}$ and $S_{ij}$ is the $(i,j)$ element of $\boldsymbol{S}$. The Hessian matrix of $(\sigma_1^{-1}, \sigma_2^{-1})$ is

$$\begin{pmatrix} n/\sigma_1^{-2}+\Omega_{11}^{-1}S_{11} & \Omega_{12}^{-1}S_{12} \\ \Omega_{12}^{-1}S_{12} & n/\sigma_2^{-2}+\Omega_{22}^{-1}S_{22}\end{pmatrix} = \begin{pmatrix} n/\sigma_1^{-2} & 0 \\ 0 & n/\sigma_2^{-2}\end{pmatrix} + \boldsymbol{\Omega}^{-1}\circ\boldsymbol{S},$$

where $\boldsymbol{\Omega}^{-1}\circ\boldsymbol{S}$ is the Hadamard product of $\boldsymbol{\Omega}^{-1}$ and $\boldsymbol{S}$. Since both $\boldsymbol{\Omega}^{-1}$ and $\boldsymbol{S}$ are positive semidefinite, $\boldsymbol{\Omega}^{-1}\circ\boldsymbol{S}$ is also positive semidefinite. Therefore, the negative log-likelihood $(-\ell)$ is convex in $(\sigma_1^{-1}, \sigma_2^{-1})$.

Given $\delta$, $\sigma_{1_{ik}}$ and $\sigma_{2_{ik}}$, for $\mathbf{b}_{ik}$, $\mathbf{u}_i$ and $\mathbf{b}$, the second-order derivatives of the negative h-likelihood function are

$$\begin{aligned}\frac{\partial^2(-\ell_h)}{\partial\mathbf{b}_{ik}\mathbf{b}_{ik}^\top} &= \frac{1}{\sigma_{2_{ik}}^2(1-\delta^2)}\boldsymbol{P}_{ik}^\top\boldsymbol{X}_{ik}^\top\boldsymbol{X}_{ik}\boldsymbol{P}_{ik} + \frac{1}{\sigma_{1_{ik}}^2}\boldsymbol{V}_{ik}^\top\mathbf{Z}_{ik}^\top\mathbf{Z}_{ik}\boldsymbol{V}_{ik} + \boldsymbol{\Lambda}^{-1}, \\ \frac{\partial^2(-\ell_h)}{\partial\mathbf{u}_i\mathbf{u}_i^\top} &= K\boldsymbol{\Lambda}^{-1} + \boldsymbol{\Psi}^{-1}, \\ \frac{\partial^2(-\ell_h)}{\partial\mathbf{b}\mathbf{b}^\top} &= NK\boldsymbol{\Lambda}^{-1},\end{aligned}$$

where

$$\boldsymbol{P}_{ik} = \begin{pmatrix} -\delta\sigma_{2_{ik}}/\sigma_{1_{ik}} & 0 & 1 \\ 0 & 1 & 0\end{pmatrix}, \quad \boldsymbol{V}_{ik} = \begin{pmatrix} 1 & 0 & 0\end{pmatrix}.$$

For a matrix $\boldsymbol{\Omega}$,

$$\partial\boldsymbol{\Omega} = -\boldsymbol{\Omega}(\partial\boldsymbol{\Omega}^{-1})\boldsymbol{\Omega},$$

therefore, for $\boldsymbol{\Psi}^{-1}$ and $\boldsymbol{\Lambda}^{-1}$, the second-order derivatives can be calculated as

$$\partial\left(\frac{\partial(-\ell_h)}{\partial\boldsymbol{\Psi}^{-1}}\right) = \frac{N}{2}\boldsymbol{\Psi}(\partial\boldsymbol{\Psi}^{-1})\boldsymbol{\Psi}, \quad \partial\left(\frac{\partial(-\ell_h)}{\partial\boldsymbol{\Lambda}^{-1}}\right) = \frac{NK}{2}\boldsymbol{\Lambda}(\partial\boldsymbol{\Lambda}^{-1})\boldsymbol{\Lambda}.$$

Both matrices $\boldsymbol{P}_{ik}^\top\boldsymbol{X}_{ik}^\top\boldsymbol{X}_{ik}\boldsymbol{P}_{ik}$ and $\boldsymbol{V}_{ik}^\top\mathbf{Z}_{ik}^\top\mathbf{Z}_{ik}\boldsymbol{V}_{ik}$ are semipositive definite, as well as the inverse covariance matrices $\boldsymbol{\Psi}^{-1}$ and $\boldsymbol{\Lambda}^{-1}$. The second-order derivatives are semipositive definite.

Therefore, given $\delta$, the negative h-likelihood function is conditional convex in parameter sets $(\sigma_{1_{ik}}^{-1}, \sigma_{2_{ik}}^{-1})$, $(\mathbf{b}_{ik})$, $(\mathbf{u}_i)$, $\mathbf{b}$, $\boldsymbol{\Psi}^{-1}$, $\boldsymbol{\Lambda}^{-1}$.

Now we derive the iterate updates for the optimization problem of maximizing the log-likelihood function. Theorems 1 and 2 provides the solutions for $(\mathbf{b}_{ik})$ and $(\sigma_{1_{ik}}, \sigma_{2_{ik}})$. $\mathbf{u}_i$'s are assumed to be distributed with mean zero. To ensure this assumption, we add a constraint to the optimization problem. Together with the constraint for the variance components, we have

$$\begin{aligned}\text{maximize} \quad & \ell_h\left(\mathbf{b}_{ik}, \sigma_{1_{ik}}, \sigma_{2_{ik}}, \mathbf{u}_i, \boldsymbol{\Psi}, \boldsymbol{\Lambda}|\delta\right) \\ \text{subject to} \quad & \sum_{i=1}^N \mathbf{u}_i = \mathbf{0}, \\ & \left(\left(\sigma_{1_{ik}}, \sigma_{2_{ik}}\right), \boldsymbol{\Lambda}, \boldsymbol{\Psi}\right) \in \mathcal{S},\end{aligned}$$

where $\mathcal{S}$ is a constraint set for the variance components. We apply the method of Lagrange multipliers for the first constraint, and yield the update for $\mathbf{u}_i$ as

$$\hat{\mathbf{u}}_i = (K\mathbf{\Lambda}^{-1} + \mathbf{\Psi}^{-1})^{-1}\mathbf{\Lambda}^{-1}\left(\sum_{k=1}^{K}(\mathbf{b}_{ik} - \mathbf{b}) - \frac{1}{N}\sum_{i=1}^{N}\sum_{k=1}^{K}(\mathbf{b}_{ik} - \mathbf{b})\right).$$

For $\mathbf{b}$, $\mathbf{\Psi}$ and $\mathbf{\Lambda}$, the updates are calculated from the first-order derivatives as,

$$\hat{\mathbf{b}} = \frac{1}{NK}\sum_{i=1}^{N}\sum_{k=1}^{K}(\mathbf{b}_{ik} - \mathbf{u}_i), \tag{A.10}$$

$$\hat{\mathbf{\Psi}} = \frac{1}{N}\sum_{i=1}^{N}\mathbf{u}_i\mathbf{u}_i^\top, \tag{A.11}$$

$$\hat{\mathbf{\Lambda}} = \frac{1}{NK}\sum_{i=1}^{N}\sum_{k=1}^{K}(\mathbf{b}_{ik} - \mathbf{u}_i - \mathbf{b})(\mathbf{b}_{ik} - \mathbf{u}_i - \mathbf{b})^\top. \tag{A.12}$$

$\mathcal{S}$ is a convex set, if the updates for the variance components calculated above are interior points of $\mathcal{S}$, the updating formula will be applied; otherwise the solutions will be projected onto the set $\mathcal{S}$. □

### A.5.2 The m-likelihood algorithm

For the mixed effects model (3), let $\mathbf{q} = (\mathbf{b}_{11}, \ldots, \mathbf{b}_{1K}, \ldots, \mathbf{b}_{N1}, \ldots, \mathbf{b}_{NK})^\top \in \mathbb{R}^{3NK}$, it can be written as

$$\mathbf{q} = (\mathbf{I}_3 \otimes \mathbf{1}_{NK})\,\mathbf{b} + \sum_{r=1}^{2}\boldsymbol{W}_r\mathbf{v}_r,$$

where $\mathbf{1}_{NK}$ is a vector of length $NK$ with all elements as one,

$$\mathbf{v}_1 \sim \mathcal{N}_{3N}\left(\mathbf{0}, \mathbf{\Psi} \otimes \mathbf{I}_N\right), \quad \mathbf{v}_2 \sim \mathcal{N}_{3NK}\left(\mathbf{0}, \mathbf{\Lambda} \otimes \mathbf{I}_{NK}\right),$$

$$\boldsymbol{W}_1 = \mathbf{I}_3 \otimes \begin{pmatrix} \mathbf{1}_K & & \\ & \ddots & \\ & & \mathbf{1}_K \end{pmatrix}_{NK\times N}, \quad \boldsymbol{W}_2 = \mathbf{I}_{3NK}.$$

Then

$$\mathrm{Cov}\left(\mathbf{q}\right) = \mathrm{Cov}\left(\sum_{r=1}^{2}\boldsymbol{W}_r\mathbf{v}_r\right) = \boldsymbol{W}_1\left(\mathbf{\Psi} \otimes \mathbf{I}_N\right)\boldsymbol{W}_1^\top + \left(\mathbf{\Lambda} \otimes \mathbf{I}_{NK}\right) \triangleq \boldsymbol{V}_{\mathbf{q}},$$

the log-likelihood function of $\mathbf{q}$ is

$$\ell^{(2)} = \ell_{\mathbf{q}} = -\frac{1}{2}\log|\boldsymbol{V}_{\mathbf{q}}| - \frac{1}{2}\left(\mathbf{q} - (\mathbf{I}_3 \otimes \mathbf{1}_{NK})\,\mathbf{b}\right)^\top \boldsymbol{V}_{\mathbf{q}}^{-1}\left(\mathbf{q} - (\mathbf{I}_3 \otimes \mathbf{1}_{NK})\,\mathbf{b}\right).$$

When $\mathbf{\Psi} = \mathrm{diag}\{\psi_\alpha, \psi_\beta, \psi_\gamma\}$, $A_{ik}$, $B_{ik}$ and $C_{ik}$ are independent, and the trivariate mixed effects model (3) can be decomposed into three univariate mixed effects models. The marginal likelihood function can be decomposed as the summation of the marginal likelihood functions of $(A_{ik})$, $(B_{ik})$ and $(C_{ik})$. Under this assumption, following the same strategy as in the proof of Theorem A.3, we can show that the negative of m-likelihood function is also conditional convex. Since $(\sigma_{1_{ik}}, \sigma_{2_{ik}})$ is independent of $\ell^{(2)}$, the updates are the same as in the h-likelihood method. The updates for the latent coefficient parameter ($\mathbf{b}_{ik}$) are, for participant $i$,

$$\begin{aligned}
\hat{\mathbf{a}}_i &= \left(\boldsymbol{Z}_i^\top \boldsymbol{\Omega}_{1i}^{-1}\boldsymbol{Z}_i + \frac{\delta^2}{1-\delta^2}\boldsymbol{Z}_i^\top \boldsymbol{\Omega}_{12i}\boldsymbol{\Omega}_{2i}^{-1}\boldsymbol{\Omega}_{12i}\boldsymbol{Z}_i + \boldsymbol{V}_\alpha^{-1}\right)^{-1} \\
&\quad \left(\boldsymbol{Z}_i^\top \boldsymbol{\Omega}_{1i}^{-1}\tilde{\mathbf{m}}_i - \frac{\delta}{1-\delta^2}\boldsymbol{Z}_i^\top \boldsymbol{\Omega}_{12i}\boldsymbol{\Omega}_{2i}^{-1}\left(\tilde{\mathbf{r}}_i - \boldsymbol{Z}_i\mathbf{c}_i - \mathbf{m}_i\mathbf{b}_i - \delta\boldsymbol{\Omega}_{12i}\tilde{\mathbf{m}}_i\right) + A\boldsymbol{V}_\alpha^{-1}\mathbf{1}_K\right), \\
\hat{\mathbf{b}}_i &= \left(\frac{1}{1-\delta^2}\boldsymbol{M}_i^\top \boldsymbol{\Omega}_{2i}^{-1}\boldsymbol{M}_i^\top + \boldsymbol{V}_\beta^{-1}\right)^{-1}\left(\frac{1}{1-\delta^2}\boldsymbol{M}_i^\top \boldsymbol{\Omega}_{2i}^{-1}\left(\tilde{\mathbf{r}}_i - \boldsymbol{Z}_i\mathbf{c}_i - \delta\boldsymbol{\Omega}_{12i}(\tilde{\mathbf{m}}_i - \boldsymbol{Z}_i\mathbf{a}_i)\right) + B\boldsymbol{V}_\beta^{-1}\mathbf{1}_K\right), \\
\hat{\mathbf{c}}_i &= \left(\frac{1}{1-\delta^2}\boldsymbol{Z}_i^\top \boldsymbol{\Omega}_{2i}^{-1}\boldsymbol{Z}_i^\top + \boldsymbol{V}_\gamma^{-1}\right)^{-1}\left(\frac{1}{1-\delta^2}\boldsymbol{Z}_i^\top \boldsymbol{\Omega}_{2i}^{-1}\left(\tilde{\mathbf{r}}_i - \boldsymbol{M}_i\mathbf{b}_i - \delta\boldsymbol{\Omega}_{12i}(\tilde{\mathbf{m}}_i - \boldsymbol{Z}_i\mathbf{a}_i)\right) + C\boldsymbol{V}_\gamma^{-1}\mathbf{1}_K\right),
\end{aligned}$$

where $\mathbf{a}_i = (A_{i1},\ldots,A_{iK})^\top$, $\mathbf{b}_i = (B_{i1},\ldots,B_{iK})^\top$, $\mathbf{c}_i = (C_{i1},\ldots,C_{iK})^\top$; $\boldsymbol{Z}_i = \mathrm{diag}\{\mathbf{Z}_{i1},\ldots,\mathbf{Z}_{iK}\}$, $\boldsymbol{M}_i = \mathrm{diag}\{\mathbf{M}_{i1},\ldots,\mathbf{M}_{iK}\}$, $\tilde{\mathbf{m}}_i = \left(\mathbf{M}_{i1}^\top,\ldots,\mathbf{M}_{iK}^\top\right)^\top$, $\tilde{\mathbf{r}}_i = \left(\mathbf{R}_{i1}^\top,\ldots,\mathbf{R}_{iK}^\top\right)^\top$; $\boldsymbol{\Omega}_{i1} = \mathrm{diag}\left\{\sigma_{1_{i1}}^2\mathbf{I}_{n_{i1}},\ldots,\sigma_{1_{iK}}^2\mathbf{I}_{n_{iK}}\right\}$, $\boldsymbol{\Omega}_{i2} = \mathrm{diag}\left\{\sigma_{2_{i1}}^2\mathbf{I}_{n_{i1}},\ldots,\sigma_{2_{iK}}^2\mathbf{I}_{n_{iK}}\right\}$, $\boldsymbol{\Omega}_{i12} = \mathrm{diag}\{\kappa_{i1}\mathbf{I}_{n_{i1}},\ldots,\kappa_{iK}\mathbf{I}_{n_{iK}}\}$, and $\kappa_{ik} = \sigma_{2_{ik}}/\sigma_{1_{ik}}$; $\boldsymbol{V}_\alpha = \psi_\alpha^2\boldsymbol{W}_{1A}\boldsymbol{W}_{1A}^\top + \lambda_\alpha^2\mathbf{I}_{NK}$ is the covariance matrix of ($\mathbf{a}_i$) with $\boldsymbol{W}_{1A} = \mathrm{diag}\{\mathbf{1}_K,\ldots,\mathbf{1}_K\}$ a $NK\times N$ matrix, $\boldsymbol{V}_\beta$ and $\boldsymbol{V}_\gamma$ are the covariance matrices for ($\mathbf{b}_i$) and ($\mathbf{c}_i$), respectively, and defined analogously. The rest parameters $\mathbf{b}$, $\boldsymbol{\Psi}$ and $\boldsymbol{\Lambda}$ can be updated by fitting $(\mathbf{a}_i,\mathbf{b}_i,\mathbf{c}_i)$ into the mixed effects models. The same constraint for the variance components will be applied.

## A.6 Proof of Theorem 3

*Proof.* In the first-level model, for participant $i$ with $n_i$ observations ($i = 1,\ldots,N$), replace $\sigma_{1_i}$ and $\sigma_{2_i}$ with their estimators from Section A.4, we have the estimators for the coefficients,

$$\begin{aligned}
\hat{A}_i &= (\mathbf{Z}_i^\top\mathbf{Z}_i)^{-1}\mathbf{Z}_i^\top\mathbf{M}_i, \\
\hat{C}_i &= (\mathbf{Z}_i^\top(\mathbf{I}_{n_i} - \boldsymbol{P}_{\mathbf{M}_i})\mathbf{Z}_i)^{-1}\mathbf{Z}_i^\top(\mathbf{I}_{n_i} - \boldsymbol{P}_{\mathbf{M}_i})\mathbf{R}_i + \frac{\delta}{\sqrt{1-\delta^2}}\frac{\sqrt{|\hat{\boldsymbol{\Sigma}}_{B_i}|}}{\hat{\boldsymbol{\Sigma}}_{B_i}(1,1)}(\mathbf{Z}_i^\top\mathbf{Z}_i)^{-1}\mathbf{Z}_i^\top\mathbf{M}_i, \\
&\triangleq \tilde{C}_i + \tau c_i, \\
\hat{B}_i &= (\mathbf{M}_i^\top\mathbf{M}_i)^{-1}\mathbf{M}_i^\top\left(\mathbf{I}_{n_i} - \mathbf{Z}_i(\mathbf{Z}_i^\top(\mathbf{I}_{n_i} - \boldsymbol{P}_{\mathbf{M}_i})\mathbf{Z}_i)^{-1}\mathbf{Z}_i^\top(\mathbf{I}_{n_i} - \boldsymbol{P}_{\mathbf{M}_i})\right)\mathbf{R}_i - \frac{\delta}{\sqrt{1-\delta^2}}\frac{\sqrt{|\hat{\boldsymbol{\Sigma}}_{B_i}|}}{\hat{\boldsymbol{\Sigma}}_{B_i}(1,1)} \\
&\triangleq \tilde{B}_i - \tau b_i,
\end{aligned}$$

where

$$\tau = \frac{\delta}{\sqrt{1-\delta^2}},$$

$b_i = \sqrt{|\hat{\boldsymbol{\Sigma}}_{B_i}|}/\hat{\boldsymbol{\Sigma}}_{B_i}(1,1)$, $c_i = \sqrt{|\hat{\boldsymbol{\Sigma}}_{B_i}|}/\hat{\boldsymbol{\Sigma}}_{B_i}(1,1)\cdot(\mathbf{Z}_i^\top\mathbf{Z}_i)^{-1}\mathbf{Z}_i^\top\mathbf{M}_i = b_i\hat{A}_i$. We note in the equations above that $b_i$, $c_i$ $\tilde{C}_i$ and $\tilde{B}_i$ are estimators computed from data but we omit the "hat" notation in these symbols and their related ones hereafter for simplicity. $\tau$ is a one-to-one function of $\delta$. Therefore, if $\tau$ is identifiable and consistently estimated, so as $\delta$.

Because $\boldsymbol{\Lambda}$ is assumed to be diagonal, we can calculate the log-likelihood function of $A_i$, $B_i$ and $C_i$ separately. The estimate of $A_i$ is independent of $\delta$. Thus, in the second stage, this part of profile likelihood function is a constant of $\delta$. Therefore, in the following proof, we will focus on $B_i$ and $C_i$. The log-likelihood function of $B_i$'s is

$$\ell_B = -\frac{N}{2}\log(\lambda_\beta^2) - \frac{1}{2\lambda_\beta^2}\sum_{i=1}^{N}(B_i - B)^2.$$

(1) If $\mathbf{\Lambda}$ is known, we have the maximum likelihood estimator of $B$ as

$$\hat{B} = \frac{1}{N}\sum_{i=1}^{N}\hat{B}_i = \frac{1}{N}\sum_{i=1}^{N}\left(\tilde{B}_i - \tau b_i\right) \triangleq \bar{\tilde{B}} - \tau\bar{b}.$$

(2) If $\mathbf{\Lambda}$ is unknown, we have

$$\begin{aligned}\hat{B} &= \frac{1}{N}\sum_{i=1}^{N}\hat{B}_i = \frac{1}{N}\sum_{i=1}^{N}\left(\tilde{B}_i - \tau b_i\right) \triangleq \bar{\tilde{B}} - \tau\bar{b},\\ \hat{\lambda}_\beta^2 &= \frac{1}{N}\left(\mathbf{\Delta}_{\tilde{B}} - 2\tau\mathbf{\Delta}_{\tilde{B},b} + \tau^2\mathbf{\Delta}_b\right),\end{aligned}$$

where

$$\mathbf{\Delta}_{\tilde{B}} = \sum_{i=1}^{N}(\tilde{B}_i - \bar{\tilde{B}})^2, \quad \mathbf{\Delta}_{\tilde{B},b} = \sum_{i=1}^{N}(\tilde{B}_i - \bar{\tilde{B}})(b_i - \bar{b}), \quad \mathbf{\Delta}_b = \sum_{i=1}^{N}(b_i - \bar{b})^2.$$

Similarly, we have the maximum likelihood estimator of $C$ and $\lambda_\gamma^2$ as

$$\begin{aligned}\hat{C} &= \frac{1}{N}\sum_{i=1}^{N}\hat{C}_i = \frac{1}{N}\sum_{i=1}^{N}\left(\tilde{C}_i + \tau c_i\right) \triangleq \bar{\tilde{C}} + \tau\bar{c},\\ \hat{\lambda}_\gamma^2 &= \frac{1}{N}\left(\mathbf{\Delta}_{\tilde{C}} + 2\tau\mathbf{\Delta}_{\tilde{C},c} + \tau^2\mathbf{\Delta}_c\right),\end{aligned}$$

where

$$\mathbf{\Delta}_{\tilde{C}} = \sum_{i=1}^{N}(\tilde{C}_i - \bar{\tilde{C}})^2, \quad \mathbf{\Delta}_{\tilde{C},c} = \sum_{i=1}^{N}(\tilde{C}_i - \bar{\tilde{C}})(c_i - \bar{c}), \quad \mathbf{\Delta}_c = \sum_{i=1}^{N}(c_i - \bar{c})^2.$$

Plug these estimators back to the likelihood function yields the profile likelihood of $\tau$, and by maximizing the profile likelihood,

(1) if $\mathbf{\Lambda}$ is known,

$$\hat{\tau} = \frac{\mathbf{\Delta}_{\tilde{B},b}/\lambda_\beta^2 - \mathbf{\Delta}_{\tilde{C},c}/\lambda_\gamma^2}{\mathbf{\Delta}_b/\lambda_\beta^2 + \mathbf{\Delta}_c/\lambda_\gamma^2};$$

(2) if $\mathbf{\Lambda}$ is unknown, the estimator of $\tau$ should satisfy the following cubic function

$$2\tau^3\mathbf{\Delta}_b\mathbf{\Delta}_c + \tau^2\left(3\mathbf{\Delta}_b\mathbf{\Delta}_{\tilde{C},c} - 3\mathbf{\Delta}_{\tilde{B},b}\mathbf{\Delta}_c\right) + \tau\left(\mathbf{\Delta}_{\tilde{B}}\mathbf{\Delta}_c + \mathbf{\Delta}_b\mathbf{\Delta}_{\tilde{C}} - 4\mathbf{\Delta}_{\tilde{B},b}\mathbf{\Delta}_{\tilde{C},c}\right) + \left(\mathbf{\Delta}_{\tilde{B}}\mathbf{\Delta}_{\tilde{C},c} - \mathbf{\Delta}_{\tilde{B},b}\mathbf{\Delta}_{\tilde{C}}\right) = 0. \tag{A.13}$$

Let $y_i = \tilde{B}_i - \bar{\tilde{B}}$, $U_i = b_i - \bar{b}$, $z_i = \tilde{C}_i - \bar{\tilde{C}}$, $V_i = c_i - \bar{c}$, and

$$\epsilon = \frac{2\boldsymbol{\Delta}_b\boldsymbol{\Delta}_c}{\boldsymbol{\Delta}_{\tilde{B}}\boldsymbol{\Delta}_c + \boldsymbol{\Delta}_b\boldsymbol{\Delta}_{\tilde{C}} - 4\boldsymbol{\Delta}_{\tilde{B},b}\boldsymbol{\Delta}_{\tilde{C},c}} = \frac{2\sum U_i^2 \sum V_i^2}{\sum y_i^2 \sum V_i^2 + \sum U_i^2 \sum z_i^2 - 4\sum y_i U_i \sum z_i V_i},$$
$$\theta = \frac{3\boldsymbol{\Delta}_b\boldsymbol{\Delta}_{\tilde{C},c} - 3\boldsymbol{\Delta}_{\tilde{B},b}\boldsymbol{\Delta}_c}{\boldsymbol{\Delta}_{\tilde{B}}\boldsymbol{\Delta}_c + \boldsymbol{\Delta}_b\boldsymbol{\Delta}_{\tilde{C}} - 4\boldsymbol{\Delta}_{\tilde{B},b}\boldsymbol{\Delta}_{\tilde{C},c}} = \frac{3\sum U_i^2 \sum z_i V_i - 3\sum y_i U_i \sum V_i^2}{\sum y_i^2 \sum V_i^2 + \sum U_i^2 \sum z_i^2 - 4\sum y_i U_i \sum z_i V_i},$$
$$\lambda = \frac{\boldsymbol{\Delta}_{\tilde{B}}\boldsymbol{\Delta}_{\tilde{C},c} - \boldsymbol{\Delta}_{\tilde{B},b}\boldsymbol{\Delta}_{\tilde{C}}}{\boldsymbol{\Delta}_{\tilde{B}}\boldsymbol{\Delta}_c + \boldsymbol{\Delta}_b\boldsymbol{\Delta}_{\tilde{C}} - 4\boldsymbol{\Delta}_{\tilde{B},b}\boldsymbol{\Delta}_{\tilde{C},c}} = \frac{\sum y_i^2 \sum z_i V_i - \sum y_i U_i \sum z_i^2}{\sum y_i^2 \sum V_i^2 + \sum U_i^2 \sum z_i^2 - 4\sum y_i U_i \sum z_i V_i},$$

it is equivalent to solve the following cubic function

$$\epsilon\tau^3 + \theta\tau^2 + \tau + \lambda = 0. \tag{A.14}$$

Now we need to find the asymptotic properties of these $\boldsymbol{\Delta}$. quantities when $n_i \to \infty$ and $N \to \infty$. For the $\boldsymbol{\Lambda}$ unknown case, we seek to show that when $n_i \to \infty$ and $N \to \infty$, both $\epsilon$ and $\theta$ converge to zero. Then we can apply the perturbation theory for finding roots to solve the problem.

We first show that under first-level model, both $\tilde{B}_i$ and $\tilde{C}_i$ are independent of $b_i$ asymptotically. Since this is the first-level property, to keep the following discussion uncluttered, we drop the participant index $i$. $\tilde{B}$ and $\tilde{C}$ are the estimators of $B$ and $C$ when assuming $\delta = 0$ in the first-level model. Under this assumption, for models 5 and (6), we have

$$\tilde{\boldsymbol{\Theta}} = \boldsymbol{P}\boldsymbol{Y},$$

where $\boldsymbol{P} = (\boldsymbol{X}^\top\boldsymbol{X})^{-1}\boldsymbol{X}^\top$. For models (5) and (7), we have the maximum likelihood estimator

$$\hat{\boldsymbol{\Sigma}}_B = \frac{1}{n}\boldsymbol{Y}^\top\boldsymbol{H}\boldsymbol{Y},$$

where $\boldsymbol{H} = \mathbf{I}_n - \boldsymbol{P}_{\mathbf{Z}}$. Since $b = \sqrt{|\hat{\boldsymbol{\Sigma}}_B|}/\hat{\boldsymbol{\Sigma}}_B(1,1)$, $\tilde{B} = \tilde{\boldsymbol{\Theta}}(2,2)$, and $\tilde{C} = \tilde{\boldsymbol{\Theta}}(1,2)$, it is sufficient to show that $\hat{\boldsymbol{\Sigma}}_B$ and $\tilde{\boldsymbol{\Theta}}$ are asymptotically independent.

$$\begin{aligned}
\boldsymbol{P}\boldsymbol{H} &= (\boldsymbol{X}^\top\boldsymbol{X})^{-1}\boldsymbol{X}^\top(\mathbf{I}_n - \mathbf{Z}(\mathbf{Z}^\top\mathbf{Z})^{-1}\mathbf{Z}^\top) \\
&= \frac{1}{(\mathbf{Z}^\top\mathbf{Z})(\mathbf{Z}^\top\mathbf{Z}\mathbf{M}^\top\mathbf{M} - \mathbf{Z}^\top\mathbf{M}\mathbf{M}^\top\mathbf{Z})}\begin{pmatrix}(\mathbf{Z}^\top\mathbf{M})(\mathbf{M}^\top\mathbf{Z}\mathbf{Z}^\top - \mathbf{Z}^\top\mathbf{Z}\mathbf{M}^\top) \\ (\mathbf{Z}^\top\mathbf{Z})(\mathbf{Z}^\top\mathbf{Z}\mathbf{M}^\top - \mathbf{M}^\top\mathbf{Z}\mathbf{Z}^\top)\end{pmatrix} \\
&= \frac{1}{(\mathbf{Z}^\top\mathbf{Z}/n)((\mathbf{Z}^\top\mathbf{Z}/n)(\mathbf{M}^\top\mathbf{M}/n) - (\mathbf{Z}^\top\mathbf{M}/n)(\mathbf{M}^\top\mathbf{Z}/n))}\begin{pmatrix}(\mathbf{Z}^\top\mathbf{M}/n)(\mathbf{M}^\top\mathbf{Z}\mathbf{Z}^\top - \mathbf{Z}^\top\mathbf{Z}\mathbf{M}^\top)/n^2 \\ (\mathbf{Z}^\top\mathbf{Z}/n)(\mathbf{Z}^\top\mathbf{Z}\mathbf{M}^\top - \mathbf{M}^\top\mathbf{Z}\mathbf{Z}^\top)/n^2\end{pmatrix}.
\end{aligned}$$

In Theorem A.1, it is assumed that $\mathbf{Z}^\top\mathbf{Z}/n \to q$ and $\mathbf{Z}^\top\mathbf{E}_1/n \to 0$, as $n \to \infty$, and more specifically, $\mathbf{Z}^\top\mathbf{Z}/n = q + \mathcal{O}_p(1/\sqrt{n})$ and $\mathbf{Z}^\top\mathbf{E}_1/n = \mathcal{O}_p(1/\sqrt{n})$. Then, we have the following results

$$\begin{aligned}
\frac{\mathbf{Z}^\top\mathbf{M}/n}{(\mathbf{Z}^\top\mathbf{Z}/n)((\mathbf{Z}^\top\mathbf{Z}/n)(\mathbf{M}^\top\mathbf{M}/n) - (\mathbf{Z}^\top\mathbf{M}/n)(\mathbf{M}^\top\mathbf{Z}/n))} &= \frac{A}{q\sigma_1^2} + \mathcal{O}_p\left(\frac{1}{\sqrt{n}}\right), \\
\frac{\mathbf{Z}^\top\mathbf{Z}/n}{(\mathbf{Z}^\top\mathbf{Z}/n)((\mathbf{Z}^\top\mathbf{Z}/n)(\mathbf{M}^\top\mathbf{M}/n) - (\mathbf{Z}^\top\mathbf{M}/n)(\mathbf{M}^\top\mathbf{Z}/n))} &= \frac{1}{q^2\sigma_1^2} + \mathcal{O}_p\left(\frac{1}{\sqrt{n}}\right).
\end{aligned}$$

For $\forall\ \mathbf{x} = (x_1, \ldots, x_n)^\top \in \mathbb{R}^n$ with $\|\ \mathbf{x}\ \|_2 < \infty$,

$$\begin{aligned}
\frac{1}{n}\mathbf{Z}^\top\mathbf{x} &\le \frac{1}{n}\ \|\ \mathbf{Z}\ \|_2 \|\ \mathbf{x}\ \|_2 = \frac{1}{\sqrt{n}}(\frac{1}{n}\mathbf{Z}^\top\mathbf{Z})^{1/2}\ \|\ \mathbf{x}\ \|_2 = \mathcal{O}_p\left(\frac{1}{\sqrt{n}}\right), \\
\frac{1}{n}\mathbf{E}_1^\top\mathbf{x} &\le \frac{1}{n}\ \|\ \mathbf{E}_1\ \|_2 \|\ \mathbf{x}\ \|_2 = \frac{1}{\sqrt{n}}(\frac{1}{n}\mathbf{E}_1^\top\mathbf{E}_1)^{1/2}\ \|\ \mathbf{x}\ \|_2 = \mathcal{O}_p\left(\frac{1}{\sqrt{n}}\right),
\end{aligned}$$

$$\Rightarrow \quad \frac{1}{n^2}\left((\mathbf{E}_1^\top \mathbf{Z})\mathbf{Z}^\top - (\mathbf{Z}^\top \mathbf{Z})\mathbf{E}_1^\top\right)\mathbf{x} = \left((\frac{1}{n}\mathbf{E}_1^\top \mathbf{Z})(\frac{1}{n}\mathbf{Z}^\top \boldsymbol{x}) - (\frac{1}{n}\mathbf{Z}^\top \mathbf{Z})(\frac{1}{n}\mathbf{E}_1^\top \mathbf{x})\right) = \mathcal{O}_p\left(\frac{1}{\sqrt{n}}\right).$$

Therefore,

$$\boldsymbol{PH} \xrightarrow{\mathcal{P}} \mathbf{0}, \quad \text{as } n \to \infty.$$

With the normality assumption, we conclude that $\hat{\mathbf{\Theta}}$ and $\hat{\mathbf{\Sigma}}_B$ are asymptotically independent.

For the two-level model, under the normality assumption of the first-level model, we have for $\mathbf{\Sigma}_{B_i}$,

$$n_i \hat{\mathbf{\Sigma}}_{B_i} \mid A_i, C_i', \mathbf{\Sigma}_{B_i} \sim W_2\left(\mathbf{\Sigma}_{B_i}, n_i - 1\right),$$

where $W_p(\mathbf{\Sigma}, n)$ is the $p$-dimensional Wishart distribution with scale matrix $\mathbf{\Sigma}$ and degrees of freedom $n$. For finite sample size $n_i$,

$$\mathbb{E}\left(\hat{\mathbf{\Sigma}}_{B_i}\right) = \frac{n_i - 1}{n_i}\mathbf{\Sigma}_{B_i}, \quad \operatorname{Var}\left(\hat{\mathbf{\Sigma}}_{B_i}(k,l)\right) = \frac{n_i - 1}{n_i^2}\left(\mathbf{\Sigma}_{B_i}^2(k,l) + \mathbf{\Sigma}_{B_i}(k,k)\mathbf{\Sigma}_{B_i}(l,l)\right).$$

Then for large $n_i$,

$$\sqrt{n_i}\left(\operatorname{vec}\left(\hat{\mathbf{\Sigma}}_{B_i}\right) - \operatorname{vec}\left(\mathbf{\Sigma}_{B_i}\right)\right) \mid A_i, C_i', \mathbf{\Sigma}_{B_i} \xrightarrow{\mathcal{D}} \mathcal{N}\left(\mathbf{0}, \mathbf{\Xi}\right),$$

where $\mathbf{\Xi}$ is the asymptotic covariance matrix.

$$b_i = \frac{\sqrt{|\hat{\mathbf{\Sigma}}_{B_i}|}}{\hat{\mathbf{\Sigma}}_{B_i}(1,1)} \triangleq g(\hat{\mathbf{\Sigma}}_{B_i}),$$

from the Delta method

$$\sqrt{n_i}\left(b_i - g(\mathbf{\Sigma}_{B_i})\right) \mid A_i, C_i', \mathbf{\Sigma}_{B_i} \xrightarrow{\mathcal{D}} \mathcal{N}\left(0, \nabla g(\mathbf{\Sigma}_{B_i})^\top \mathbf{\Xi} \nabla g(\mathbf{\Sigma}_{B_i})\right),$$

where $g(\mathbf{\Sigma}_{B_i}) = \sqrt{1-\delta^2}\kappa_i$, and therefore

$$b_i \mid A_i, C_i', \mathbf{\Sigma}_{B_i} = \sqrt{1-\delta^2}\kappa_i + \mathcal{O}_p\left(\frac{1}{\sqrt{n_i}}\right).$$

$$U_i^2 = (1-\delta^2)(\kappa_i - \bar{\kappa})^2 + \mathcal{O}_p\left(\frac{1}{\sqrt{n_i}}\right),$$

where $\bar{\kappa} = (1/N)\sum \kappa_i$. Therefore

$$\frac{1}{N}\sum U_i^2 = (1-\delta^2)\varrho^2 + \mathcal{O}_p\left(\frac{1}{\sqrt{Nn}}\right),$$

where $\varrho^2 = (1/N)\sum(\kappa_i - \bar{\kappa})^2$ and $n = \min_i n_i$. Similarly, we have

$$\frac{1}{N}\sum_{i=1}^{N} y_i U_i \quad = \quad \delta\sqrt{1-\delta^2}\varrho^2 + \mathcal{O}_p\left(\frac{1}{\sqrt{Nn}}\right).$$

To find the asymptotics of $\sum_i y_i^2/N$, where $y_i = \tilde{B}_i - \bar{\tilde{B}}$, $\tilde{B}_i = \hat{B}_i + \tau b_i$, we follow the same strategy as before to derive

$$\frac{1}{N}\sum_{i=1}^{N} y_i^2 \;=\; \lambda_\beta^2 + \delta^2\varrho^2 + \mathcal{O}_p\left(\frac{1}{\sqrt{Nn}}\right).$$

For the $\mathbf{\Delta}$.'s related to $\tilde{C}_i$ and $c_i$, since $c_i = b_i\hat{A}_i$, we need to use the conclusion that $\hat{\mathbf{\Sigma}}_{B_i}$ and $\tilde{\mathbf{\Theta}}$ are asymptotically independent, and further as $n_i \to \infty$,

$$\hat{A}_i \perp b_i \mid \mathbf{\Theta}_i, \mathbf{\Sigma}_{B_i}, \quad \tilde{C}_i \perp b_i \mid \mathbf{\Theta}_i, \mathbf{\Sigma}_{B_i}.$$

Based on these results, we use the same strategy to derive that

$$\begin{aligned}
\frac{1}{N}\sum_{i=1}^{N} z_i V_i &= -\delta\sqrt{1-\delta^2}\left((A^2+\lambda_\alpha^2)\varrho^2 + \lambda_\alpha^2\bar{\kappa}^2\right) + \mathcal{O}_p\left(\frac{1}{\sqrt{Nn}}\right),\\
\frac{1}{N}\sum_{i=1}^{N} V_i^2 &= (1-\delta^2)\left((A^2+\lambda_\alpha^2)\varrho^2 + \lambda_\alpha^2\bar{\kappa}^2\right) + \mathcal{O}_p\left(\frac{1}{\sqrt{Nn}}\right).
\end{aligned}$$

Similarly, we have

$$\frac{1}{N}\sum_{i=1}^{N} V_i^2 = (1-\delta^2)\left((A^2+\lambda_\alpha^2)\varrho^2 + \lambda_\alpha^2\bar{\kappa}^2\right) + \mathcal{O}_p\left(\frac{1}{\sqrt{Nn}}\right),$$

and

$$\frac{1}{N}\sum_{i=1}^{N} z_i^2 = \lambda_\gamma^2 + \delta^2\left((A^2+\lambda_\alpha^2)\varrho^2 + \lambda_\alpha^2\bar{\kappa}^2\right) + \mathcal{O}_p\left(\frac{1}{\sqrt{Nn}}\right).$$

When $\mathbf{\Lambda}$ is known,

$$\hat{\tau} = \frac{\mathbf{\Delta}_{\tilde{B},b}/\lambda_b^2 - \mathbf{\Delta}_{\tilde{C},c}/\lambda_c^2}{\mathbf{\Delta}_b/\lambda_\beta^2 + \mathbf{\Delta}_c/\lambda_\gamma^2} = \frac{\delta}{\sqrt{1-\delta^2}} + \mathcal{O}_p\left(\frac{1}{\sqrt{Nn}}\right) = \tau + \mathcal{O}_p\left(\frac{1}{\sqrt{Nn}}\right).$$

When $\mathbf{\Lambda}$ is unknown, in the cubic function $\epsilon\tau^3 + \theta\tau^2 + \tau + \lambda$,

$$\begin{aligned}
\epsilon &= \frac{(1-\delta^2)^2\xi\varrho^2}{\mathbf{\Delta}} + \mathcal{O}_p\left(\frac{1}{\sqrt{Nn}}\right),\\
\theta &= -\frac{6\delta(1-\delta^2)^{3/2}\xi\varrho^2}{\mathbf{\Delta}} + \mathcal{O}_p\left(\frac{1}{\sqrt{Nn}}\right),\\
\lambda &= -\frac{\delta(1-\delta^2)^{1/2}}{\mathbf{\Delta}}\left(\lambda_\beta^2\xi + \lambda_\gamma^2\varrho^2 + \delta^2\varrho^2\xi\right) + \mathcal{O}_p\left(\frac{1}{\sqrt{Nn}}\right),
\end{aligned}$$

where

$$\begin{aligned}
\xi &= (A^2+\lambda_\alpha^2)\varrho^2 + \lambda_\alpha^2\bar{\kappa}^2,\\
\mathbf{\Delta} &= (1-\delta^2)\left[\lambda_\beta^2\xi + \lambda_\gamma^2\varrho^2 + 6\delta^2\varrho^2\xi\right] + \mathcal{O}_p\left(\frac{1}{\sqrt{Nn}}\right).
\end{aligned}$$

Under the assumption that $\varrho^2 = \bar{\kappa}^2/\varpi$ and $1/\varpi = \mathcal{O}_p(1/\sqrt{Nn})$,

$$\begin{aligned}
\xi &= \left(\frac{A^2+\lambda_\alpha^2}{\varpi}+\lambda_\alpha^2\right)\bar{\kappa}^2 = \lambda_\alpha^2\bar{\kappa}^2 + \mathcal{O}_p\left(\frac{1}{\sqrt{Nn}}\right) \\
\boldsymbol{\Delta} &= (1-\delta^2)\lambda_\alpha^2\lambda_\beta^2\bar{\kappa}^2 + \mathcal{O}_p\left(\frac{1}{\sqrt{Nn}}\right) \\
\epsilon &= \mathcal{O}_p\left(\frac{1}{\sqrt{Nn}}\right) \\
\theta &= \mathcal{O}_p\left(\frac{1}{\sqrt{Nn}}\right) \\
\lambda &= -\frac{\delta}{\sqrt{1-\delta}} + \mathcal{O}_p\left(\frac{1}{\sqrt{Nn}}\right)
\end{aligned}$$

By applying the perturbation theory, we have the unique solution for $\tau$ as

$$\tilde{\tau} = \tau_0 + \epsilon\tau_1 + \epsilon^2\tau_2 + \mathcal{O}(\epsilon^3), \tag{A.15}$$

where

$$\begin{cases}\tau_0 = -\lambda - \theta\lambda^2 - 2\theta^2\lambda^3 + \mathcal{O}(\theta^3) \\ \tau_1 = -\tau_0^3/(2\theta\tau_0+1) \\ \tau_2 = -(3\tau_0^2\tau_1 + \theta\tau_1^2)/(2\theta\tau_0+1)\end{cases}.$$

Therefore, using the two-stage approach, asymptotically, $\delta$ is identifiable under our two-level model. This also proves the consistency of the two-stage estimator as both the sample size and the number of trials of each participant go to infinity. □

### A.7 Proof of Theorem 5

*Proof.* In the mixed effects model, both $\boldsymbol{\Psi}$ and $\boldsymbol{\Lambda}$ are assumed to be diagonal. Analogous to the proof of Theorem 3, we decompose the trivariate mixed effects model into three univariate mixed effects models and focus on $B_{ik}$ and $C_{ik}$ only. By concatenating the first stage estimates of all the participants, the higher level mixed effects model for $\hat{B}_{ik}$'s can be written into the following matrix form

$$\hat{\mathbf{B}} = \mathbf{1}_{NK}B + \mathbf{J}\boldsymbol{\beta} + \boldsymbol{\epsilon}^B,$$

where $\hat{\mathbf{B}}$, $\boldsymbol{\beta}$ and $\boldsymbol{\epsilon}^B$ are the vectorization of $\hat{B}_{ik}$, $\beta_i$ and $\epsilon_{ik}^B$, respectively; $\mathbf{J} = \mathrm{diag}\left\{\mathbf{1}_K,\ldots,\mathbf{1}_K\right\}_{NK\times N}$. Under the assumption that $\boldsymbol{\beta}$ is independent of $\boldsymbol{\epsilon}^B$, the log-likelihood function is

$$\ell_B = -\frac{1}{2}\log\left(|\boldsymbol{\Theta}_\beta|\right) - \frac{1}{2}(\hat{\mathbf{B}} - \mathbf{1}_{NK}B)^\top\boldsymbol{\Omega}_\beta(\hat{\mathbf{B}} - \mathbf{1}_{NK}B),$$

where $\boldsymbol{\Theta}_\beta = \psi_\beta^2\mathbf{J}\mathbf{J}^\top + \lambda_\beta^2\mathbf{I}_{NK}$, and $\boldsymbol{\Omega}_\beta = \boldsymbol{\Theta}_\beta^{-1}$.The maximum likelihood estimator of $B$ is

$$\bar{B} = \frac{1}{NK}\mathbf{1}_{NK}^\top\hat{\mathbf{B}} = \frac{1}{NK}\mathbf{1}_{NK}^\top(\tilde{\mathbf{B}} - \tau\mathbf{b}) \triangleq \bar{\tilde{B}} - \tau\bar{b}, \tag{A.16}$$

where $\tilde{\mathbf{B}} = \left(\tilde{B}_{ik}\right)$, $\boldsymbol{b} = (b_{ik})$, $\tilde{B}_{ik}$ and $b_{ik}$ follow the same definition as in the proof of Theorem 3. $\boldsymbol{\Omega}_\beta$ is assumed to be known, replacing $B$ with $\bar{B}$ in the log-likelihood function, we have

$$\ell_B(\tau) \propto -\frac{1}{2}\tau^2(\mathbf{b}-\mathbf{1}_{NK}\bar{b})^\top\boldsymbol{\Omega}_\beta(\mathbf{b}-\mathbf{1}_{NK}\bar{b}) + \tau(\tilde{\mathbf{B}}-\mathbf{1}_{NK}\bar{\tilde{B}})^\top\boldsymbol{\Omega}_\beta(\mathbf{b}-\mathbf{1}_{NK}\bar{b}) - \frac{1}{2}(\tilde{\mathbf{B}}-\mathbf{1}_{NK}\bar{\tilde{B}})^\top\boldsymbol{\Omega}_\beta(\tilde{\mathbf{B}}-\mathbf{1}_{NK}\bar{\tilde{B}}).$$

For $C$, we have

$$\ell_C(\tau) \propto -\frac{1}{2}\tau^2(\mathbf{c}-\mathbf{1}_{NK}\bar{c})^\top\mathbf{\Omega}_\gamma(\mathbf{c}-\mathbf{1}_{NK}\bar{c})-\tau(\tilde{\mathbf{C}}-\mathbf{1}_{NK}\bar{\tilde{C}})^\top\mathbf{\Omega}_\gamma(\mathbf{c}-\mathbf{1}_{NK}\bar{c})-\frac{1}{2}(\tilde{\mathbf{C}}-\mathbf{1}_{NK}\bar{\tilde{C}})^\top\mathbf{\Omega}_\gamma(\tilde{\mathbf{C}}-\mathbf{1}_{NK}\bar{\tilde{C}}),$$

where $\mathbf{\Omega}_\gamma = \mathbf{\Theta}_\gamma^{-1}$, $\mathbf{\Theta}_\gamma^{-1} = \psi_\gamma^2\mathbf{J}\mathbf{J}^\top + \lambda_\gamma^2\mathbf{I}_{NK}$; and

$$\bar{C} = \frac{1}{NK}\mathbf{1}_{NK}^\top\hat{\mathbf{C}} = \frac{1}{NK}\mathbf{1}_{NK}^\top(\tilde{\mathbf{C}}+\tau\mathbf{c}) = \bar{\tilde{C}} + \tau\bar{c}. \tag{A.17}$$

The profile likelihood related to $\tau$ is

$$\ell(\tau) = \ell_B(\tau) + \ell_C(\tau). \tag{A.18}$$

$$\frac{\partial\ell(\tau)}{\partial\tau} = 0 \quad\Rightarrow\quad \hat{\tau} = \frac{(\tilde{\mathbf{B}}-\mathbf{1}_{NK}\bar{\tilde{B}})^\top\mathbf{\Omega}_\beta(\mathbf{b}-\mathbf{1}_{NK}\bar{b}) - (\tilde{\mathbf{C}}-\mathbf{1}_{NK}\bar{\tilde{C}})^\top\mathbf{\Omega}_\gamma(\mathbf{c}-\mathbf{1}_{NK}\bar{c})}{(\mathbf{b}-\mathbf{1}_{NK}\bar{b})^\top\mathbf{\Omega}_\beta(\mathbf{b}-\mathbf{1}_{NK}\bar{b}) + (\mathbf{c}-\mathbf{1}_{NK}\bar{c})^\top\mathbf{\Omega}_\gamma(\mathbf{c}-\mathbf{1}_{NK}\bar{c})},$$
$$\frac{\partial^2\ell(\tau)}{\partial\tau^2} = -\left[(\mathbf{b}-\mathbf{1}_{NK}\bar{b})^\top\mathbf{\Omega}_\beta(\mathbf{b}-\mathbf{1}_{NK}\bar{b}) + (\mathbf{c}-\mathbf{1}_{NK}\bar{c})^\top\mathbf{\Omega}_\gamma(\mathbf{c}-\mathbf{1}_{NK}\bar{c})\right] < 0,$$

therefore, $\hat{\tau}$ uniquely maximizes the likelihood function.

Denote $y_{ik} = \tilde{B}_{ik} - \bar{\tilde{B}}$, $U_{ik} = b_{ik} - \bar{b}$, $z_{ik} = \tilde{C}_{ik} - \bar{\tilde{C}}$, $V_{ik} = c_{ik} - \bar{c}$, and $\mathbf{\Omega}_\beta = \left\{\omega_{ik}^\beta\right\}$, $\mathbf{\Omega}_\gamma = \left\{\omega_{ik}^\gamma\right\}$, where the diagonal elements of $\mathbf{\Omega}_\beta$ are identical, so as $\mathbf{\Omega}_\gamma$, i.e., $\omega_{ii}^\beta = \omega_\beta$ and $\omega_{ii}^\gamma = \omega_\gamma$, $i = 1, \ldots, NK$. Let

$$\begin{aligned}
\mathbf{\Delta}_{\tilde{B},b} &\triangleq (\tilde{\mathbf{B}}-\mathbf{1}_{NK}\bar{\tilde{B}})^\top\mathbf{\Omega}_\beta(\mathbf{b}-\mathbf{1}_{NK}\bar{b}) = \sum_i\sum_k\sum_j\sum_l y_{ik}U_{jl}\omega^\beta_{((i-1)K+k)((j-1)K+l)},\\
\mathbf{\Delta}_{\tilde{C},c} &\triangleq (\tilde{\mathbf{C}}-\mathbf{1}_{NK}\bar{\tilde{C}})^\top\mathbf{\Omega}_\gamma(\mathbf{c}-\mathbf{1}_{NK}\bar{c}) = \sum_i\sum_k\sum_j\sum_l z_{ik}V_{jl}\omega^\gamma_{((i-1)K+k)((j-1)K+l)},\\
\mathbf{\Delta}_b &\triangleq (\mathbf{b}-\mathbf{1}_{NK}\bar{b})^\top\mathbf{\Omega}_\beta(\mathbf{b}-\mathbf{1}_{NK}\bar{b}) = \sum_i\sum_k\sum_j\sum_l U_{jl}^2\omega^\beta_{((i-1)K+k)((j-1)K+l)},\\
\mathbf{\Delta}_c &\triangleq (\mathbf{c}-\mathbf{1}_{NK}\bar{c})^\top\mathbf{\Omega}_\gamma(\mathbf{c}-\mathbf{1}_{NK}\bar{c}) = \sum_i\sum_k\sum_j\sum_l V_{jl}^2\omega^\gamma_{((i-1)K+k)((j-1)K+l)}.
\end{aligned}$$

Since

$$\begin{aligned}
&y_{ik} \perp U_{il},\ \text{for } k \neq l, \quad y_{ik} \perp U_{jl},\ \text{for } i \neq j,\\
&z_{ik} \perp V_{il},\ \text{for } k \neq l, \quad z_{ik} \perp V_{jl},\ \text{for } i \neq j,\\
&U_{ik} \perp U_{jl},\ \text{for } k \neq l, \quad V_{ik} \perp V_{jl},\ \text{for } k \neq l,
\end{aligned}$$

using the results from the two-level model, we have, for the three-level model,

$$\begin{aligned}
\frac{1}{NK}\mathbf{\Delta}_{\tilde{B},b} &= \omega_\beta\delta\sqrt{1-\delta^2}\varrho^2 + \mathcal{O}_p\left(\frac{1}{\sqrt{Nn}}\right),\\
\frac{1}{NK}\mathbf{\Delta}_{\tilde{C},c} &= -\omega_\gamma\delta\sqrt{1-\delta^2}\left((A^2+\lambda_\alpha^2)\varrho^2 + \lambda_\alpha^2\bar{\kappa}^2\right) + \mathcal{O}_p\left(\frac{1}{\sqrt{Nn}}\right),\\
\frac{1}{NK}\mathbf{\Delta}_b &= \omega_\beta(1-\delta^2)\varrho^2 + \mathcal{O}_p\left(\frac{1}{\sqrt{Nn}}\right),\\
\frac{1}{NK}\mathbf{\Delta}_c &= \omega_\gamma(1-\delta^2)\left((A^2+\lambda_\alpha^2)\varrho^2 + \lambda_\alpha^2\bar{\kappa}^2\right) + \mathcal{O}_p\left(\frac{1}{\sqrt{Nn}}\right).
\end{aligned}$$

where $\kappa_{ik} = \sigma_{2_{ik}}/\sigma_{1_{ik}}$ and $n = \min_{i,k} n_{ik}$. With an abuse of notations, we let $\varrho^2 = (NK)^{-1}\sum(\kappa_{ik} - \bar{\kappa})^2$ and $\bar{\kappa} = (NK)^{-1}\sum \kappa_{ik}$. Therefore,

$$\hat{\tau} = \frac{\mathbf{\Delta}_{\tilde{B},b} - \mathbf{\Delta}_{\tilde{C},c}}{\mathbf{\Delta}_b + \mathbf{\Delta}_c} = \frac{\delta}{\sqrt{1-\delta^2}} + \mathcal{O}_p\left(\frac{1}{\sqrt{Nn}}\right) = \tau + \mathcal{O}_p\left(\frac{1}{\sqrt{Nn}}\right).$$

□

### A.8 Proof of Theorem 6

*Proof.* When $\mathbf{\Psi}$ and $\mathbf{\Lambda}$ are unknown, we first need to find the maximum likelihood estimators, which can be obtained by following the standard approach for a one-way mixed effects model (Searle et al., 2009). Let

$$Q_1 = \sum_i \sum_k \left(\hat{B}_{ik} - \bar{B}\right)^2, \quad Q_2 = \sum_i \left[\sum_k \left(\hat{B}_{ik} - \bar{B}\right)\right]^2, \quad Q_3 = \sum_i \sum_{j \neq k} \left(\hat{B}_{ik} - \bar{B}\right)\left(\hat{B}_{ij} - \bar{B}\right),$$

where $Q_2 = Q_1 + Q_3$, the profile likelihood function is

$$\ell_B(\tau) = -\frac{N(K-1)}{2}\log\left(KQ_1 - Q_2\right) - \frac{N}{2}\log Q_2.$$

For $C$, the profile likelihood funciton is

$$\ell_C(\tau) = -\frac{N(K-1)}{2}\log\left(KS_1 - S_2\right) - \frac{N}{2}\log S_2,$$

where

$$S_1 = \sum_i \sum_k \left(\hat{C}_{ik} - \bar{C}\right)^2, \quad S_2 = \sum_i \left[\sum_k \left(\hat{C}_{ik} - \bar{C}\right)\right]^2, \quad S_3 = \sum_i \sum_{j \neq k} \left(\hat{C}_{ik} - \bar{C}\right)\left(\hat{C}_{ik} - \bar{C}\right),$$

and $S_2 = S_1 + S_3$. Therefore, the profile likelihood function of $\tau$ is

$$\begin{aligned} \ell(\tau) &= \ell_B(\tau) + \ell_C(\tau) \\ &= -\frac{N(K-1)}{2}\log\left(KQ_1 - Q_2\right) - \frac{N}{2}\log Q_2 - \frac{N(K-1)}{2}\log\left(KS_1 - S_2\right) - \frac{N}{2}\log S_2. \end{aligned}$$

By taking derivative with respect to $\tau$ and set to zero, the solution of $\tau$ should satisfy the following

$$\frac{K-1}{KQ_1 - Q2}\frac{\partial Q_1}{\partial \tau} + \frac{K-1}{KS_1 - S_2}\frac{\partial S_1}{\partial \tau} = \frac{Q_2 - Q_1}{(KQ_1 - Q_2)Q_2}\frac{\partial Q_2}{\partial \tau} + \frac{S_2 - S_1}{(KS_1 - S_2)S_2}\frac{\partial S_2}{\partial \tau},$$

which is equivalent to solve the following seventh order polynomial

$$\begin{aligned} 0 \;=\; & \left[(K-2)S_1S_3\left((K-1)Q_1 + (K-2)Q_3\right) + (K-1)^2S_1^2Q_1 - (K-2)S_3^2Q_3 - (K-1)Q_1S_3^2 + (K-2)(K-1)Q_3S_1^2\right]\frac{\partial Q_1}{\partial \tau} \\ & + \left[(K-2)Q_1Q_3\left((K-1)S_1 + (K-2)S_3\right) + (K-1)^2S_1Q_1^2 - (K-2)S_3Q_3^2 - (K-1)S_1Q_3^2 + (K-2)(K-1)S_3Q_1^2\right]\frac{\partial S_1}{\partial \tau} \\ & + \left[Q_3S_3^2 - (K-2)S_1S_3Q_3 - (K-1)Q_3S_1^2\right]\frac{\partial Q_3}{\partial \tau} + \left[Q_3^2S_3 - (K-2)Q_1Q_3S_3 - (K-1)S_3Q_1^2\right]\frac{\partial S_3}{\tau} \end{aligned}$$

Using the analogous definitions in the proof of Theorem 3 in Section A.6 and Theorem 5 in Section A.7,

$$Q_1 = \mathbf{\Delta}_{\tilde{B}} - 2\tau\mathbf{\Delta}_{\tilde{B},b} + \tau^2\mathbf{\Delta}_b, \quad Q_3 = \mathbf{\Gamma}_{\tilde{B}} - 2\tau\mathbf{\Gamma}_{\tilde{B},b} + \tau^2\mathbf{\Gamma}_b,$$

$$S_1 = \mathbf{\Delta}_{\tilde{C}} + 2\tau\mathbf{\Delta}_{\tilde{C},c} + \tau^2\mathbf{\Delta}_c, \quad S_3 = \mathbf{\Gamma}_{\tilde{C}} + 2\tau\mathbf{\Gamma}_{\tilde{C},c} + \tau^2\mathbf{\Gamma}_c,$$

where $\mathbf{\Gamma}_{\tilde{B}} = \sum_i \sum_{j\neq k} y_{ik}y_{ij}$, $\mathbf{\Gamma}_{\tilde{B},b} = \sum_i \sum_{j\neq k} y_{ik}U_{ij}$, $\mathbf{\Gamma}_b = \sum_i \sum_{j\neq k} U_{ik}U_{ij}$, $\mathbf{\Gamma}_{\tilde{C}} = \sum_i \sum_{j\neq k} z_{ik}z_{ij}$, $\mathbf{\Gamma}_{\tilde{C},c} = \sum_i \sum_{j\neq k} z_{ik}V_{ij}$, and $\mathbf{\Gamma}_c = \sum_i \sum_{j\neq k} V_{ik}V_{ij}$. To solve the above polynomial function, we first look at the asymptotic behavior of the coefficients. Extending the results in the proof of Theorem 3, for the three-level model, we have the following results.

$$\begin{aligned}
\frac{1}{NK}\mathbf{\Delta}_{\tilde{B}} &= \psi_\beta^2 + \lambda_\beta^2 + \delta^2\varrho^2 + \mathcal{O}_p\left(\frac{1}{\sqrt{Nn}}\right), \\
\frac{1}{NK}\mathbf{\Delta}_{\tilde{B},b} &= \delta\sqrt{1-\delta^2}\varrho^2 + \mathcal{O}_p\left(\frac{1}{\sqrt{Nn}}\right), \\
\frac{1}{NK}\mathbf{\Delta}_b &= (1-\delta^2)\varrho^2 + \mathcal{O}_p\left(\frac{1}{\sqrt{Nn}}\right), \\
\frac{1}{NK}\mathbf{\Delta}_{\tilde{C}} &= \psi_\gamma^2 + \lambda_\gamma^2 + \delta^2\left((A^2+\lambda_\alpha^2)\varrho^2 + \lambda_\alpha^2\bar{\kappa}^2\right) + \mathcal{O}_p\left(\frac{1}{\sqrt{Nn}}\right), \\
\frac{1}{NK}\mathbf{\Delta}_{\tilde{C},c} &= -\delta\sqrt{1-\delta^2}\left((A^2+\lambda_\alpha^2)\varrho^2 + \lambda_\alpha^2\bar{\kappa}^2\right) + \mathcal{O}_p\left(\frac{1}{\sqrt{Nn}}\right), \\
\frac{1}{NK}\mathbf{\Delta}_c &= (1-\delta^2)\left((A^2+\lambda_\alpha^2)\varrho^2 + \lambda_\alpha^2\bar{\kappa}^2\right) + \mathcal{O}_p\left(\frac{1}{\sqrt{Nn}}\right), \\
\frac{1}{NK(K-1)}\mathbf{\Gamma}_{\tilde{B}} &= \psi_\beta^2 + \mathcal{O}_p\left(\frac{1}{\sqrt{Nn}}\right), \\
\frac{1}{NK(K-1)}\mathbf{\Gamma}_{\tilde{B},b} &= \mathcal{O}_p\left(\frac{1}{\sqrt{Nn}}\right), \\
\frac{1}{NK(K-1)}\mathbf{\Gamma}_b &= \mathcal{O}_p\left(\frac{1}{\sqrt{Nn}}\right), \\
\frac{1}{NK(K-1)}\mathbf{\Gamma}_{\tilde{C}} &= \psi_\gamma^2 + \delta^2\psi_\alpha^2\bar{\kappa}^2 + \mathcal{O}_p\left(\frac{1}{\sqrt{Nn}}\right), \\
\frac{1}{NK(K-1)}\mathbf{\Gamma}_{\tilde{C},c} &= -\delta\sqrt{1-\delta^2}\psi_\alpha^2\bar{\kappa}^2 + \mathcal{O}_p\left(\frac{1}{\sqrt{Nn}}\right), \\
\frac{1}{NK(K-1)}\mathbf{\Gamma}_c &= (1-\delta^2)\psi_\alpha^2\bar{\kappa}^2 + \mathcal{O}_p\left(\frac{1}{\sqrt{Nn}}\right).
\end{aligned}$$

For the $\mathbf{\Gamma}$.'s, we use the conclusion that for one-way mixed effects model with fixed $K$, all the MLEs are $\sqrt{N}$-consistent (Nie, 2007).

We represent the seventh order polynomial using the following

$$\theta_7\tau^7 + \theta_6\tau^6 + \theta_5\tau^5 + \theta_4\tau^4 + \theta_3\tau^3 + \theta_2\tau^2 + \theta_1\tau + \theta_0 = 0.$$

Under the assumption that $\varrho^2/\bar{\kappa}^2 = 1/\varpi$ and $1/\varpi = \mathcal{O}_p\left(1/\sqrt{Nn}\right)$, we have

$$\theta_7 \to 0, \quad \theta_6 \to 0, \quad \theta_5 \to 0, \quad \theta_4 \to 0,$$

as $N \to \infty$ and $n \to \infty$.

Let

$$\begin{aligned}
\Theta_1 &= (K-1)^2\lambda_\beta^2(\lambda_\beta^2 + K\psi_\beta^2)(\lambda_\alpha^2 - \psi_\alpha^2)(\lambda_\alpha^2 + (K-1)\psi_\alpha^2), \\
\Theta_2 &= (K-1)^2\lambda_\beta^2(\lambda_\beta^2 + K\psi_\beta^2)((K-1)\psi_\gamma^2(\lambda_\alpha^2 - \psi_\alpha^2) + \lambda_\alpha^2\lambda_\gamma^2),
\end{aligned}$$

then

$$
\begin{aligned}
\theta_3 &\to 2(1-\delta^2)^2\bar{\kappa}^2\Theta_1, \\
\theta_2 &\to -6(1-\delta^2)\delta\sqrt{1-\delta^2}\bar{\kappa}^2\Theta_1, \\
\theta_1 &\to (1-\delta^2)\left(2\Theta_2+6\delta^2\bar{\kappa}^2\Theta_1\right), \\
\theta_0 &\to -2\delta\sqrt{1-\delta^2}\left(\delta^2\bar{\kappa}^2\Theta_1-\Theta_2\right).
\end{aligned}
$$

Solve equation

$$
\theta_3\tau^3+\theta_2\tau^2+\theta_1\tau+\theta_0=0,
$$

we have

$$
\begin{aligned}
\tau_1 &= \tau^* = \frac{\delta}{\sqrt{1-\delta^2}}, \\
\tau_2 &= \tau^* - \frac{\sqrt{-\bar{\kappa}^2\Theta_1\Theta_2(1+\tau^{*2})}}{\bar{\kappa}^2\Theta_1}, \\
\tau_3 &= \tau^* + \frac{\sqrt{-\bar{\kappa}^2\Theta_1\Theta_2(1+\tau^{*2})}}{\bar{\kappa}^2\Theta_1}.
\end{aligned}
$$

For $\Theta_1$ and $\Theta_2$,

(1) if $\lambda_\alpha^2 \geq \psi_\alpha^2$, then $\Theta_1\Theta_2 \geq 0$, for $K \geq K_0 = 2$, $\tau = \tau^*$ is the unique real solution;

(2) if $\lambda_\alpha^2 < \psi_\alpha^2$, for $K \geq K_0$, $\Theta_1\Theta_2 \geq 0$, and $\tau = \tau^*$ is the unique real solution, where

$$
K_0 = \frac{\lambda_\gamma^2}{\psi_\gamma^2}\cdot\frac{\lambda_\alpha^2}{\psi_\alpha^2-\lambda_\alpha^2}+1.
$$

Therefore, asymptotically, the profile likelihood is maximized at a unique $\hat{\tau}$ value, and this $\hat{\tau}$ is a $\sqrt{Nn}$-consistent estimator of $\tau$. □

# B An example of non-identifiability

Here we give an example to explain the non-identifiability of the parameters in the first-level models (5) and (6). Under our proposed model, the joint distribution of $M$ and $R$ is

$$
\begin{pmatrix} M \\ R \end{pmatrix} \sim \mathcal{N}\left(\begin{pmatrix} A \\ (C+AB) \end{pmatrix}, \begin{pmatrix} \sigma_1^2 & B\sigma_1^2+\delta\sigma_1\sigma_2 \\ B\sigma_1^2+\delta\sigma_1\sigma_2 & B^2\sigma_1^2+2B\delta\sigma_1\sigma_2+\sigma_2^2 \end{pmatrix}\right). \tag{B.1}
$$

For a normal distribution, it is uniquely determined by mean and variance. For a bivariate normal distribution, the parameter space $\mathcal{S}$ is a subset of $\mathbb{R}^5$. Therefore, to have the same joint distribution of $(M, R)$, for given $\boldsymbol{c} = (c_1, c_2, c_3, c_4, c_5)^\top \in \mathcal{S}$, model parameter $\boldsymbol{\theta} = (A, B, C, \delta, \sigma_1, \sigma_2)^\top$ should satisfy the following equations,

$$
\begin{cases}
A = c_1, \\
C + AB = c_2, \\
\sigma_1^2 = c_3, \\
B\sigma_1^2+\delta\sigma_1\sigma_2 = c_4, \\
B^2\sigma_1^2+2B\delta\sigma_1\sigma_2+\sigma_2^2 = c_5.
\end{cases} \tag{B.2}
$$

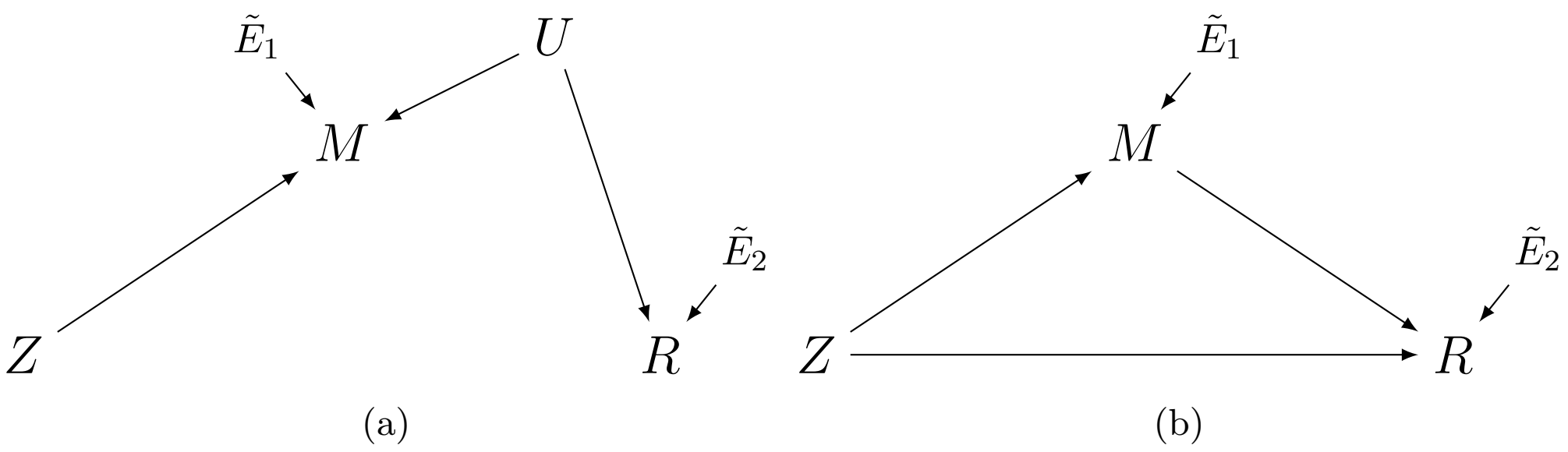

Figure B.1: Causal diagram under parameters (a) $\boldsymbol{\theta}_1$ and (b) $\boldsymbol{\theta}_2$, where both cases yield the same joint distribution of $M$ and $R$.

Notice that there are only five equations for six parameters, and thus $\boldsymbol{\theta} = (A, B, C, \delta, \sigma_1, \sigma_2)^\top$ cannot be uniquely determined. For example, both $\boldsymbol{\theta}_1 = (1, 0, 0, 1/2, 1, 2)^\top$ and $\boldsymbol{\theta}_2 = \left(1, 1, -1, 0, 1, \sqrt{3}\right)^\top$ yield the same joint distribution

$$\begin{pmatrix} M \\ R \end{pmatrix} \sim \mathcal{N}\left(Z\begin{pmatrix} 1 \\ 0 \end{pmatrix}, \begin{pmatrix} 1 & 1 \\ 1 & 4 \end{pmatrix}\right).$$

Therefore, these two distinct sets of parameters $\boldsymbol{\theta}_1$ and $\boldsymbol{\theta}_2$ will give the same observed data. However, under $\boldsymbol{\theta}_1$, there exits unmeasured confounding effect, and both direct and indirect effect are zero; while under $\boldsymbol{\theta}_2$, the ignorability assumption of $M$ holds, and the direct effect is $-1$ and the indirect effect is 1, which ends up with a zero total effect as well. The corresponding causal diagram under these two cases are shown in Figure B.1. From the figure, though the $(M, R)$ we observe are from the same joint distribution, the causal mechanisms are different.

Though the parameters cannot be uniquely determined, we can find the solution by fixing one of the parameters. Suppose $\delta$ is fixed, the solution for the rest parameters is

$$\begin{cases} A = c_1, \\ B = \frac{c_4}{c_3} - \frac{\delta}{c_3\sqrt{1-\delta^2}}\sqrt{c_3c_5 - c_4^2}, \\ C = c_2 - \frac{c_1c_4}{c_3} + \frac{c_1\delta}{c_3\sqrt{1-\delta^2}}\sqrt{c_3c_5 - c_4^2}, \\ \sigma_1^2 = c_3, \\ \sigma_2^2 = \frac{1}{c_3(1-\delta^2)}(c_3c_5 - c_4^2). \end{cases} \tag{B.3}$$

This also provides us the method of moments estimator. By plugging in the sample covariance, the solution for $\left(B, \sigma_1^2, \sigma_2^2\right)$ coincides with the results in Section A.4.

# C First-level model with covariates

Here we extend our first-level models (5) and (6) by considering $q$ measured covariates, denoted by $\boldsymbol{W}$ which is an $n \times q$ matrix, for both mediator $M$ and outcome $R$. The two linear equations are

$$\mathbf{M} = \mathbf{Z}A + \boldsymbol{W}\boldsymbol{\alpha} + \mathbf{E}_1, \tag{C.1}$$
$$\mathbf{R} = \mathbf{Z}C + \mathbf{M}B + \boldsymbol{W}\boldsymbol{\beta} + \mathbf{E}_2, \tag{C.2}$$

where $\boldsymbol{\alpha}$ and $\boldsymbol{\beta}$ are $q \times 1$ vectors; and $\mathbf{E}_1$ and $\mathbf{E}_2$ have the same joint distribution as introduced in the main text. For observational studies, with the existence of these "pre-treatment" covariates, the causal assumption (A4) in Section 2.1.1 is modified as,

- (A4*) $\{\mathbf{R}(\mathbf{z}', \mathbf{m}), \mathbf{M}(\mathbf{z})\} \perp \mathbf{Z} \mid \boldsymbol{W} = \boldsymbol{w}$; $\{\mathbf{E}_1(\mathbf{z}), \mathbf{E}_2(\mathbf{z}')\} \perp \mathbf{Z} \mid \boldsymbol{W} = \boldsymbol{w}$.

Let

$$\boldsymbol{X} = \begin{pmatrix} \mathbf{Z} & \boldsymbol{W} \end{pmatrix}, \quad \boldsymbol{\theta}_1 = \begin{pmatrix} A \\ \boldsymbol{\alpha} \end{pmatrix}, \quad \boldsymbol{\theta}_2 = \begin{pmatrix} C \\ \boldsymbol{\beta} \end{pmatrix},$$

under model (C.1), the conditional distribution of $\mathbf{M}$ and $\mathbf{R}$ are

$$\mathbf{M} \mid \boldsymbol{X} \sim \mathcal{N}\left(\mu_M, \sigma_1^2 \mathbf{I}_n\right),$$
$$\mathbf{R} \mid \mathbf{M}, \boldsymbol{X} \sim \mathcal{N}\left(\mu_{R|M}, \sigma_2^2(1-\delta^2)\mathbf{I}_n\right),$$

where $\mu_M = \boldsymbol{X}\boldsymbol{\theta}_1$, $\mu_{R|M} = \mathbf{M}B + \boldsymbol{X}\boldsymbol{\theta}_2 + \kappa(\mathbf{M} - \boldsymbol{X}\boldsymbol{\theta}_1)$, and $\kappa = \delta\sigma_2/\sigma_1$. The log-likelihood function of model (C.1) is

$$\begin{aligned} & \ell\left(\boldsymbol{\theta}_1, \boldsymbol{\theta}_2, B, \boldsymbol{\Sigma}\right) \\ = \; & -\frac{n}{2}\log\sigma_1^2\sigma_2^2(1-\delta^2) - \frac{1}{2\sigma_1^2}\left(\mathbf{M} - \boldsymbol{X}\boldsymbol{\theta}_1\right)^\top\left(\mathbf{M} - \boldsymbol{X}\boldsymbol{\theta}_1\right) \\ & - \frac{1}{2\sigma_2^2(1-\delta^2)}\left((\mathbf{R} - \mathbf{M}B - \boldsymbol{X}\boldsymbol{\theta}_2) - \kappa(\mathbf{M} - \boldsymbol{X}\boldsymbol{\theta}_1)\right)^\top\left((\mathbf{R} - \mathbf{M}B - \boldsymbol{X}\boldsymbol{\theta}_2) - \kappa(\mathbf{M} - \boldsymbol{X}\boldsymbol{\theta}_1)\right). \end{aligned}$$

**Theorem C.1.** *For a given $\delta$ value, the maximum likelihood estimator of $(\boldsymbol{\theta}_1, \boldsymbol{\theta}_2, B, \sigma_1, \sigma_2)$ in model* (C.1) *are given by*

$$\begin{aligned} \hat{\boldsymbol{\theta}}_1 &= (\boldsymbol{X}^\top\boldsymbol{X})^{-1}\boldsymbol{X}^\top\mathbf{M}, \\ \hat{\boldsymbol{\theta}}_2 &= (\boldsymbol{X}^\top(\mathbf{I}_n - \boldsymbol{P}_\mathbf{M})\boldsymbol{X})^{-1}\boldsymbol{X}^\top(\mathbf{I}_n - \boldsymbol{P}_\mathbf{M})\mathbf{R} + \hat{\kappa}(\boldsymbol{X}^\top\boldsymbol{X})^{-1}\boldsymbol{X}^\top\mathbf{M}, \\ \hat{B} &= (\mathbf{M}^\top\mathbf{M})^{-1}\mathbf{M}\left(\mathbf{I}_n - \boldsymbol{X}(\boldsymbol{X}^\top(\mathbf{I}_n - \boldsymbol{P}_\mathbf{M})\boldsymbol{X})^{-1}\boldsymbol{X}^\top(\mathbf{I}_n - \boldsymbol{P}_\mathbf{M})\right)\mathbf{R} - \hat{\kappa}, \\ \hat{\sigma}_1^2 &= \frac{1}{n}\mathbf{M}^\top(\mathbf{I}_n - \boldsymbol{P}_{\boldsymbol{X}})\mathbf{M}, \\ \hat{\sigma}_2^2 &= \frac{1}{n(1-\delta^2)}\mathbf{R}^\top(\mathbf{I}_n - \boldsymbol{P}_{\mathbf{M}\boldsymbol{X}} - \boldsymbol{P}_\mathbf{M})\mathbf{R}, \end{aligned}$$

*where $\hat{\kappa} = \delta\hat{\sigma}_2/\hat{\sigma}_1$; $\boldsymbol{P}_{\boldsymbol{X}} = \boldsymbol{X}(\boldsymbol{X}^\top\boldsymbol{X})^{-1}\boldsymbol{X}^\top$, $\boldsymbol{P}_\mathbf{M} = \mathbf{M}(\mathbf{M}^\top\mathbf{M})^{-1}\mathbf{M}^\top$, and $\boldsymbol{P}_{\mathbf{M}\boldsymbol{X}} = (\mathbf{I}_n - \boldsymbol{P}_\mathbf{M})\boldsymbol{X}(\boldsymbol{X}^\top(\mathbf{I}_n - \boldsymbol{P}_\mathbf{M})\boldsymbol{X})^{-1}\boldsymbol{X}^\top(\mathbf{I}_n - \boldsymbol{P}_\mathbf{M})$ are projection matrices.*

From the theorem, estimators of $(\boldsymbol{\theta}_2, B, \sigma_2)$ are functions of $\delta$. The bias correction of the BK estimator has the same form as in Theorem 1. For the three-level mediation model, one can consider a mixed effects model for all the coefficients including $A$, $B$, $C$, as well as $\boldsymbol{\alpha}$ and $\boldsymbol{\beta}$. We will leave this to our future research.

# D Bias analysis of KKB, BK and CMA-$\delta$ under the two-level model

In this section, we present some bias analysis for the existing methods under two-level model, including the linear mixed effects SEM (KKB) method, the Baron-Kenny (BK) method considering all the observations as independent trials and ignoring the between participant variation, as well as our first-level method (CMA-$\delta$) using the true $\delta$. Figure D.2 shows the estimation bias from a simulation study. In the simulation study, we use the same parameter settings as in Section 4 but under two-level model ($K = 1$ case). We observe the same pattern as in the three-level model simulation, i.e., KKB method yields the highest magnitude of bias, followed by BK and CMA-$\delta$.

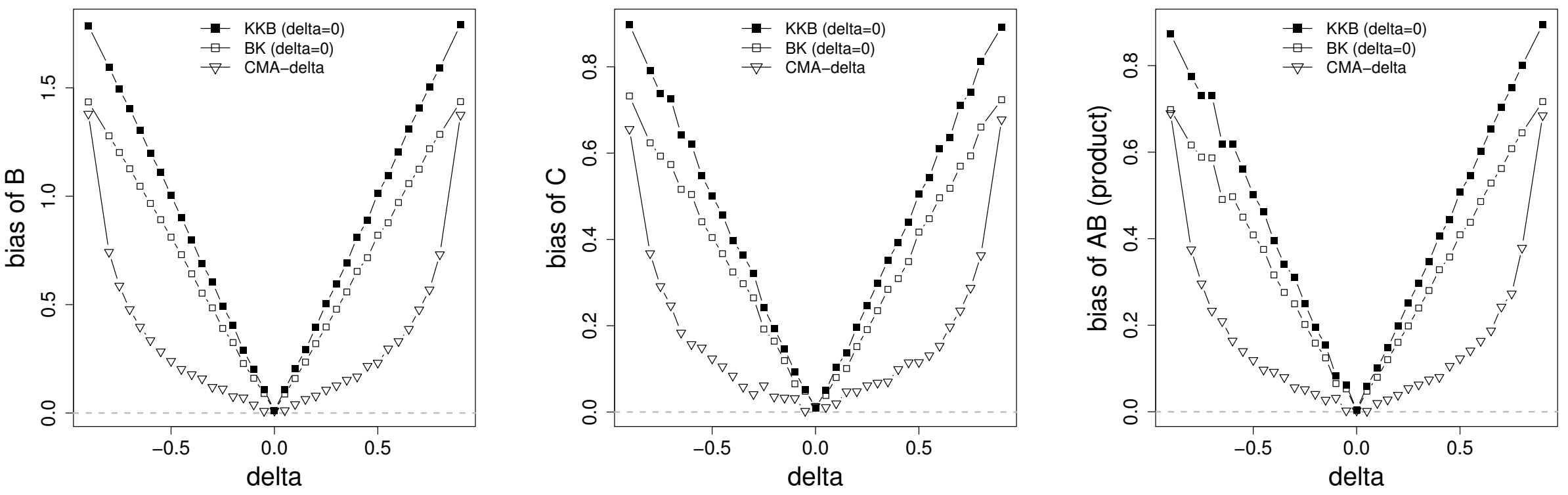

(a) (b) (c)

Figure D.2: Biases of KKB, BK and CMA-$\delta$ estimators in estimating (a) $B$, (b) $C$ and (c) $AB$ (product method). The solid squares are from KKB, the squares are from BK, and the triangles are from CMA-$\delta$.

From Theorem 1, the bias of BK estimator of $B_i$ under first-level model is

$$\text{Bias}\left(\hat{B}_i^{(\text{BK})} \mid B_i, \mathbf{\Sigma}_i\right) = \frac{\delta\sigma_{1_i}\sigma_{2_i}}{\sigma_{1_i}^2}, \tag{D.1}$$

since

$$\mathbb{E}\left(\hat{B}_i^{(BK)} \mid B_i, \mathbf{\Sigma}_i\right) = B_i + \frac{\delta\sigma_{1_i}\sigma_{2_i}}{\sigma_{1_i}^2}.$$

For simplicity, if we assume that for $\forall\ i$, $\sigma_{1_i} = \sigma_1$ and $\sigma_{2_i} = \sigma_2$, the bias of KKB estimator of $B$ is

$$\text{Bias}\left(\hat{B}^{(\text{KKB})}\right) = \frac{\delta\sigma_1\sigma_2}{\sigma_1^2}. \tag{D.2}$$

Both BK and our CMA-$\delta$ methods concatenate the observations from all the participants and treat them as independent samples. Under the two-level model, i.e., under models (1) and (3) with $K = 1$,

$$\begin{cases} \mathbf{M}_i = \mathbf{Z}_i A + \left(\mathbf{Z}_i \epsilon_i^A + \mathbf{E}_{1_i}\right) \triangleq \mathbf{Z}_i A + \boldsymbol{\xi}_{1_i}, \\ \mathbf{R}_i = \mathbf{Z}_i C + \mathbf{M}_i B + \left(\mathbf{Z}_i\left(\boldsymbol{\epsilon}_i^C + A\boldsymbol{\epsilon}_i^B + \boldsymbol{\epsilon}_i^A\boldsymbol{\epsilon}_i^B\right) + \mathbf{E}_{1_i}\boldsymbol{\epsilon}_i^B + \mathbf{E}_{2_i}\right) \triangleq \mathbf{Z}_i C + \mathbf{M}_i B + \boldsymbol{\xi}_{2_i}. \end{cases} \tag{D.3}$$

For simplicity, if we assume $\mathbf{\Lambda} = \text{diag}\{\lambda_\alpha^2, \lambda_\beta^2, \lambda_\gamma^2\}$, then we have

$$\begin{aligned} \text{Var}(\boldsymbol{\xi}_{1_i}) &= \lambda_\alpha^2 \mathbb{E}\left(\mathbf{Z}_i\mathbf{Z}_i^\top\right) + \sigma_{1_i}^2 \mathbf{I}_{n_i}, \\ \text{Var}(\boldsymbol{\xi}_{2_i}) &= (\lambda_\gamma^2 + A^2\lambda_\beta^2 + \lambda_\alpha^2\lambda_\beta^2)\mathbb{E}\left(\mathbf{Z}_i\mathbf{Z}_i^\top\right) + (\lambda_\beta^2\sigma_{1_i}^2 + \sigma_{2_i}^2)\mathbf{I}_{n_i}, \\ \text{Cov}(\boldsymbol{\xi}_{1_i}, \boldsymbol{\xi}_{2_i}) &= \delta\sigma_{1_i}\sigma_{2_i}\mathbf{I}_{n_i}. \end{aligned}$$

In our analysis, $\mathbf{Z}_i$'s are centralized first, if we assume $Z_{i1}, \ldots, Z_{in_i}$ are independently identically distributed with mean zero and variance $\tau^2$, $\mathbb{E}\left(\mathbf{Z}_i\mathbf{Z}_i^\top\right) = \tau^2\mathbf{I}_{n_i}$, $i = 1, \ldots, N$. Thus, we can approximate the variance covariance of $\left(\boldsymbol{\xi}_{1_i}, \boldsymbol{\xi}_{2_i}\right)$ as

$$\begin{aligned} \text{Var}(\boldsymbol{\xi}_{1_i}) &= \left(\lambda_\alpha^2\tau^2 + \sigma_{1_i}^2\right)\mathbf{I}_{n_i}, \\ \text{Var}(\boldsymbol{\xi}_{2_i}) &= \left((\lambda_\gamma^2 + A^2\lambda_\beta^2 + \lambda_\alpha^2\lambda_\beta^2)\tau^2 + \lambda_\beta^2\sigma_{1_i}^2 + \sigma_{2_i}^2\right)\mathbf{I}_{n_i}. \end{aligned}$$

Under the assumption that for $\forall\ i$, $\sigma_{1_i} = \sigma_1$ and $\sigma_{2_i} = \sigma_2$, the bias of BK estimator of $B$ is

$$\text{Bias}\left(\hat{B}^{(\text{BK})}\right) \approx \frac{\delta\sigma_1\sigma_2}{\sigma_1^2 + \lambda_\alpha^2\tau^2}. \tag{D.4}$$

Therefore, under our simulation parameter settings, the bias of KKB and BK estimators of $B$ is a linear function of $\delta$. Comparing (D.2) and (D.4), we see that the magnitude of $BK$ bias is lower.

For our CMA-$\delta$ method, after concatenating the observations from all the participants and assume that for $\forall\ i$, $\sigma_{1_i} = \sigma_1$ and $\sigma_{2_i} = \sigma_2$, the model can be written as

$$\begin{aligned} \mathbf{M} &= \mathbf{Z}A + \boldsymbol{\xi}_1, \\ \mathbf{R} &= \mathbf{Z}C + \mathbf{M}B + \boldsymbol{\xi}_2, \end{aligned}$$

where

$$\text{vec}\left[\begin{pmatrix}\boldsymbol{\xi}_1 & \boldsymbol{\xi}_2\end{pmatrix}\right] \dot{\sim} \mathcal{N}\left(\mathbf{0}, \begin{pmatrix} \tilde{\sigma}_1^2 & \tilde{\delta}\tilde{\sigma}_1^2\tilde{\sigma}_2^2 \\ \tilde{\delta}\tilde{\sigma}_1^2\tilde{\sigma}_2^2 & \tilde{\sigma}_2^2 \end{pmatrix} \otimes \mathbf{I}_n\right), \tag{D.5}$$

$$\begin{cases} \tilde{\sigma}_1^2 = \lambda_\alpha^2\tau^2 + \sigma_1^2, \\ \tilde{\sigma}_2^2 = (\lambda_\gamma^2 + A^2\lambda_\beta^2 + \lambda_\alpha^2\lambda_\beta^2)\tau^2 + \lambda_\beta^2\sigma_1^2 + \sigma_2^2, \\ \tilde{\delta} = \delta\sigma_1\sigma_2/\tilde{\sigma}_1\tilde{\sigma}_2, \end{cases}$$

and $n = \sum_i n_i$. From Theorem 1, for given $(\tilde{\delta}, \tilde{\sigma}_1, \tilde{\sigma}_2)$, CMA-$\delta$ estimator of $B$ is

$$\hat{B} = (\mathbf{M}^\top\mathbf{M})^{-1}\mathbf{M}^\top\left(\mathbf{I}_n - \mathbf{Z}(\mathbf{Z}^\top(\mathbf{I}_n - P_\mathbf{M})\mathbf{Z})^{-1}\mathbf{Z}^\top(\mathbf{I}_n - P_\mathbf{M})\right)\mathbf{R} - \frac{\tilde{\delta}\tilde{\sigma}_1\tilde{\sigma}_2}{\tilde{\sigma}_1^2}.$$

When $(\tilde{\sigma}_1, \tilde{\sigma}_2)$ is unknown, for a given $\delta$, the estimator of $\tilde{\sigma}_1$ is unrelated to $\delta$ while the estimator of $\tilde{\sigma}_2$ is a function of $\delta$ using our method. Thus, the CMA-$\delta$ estimator of $B$ with true $\delta$ is

$$\hat{B}(\delta) = (\mathbf{M}^\top\mathbf{M})^{-1}\mathbf{M}^\top\left(\mathbf{I}_n - \mathbf{Z}(\mathbf{Z}^\top(\mathbf{I}_n - P_\mathbf{M})\mathbf{Z})^{-1}\mathbf{Z}^\top(\mathbf{I}_n - P_\mathbf{M})\right)\mathbf{R} - \frac{\delta\hat{\tilde{\sigma}}_1\hat{\tilde{\sigma}}_2(\delta)}{\hat{\tilde{\sigma}}_1^2}, \tag{D.6}$$

where $\hat{\tilde{\sigma}}_1^2 = \hat{\tilde{\boldsymbol{\Sigma}}}_B(1,1)$,

$$\hat{\tilde{\sigma}}_2(\delta) = \frac{1}{\hat{\tilde{\sigma}}_1\sqrt{1-\delta^2}}\sqrt{\det(\hat{\tilde{\boldsymbol{\Sigma}}}_B)}, \quad \tilde{\boldsymbol{\Sigma}}_B = \begin{pmatrix} \tilde{\sigma}_1^2 & B\tilde{\sigma}_1^2 + \tilde{\delta}\tilde{\sigma}_1\tilde{\sigma}_2 \\ B\tilde{\sigma}_1^2 + \tilde{\delta}\tilde{\sigma}_1\tilde{\sigma}_2 & B^2\tilde{\sigma}_1^2 + 2B\tilde{\delta}\tilde{\sigma}_1\tilde{\sigma}_2 + \tilde{\sigma}_2^2 \end{pmatrix},$$

and $\det(\tilde{\boldsymbol{\Sigma}}_B) = \tilde{\sigma}_1^2\tilde{\sigma}_2^2 - \delta^2\sigma_1^2\sigma_2^2$; $\hat{\tilde{\boldsymbol{\Sigma}}}_B$ is obtained using (A.3). Therefore, for large $n$,

$$\mathbb{E}\left(\hat{B}(\delta)\right) \to B + \frac{\delta\sigma_1\sigma_2}{\tilde{\sigma}_1^2} - \frac{\delta}{\sqrt{1-\delta^2}}\frac{\sqrt{\tilde{\sigma}_1^2\tilde{\sigma}_2^2 - \delta^2\sigma_1^2\sigma_2^2}}{\tilde{\sigma}_1^2}, \tag{D.7}$$

and

$$\text{Bias}\left(\hat{B}(\delta)\right) \approx \frac{\delta\sigma_1\sigma_2}{\sigma_1^2 + \lambda_\alpha^2\tau^2} - \frac{\delta}{\sqrt{1-\delta^2}}\frac{\sqrt{\tilde{\sigma}_1^2\tilde{\sigma}_2^2 - \delta^2\sigma_1^2\sigma_2^2}}{\sigma_1^2 + \lambda_\alpha^2\tau^2}. \tag{D.8}$$

Comparing these three estimators, we have the following relationship that

$$\left|\text{Bias}\left(\hat{B}(\delta)\right)\right| < \left|\text{Bias}\left(\hat{B}^{(BK)}\right)\right| < \left|\text{Bias}\left(\hat{B}^{(KKB)}\right)\right|. \tag{D.9}$$

Analogously, we have the same conclusion for $C$ and $AB$ estimators.

# E Additional simulation results

In this section, we present the simulation results of our proposed first-level model (Section E.1), two-level model (Section E.2) and three-level model (Section E.3).

## E.1 First-level model simulation study

For the first-level model, we compare our estimators (CMA-$\delta$) introduced in Section 3.1, the Baron-Kenny (BK) method (Baron and Kenny, 1986), and the causal mediation (TYHKI) method (Imai et al., 2010). Our CMA-$\delta$ is implemented using our developed R package macc, BK method via standard regression, and TYHKI using R package mediation.

We simulate 100 independent and identically distributed (iid) samples from the following models. $Z$ is generated from a Bernoulli distribution with probability 0.5 to be one. In mediation analysis, the main objective is to identify the direct effect (denoted by $C$) and the indirect effect (denoted by $AB$ or $C' - C$). Using the product definition of the indirect effect, the null hypothesis for indirect effect is $\text{H}_0 : AB = 0$, which includes three scenarios for $A$ and $B$, i.e., *a)* $A = 0$, $B \neq 0$; *b)* $A \neq 0$, $B = 0$; and *c)* $A = B = 0$. We include all three cases in our simulation study and set $A = 0.5$, $B = -1$, and $C = 0.5$ under the alternatives. The errors are generated from a bivariate normal distribution with mean zero. The marginal variances of $E_1$ and $E_2$ are $\sigma_1^2 = 1$ and $\sigma_2^2 = 2^2$, respectively. The correlation $\delta$ is set to be one of the two scenarios: $\delta = 0$ and $\delta = 0.5$. All simulations are repeated 1000 times.

Table E.1 compares the estimates for $(A, C, B, C', AB, C' - C)$. In CMA-$\delta$ and TYHKI (when setting $\delta$ equal to the truth), the estimates for $(B, C, AB, C' - C)$ are unbiased and almost identical. The product and difference estimates for the indirect effect by CMA-$\delta$ are almost identical. As demonstrated in Theorem 1, ignoring nonzero $\delta$, BK and TYHKI can yield large biases in $B$ and $C$, and thus in the indirect and direct effect estimates. For example, under the third scenario in Table E.1 with $A = 0.5$ and $B = 0$, the BK and TYHKI ($\delta = 0$) estimates are about one, resulting in a nonzero indirect effect estimates, while the true value is the null $AB = 0$.

Table E.2 shows the confidence intervals and coverage probabilities for the methods compared under the first-level mediation model. From the table, CMA-$\delta$ obtains similar confidence intervals as well as coverage probabilities as the TYHKI bootstrap method with the true $\delta$ value does. All these coverage probabilities are close to the designated level. For the BK method and the TYHKI with $\delta = 0$, for some of the cases, the coverage probabilities are zero, indicating that the estimators from these two methods have large biases.

## E.2 Two-level model simulation study

For two-level model, the methods include our two-stage method (CMA-ts), our coordinate-descent method (CMA-h), our first-level method (CMA-$\delta$), the linear SEM (KKB) method (Kenny et al., 2003), and the Baron-Kenny (BK) method. For KKB method, we use the same trivariate regression model as ours as the higher-level model. Both KKB and BK methods assume there is no unmeasured confounding ($\delta = 0$). For our CMA-$\delta$ method, we will use the true $\delta$ value to estimate the rest parameters. Both BK and CMA-$\delta$ are for the first-level model, and thus we concatenate the data from all participants when applying these two methods. The CMA methods are implemented using our developed R package macc, KKB using the lem4 package, and BK via standard regression.

The total number of participants is set to be $N = 50$. For each participant, the number of trials is a random draw from the Poisson distribution with mean 100. We set the population level $A = 0.5$, $B = -1$, and $C = 0.5$. $\mathbf{\Lambda}$ is set to be diagonal with $\lambda_\alpha^2 = \lambda_\beta^2 = \lambda_\gamma^2 = 0.5$. For Table E.1: Average point estimates under the first-level models with a sample size of 100 over 1000 runs. The number in $(\cdot)$ is the empirical standard error of the estimates, and the number in $[\cdot]$ is the value that the BK and TYHKI estimates converge to theoretically, if different from the truth.

| *True* $\delta$ | *Method* | $A$ | $C$ $[C - A\delta\sigma_2/\sigma_1]$ | $B$ $[B + \delta\sigma_2/\sigma_1]$ | $C'$ | AB $[AB + A\delta\sigma_2/\sigma_1]$ | $C' - C$ $[C' - C + A\delta\sigma_2/\sigma_1]$ |
|---|---|---|---|---|---|---|---|
| | True value | 0.5 | 0.5 [0] | -1 [0] | 0 | -0.5 [0] | -0.5 [0] |
| | CMA-$\delta$ ($\delta = 0.5$) | 0.499 (0.198) | 0.513 (0.397) | -1.000 (0.201) | 0.013 (0.334) | -0.499 (0.224) | -0.499 (0.224) |
| 0.5 | BK | 0.499 (0.198) | 0.013 (0.346) | 0.003 (0.182) | 0.013 (0.334) | 0.001 (0.098) | 0.001 (0.098) |
| | TYHKI ($\delta = 0$) | 0.499 (0.198) | 0.013 (0.346) | 0.003 (0.182) | 0.013 (0.334) | 0.001 (0.098) | 0.001 (0.098) |
| | TYHKI ($\delta = 0.5$) | 0.499 (0.198) | 0.512 (0.397) | -1.000 (0.202) | 0.013 (0.334) | -0.499 (0.224) | -0.499 (0.224) |
| | True value | 0 | 0.5 [0.5] | -1 [0] | 0.5 | 0 [0] | 0 [0] |
| | CMA ($\delta = 0.5$) | 0.005 (0.202) | 0.496 (0.408) | -0.998 (0.196) | 0.492 (0.353) | -0.004 (0.207) | -0.004 (0.207) |
| 0.5 | BK | 0.005 (0.202) | 0.491 (0.355) | 0.002 (0.171) | 0.492 (0.353) | 0.001 (0.034) | 0.001 (0.034) |
| | TYHKI ($\delta = 0$) | 0.005 (0.202) | 0.491 (0.355) | 0.002 (0.171) | 0.492 (0.353) | 0.001 (0.034) | 0.001 (0.034) |
| | TYHKI ($\delta = 0.5$) | 0.005 (0.202) | 0.496 (0.408) | -0.999 (0.197) | 0.492 (0.353) | -0.004 (0.207) | -0.004 (0.207) |
| | True value | 0.5 | 0.5 [0] | 0 [1] | 0.5 | 0 [0.5] | 0 [0.5] |
| | CMA ($\delta = 0.5$) | 0.500 (0.206) | 0.506 (0.417) | 0.003 (0.203) | 0.507 (0.400) | 0.001 (0.111) | 0.001 (0.111) |
| 0.5 | BK | 0.500 (0.206) | 0.005 (0.357) | 1.004 (0.178) | 0.507 (0.400) | 0.501 (0.226) | 0.501 (0.226) |
| | TYHKI ($\delta = 0$) | 0.500 (0.206) | 0.005 (0.357) | 1.004 (0.178) | 0.507 (0.400) | 0.501 (0.226) | 0.501 (0.226) |
| | TYHKI ($\delta = 0.5$) | 0.500 (0.206) | 0.507 (0.417) | 0.002 (0.205) | 0.507 (0.400) | 0.000 (0.112) | 0.001 (0.111) |
| | True value | 0 | 0.5 [0.5] | 0 [1] | 0.5 | 0 [0] | 0 [0] |
| | CMA ($\delta = 0.5$) | 0.012 (0.202) | 0.495 (0.399) | -0.005 (0.205) | 0.496 (0.396) | 0.001 (0.042) | 0.001 (0.042) |
| 0.5 | BK | 0.012 (0.202) | 0.482 (0.336) | 1.000 (0.179) | 0.496 (0.396) | 0.014 (0.205) | 0.014 (0.205) |
| | TYHKI ($\delta = 0$) | 0.012 (0.202) | 0.482 (0.336) | 1.000 (0.179) | 0.496 (0.396) | 0.014 (0.205) | 0.014 (0.205) |
| | TYHKI ($\delta = 0.5$) | 0.012 (0.202) | 0.495 (0.399) | -0.006 (0.207) | 0.496 (0.396) | 0.001 (0.042) | 0.001 (0.042) |
| | True value | 0.5 | 0.5 [0.5] | -1 [-1] | 0 | -0.5 [-0.5] | -0.5 [-0.5] |
| 0 | CMA ($\delta = 0$) | 0.500 (0.203) | 0.486 (0.416) | -1.004 (0.201) | -0.016 (0.455) | -0.503 (0.236) | -0.503 (0.236) |
| | BK | 0.500 (0.203) | 0.486 (0.417) | -1.004 (0.201) | -0.016 (0.455) | -0.503 (0.237) | -0.503 (0.237) |
| | TYHKI ($\delta = 0$) | 0.500 (0.203) | 0.486 (0.417) | -1.004 (0.201) | -0.016 (0.455) | -0.503 (0.237) | -0.503 (0.237) |

Table E.2: 95% confidence intervals (CI) and coverage probabilities (CP) of the estimates with a sample size of 100 over 1000 runs.

| *True* $\delta$ | *Method* | $A$ | | $C$ | | $B$ | | $AB_p$ | | $AB_d$ | |
|---|---|---|---|---|---|---|---|---|---|---|---|
| | | *CI* | *CP* | *CI* | *CP* | *CI* | *CP* | *CI* | *CP* | *CI* | *CP* |
| | True value | 0.5 | | 0.5 | | -1 | | -0.5 | | -0.5 | |
| | CMA ($\delta = 0.5$) | (0.109, 0.888) | 0.952 | (-0.286, 1.311) | 0.950 | (-1.340, -0.659) | 0.917 | (-0.928, -0.070) | 0.933 | (-0.928, -0.070) | 0.933 |
| $\delta = 0.5$ | BK | (0.111, 0.886) | 0.951 | (-0.682, 0.707) | 0.727 | (-0.337, 0.344) | 0 | (-0.184, 0.186) | 0.012 | (-0.184, 0.186) | 0.012 |
| | TYHKI ($\delta = 0$) | (0.107, 0.890) | 0.953 | (-0.684, 0.710) | 0.734 | (-0.337, 0.344) | 0 | (-0.209, 0.210) | 0.017 | (-0.203, 0.205) | 0.031 |
| | TYHKI ($\delta = 0.5$) | (0.107, 0.890) | 0.953 | (-0.286, 1.311) | 0.952 | (-1.340, -0.659) | 0.917 | (-0.928, -0.070) | 0.929 | (-0.928, -0.070) | 0.933 |
| | True value | 0 | | 0.5 | | -1 | | 0 | | 0 | |
| | CMA ($\delta = 0.5$) | (-0.390, 0.401) | 0.951 | (-0.294, 1.286) | 0.943 | (-1.338, -0.659) | 0.921 | (-0.402, 0.395) | 0.961 | (-0.402, 0.395) | 0.961 |
| $\delta = 0.5$ | BK | (-0.388, 0.398) | 0.950 | (-0.191, 1.172) | 0.946 | (-0.338, 0.341) | 0 | (-0.083, 0.086) | 1.000 | (-0.083, 0.086) | 1 |
| | TYHKI ($\delta = 0$) | (-0.392, 0.402) | 0.952 | (-0.188, 1.170) | 0.939 | (-0.338, 0.341) | 0 | (-0.122, 0.125) | 0.989 | (-0.118, 0.120) | 0.999 |
| | TYHKI ($\delta = 0.5$) | (-0.392, 0.402) | 0.952 | (-0.294, 1.285) | 0.943 | (-1.338, -0.659) | 0.921 | (-0.402, 0.395) | 0.960 | (-0.402, 0.395) | 0.961 |
| | True value | 0.5 | | 0.5 | | 0 | | 0 | | 0 | |
| | CMA ($\delta = 0.5$) | (0.106, 0.894) | 0.937 | (-0.301, 1.312) | 0.938 | (-0.337, 0.343) | 0.901 | (-0.188, 0.190) | 0.980 | (-0.188, 0.190) | 0.980 |
| $\delta = 0.5$ | BK | (0.108, 0.892) | 0.937 | (-0.696, 0.706) | 0.716 | (0.664, 1.344) | 0 | (0.066, 0.937) | 0.355 | (0.066, 0.937) | 0.355 |
| | TYHKI ($\delta = 0$) | (0.104, 0.896) | 0.942 | (-0.698, 0.709) | 0.717 | (0.664, 1.344) | 0 | (0.052, 0.927) | 0.394 | (0.094, 0.974) | 0.291 |
| | TYHKI ($\delta = 0.5$) | (0.104, 0.896) | 0.942 | (-0.300, 1.313) | 0.938 | (-0.337, 0.343) | 0.901 | (-0.190, 0.190) | 0.981 | (-0.188, 0.190) | 0.980 |
| | True value | 0 | | 0.5 | | 0 | | 0 | | 0 | |
| | CMA ($\delta = 0.5$) | (-0.377, 0.400) | 0.948 | (-0.286, 1.275) | 0.950 | (-0.347, 0.336) | 0.901 | (-0.092, 0.094) | 1 | (-0.092, 0.094) | 1 |
| $\delta = 0.5$ | BK | (-0.375, 0.398) | 0.946 | (-0.191, 1.155) | 0.952 | (0.659, 1.341) | 0 | (-0.379, 0.407) | 0.964 | (-0.379, 0.407) | 0.964 |
| | TYHKI ($\delta = 0$) | (-0.379, 0.402) | 0.947 | (-0.191, 1.154) | 0.950 | (0.659, 1.341) | 0 | (-0.388, 0.414) | 0.968 | (-0.387, 0.416) | 0.963 |
| | TYHKI ($\delta = 0.5$) | (-0.379, 0.402) | 0.947 | (-0.286, 1.275) | 0.951 | (-0.347, 0.336) | 0.901 | (-0.092, 0.095) | 1 | (-0.092, 0.094) | 1 |
| | True value | 0.5 | | 0.5 | | -1 | | -0.5 | | -0.5 | |
| $\delta = 0$ | CMA ($\delta = 0$) | (0.107, 0.893) | 0.950 | (-0.324, 1.296) | 0.947 | (-1.396, -0.612) | 0.951 | (-0.951, -0.054) | 0.939 | (-0.951, -0.054) | 0.939 |
| | BK | (0.109, 0.891) | 0.949 | (-0.320, 1.293) | 0.945 | (-1.396, -0.612) | 0.949 | (-0.949, -0.056) | 0.936 | (-0.949, -0.056) | 0.936 |
| | TYHKI ($\delta = 0$) | (0.105, 0.895) | 0.951 | (-0.320, 1.290) | 0.947 | (-1.396, -0.612) | 0.949 | (-0.935, -0.039) | 0.938 | (-0.994, -0.093) | 0.943 |

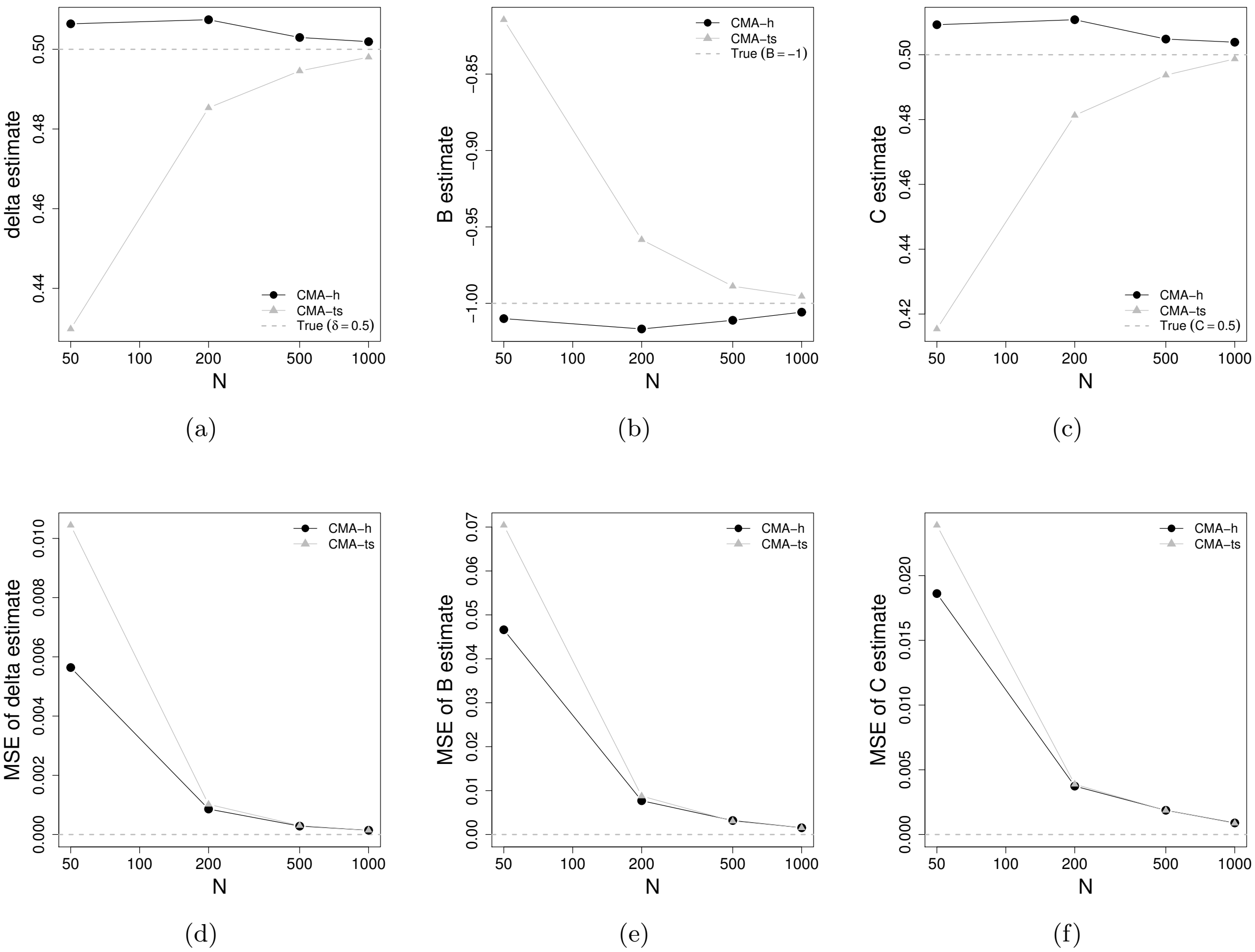

Figure E.3: Average point estimate of (a) $\delta$, (b) $B$, and (c) $C$, and the mean squared errors of (d) $\hat{\delta}$, (e) $\hat{B}$ and (f) $\hat{C}$. The solid circles are from the coordinate-descent algorithm, and the solid diamonds are from the two-stage algorithm. The true values is shown by the dashed lines.

each participant, the variances of the measurement errors in the mediation model are $\sigma_{1_i} = 1$ and $\sigma_{2_i} = 2$ for $i = 1, \ldots, N$. The correlation between the errors is 0.5. The simulation is repeated 200 times.

Figure E.3 presents the finite sample performance of our CMA-ts and CMA-h methods with $N = n_i = 50, 200, 500, 100$. From the figure, the estimates of $\delta$, as well as the estimates of $B$ and $C$, approach the true values as the number of trials and the number of participants increases. Consistent with the results under three-level model, CMA-h achieves lower bias.

Figure E.4 tests the robustness of our method to the strength of the unmeasured confounding effect. From the figure, our three-level methods (CMA-ts and CMA-h) yield good estimates of $\delta$, and lower biases in estimating $B$ and $C$ than the competing methods with $\delta$ value varying in range $(-1, 1)$. CMA-h has the lowest bias among all methods. The biases in KKB, BK and CMA-$\delta$ methods increase as $\delta$ magnitude increases. As explained in Section D, KKB has the largest bias, followed by BK and CMA-$\delta$, and the bias in estimating $B$ and $C$ of BK and KKB methods are linear function of $\delta$ under our parameter setting.

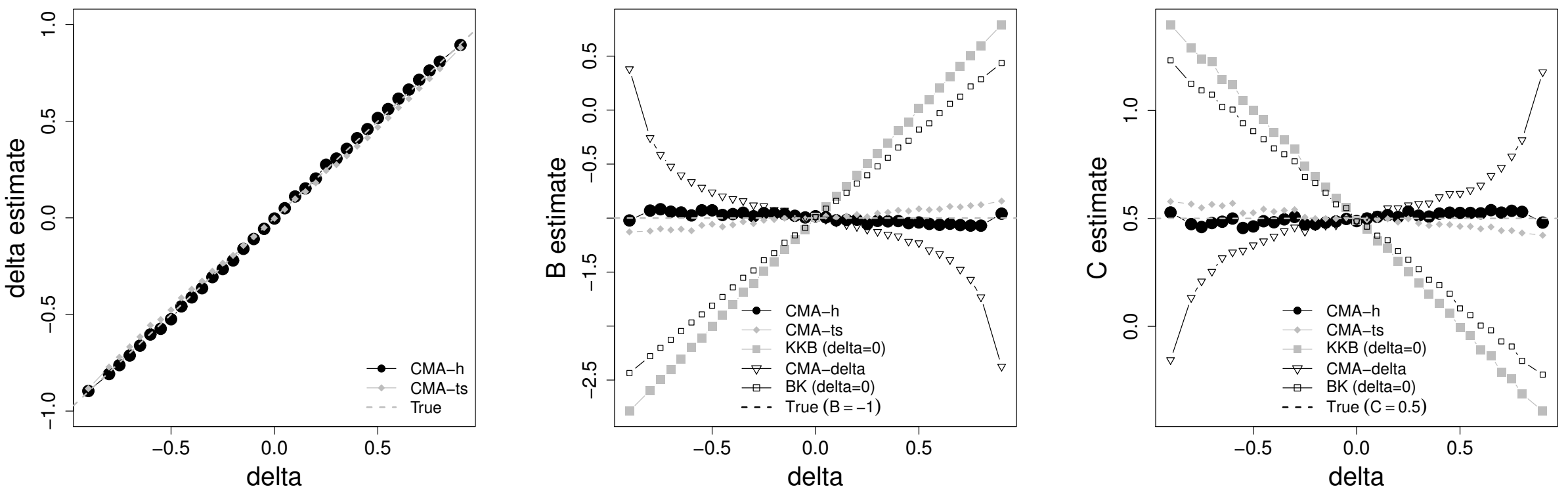

(a) (b) (c)

Figure E.4: Point estimate of (a) $\delta$, (b) $B$, and (c) $C$. The solid circles are from the coordinate-descent algorithm, the solid diamonds are from the two-stage algorithm, the solid squares are from KKB, the squares are from BK, and the triangles are from CMA-$\delta$. The true values is shown by the dashed lines.

### E.3 Three-level model simulation study

In this section, we present some additional simulation results of three-level model. The simulation settings are described in Section 4.

Table E.3 shows the estimates of the variance components from each method with unknown $\delta$ value. Since BK and CMA-$\delta$ are methods for one-level data, the estimate of the variance components are not available. CMA-ts slightly overestimates $\lambda_\alpha^2$, $\lambda_\beta^2$ and $\lambda_\gamma^2$ since it does not take the extra variation in estimating $b_{ik}$ in the first-level model into consideration. CMA-h underestimates the variances, probably due to the known bias in h-likelihood. CMA-m yields a better estimation in the variance components. When the true $\delta$ is 0.5, KKB method estimates about twice of the true variation in $C_{ik}$'s. A possible reason is that when $\delta \neq 0$, the coefficient estimates from KKB are biased.

Figure E.5 shows the estimates of $B$, $C$ and $AB_p$ by our CMA methods. As the number of participants $N$ and the number of sessions $K$ increase, the estimates approach to the true value. Similar to the estimate of $\delta$, CMA-h yields lower bias than the other two approaches.

The estimates of $B$ and $C$ with varying $\delta$ are shown in Figure E.6. Compared to BK, KKB and CMA-$\delta$ methods, the CMA-ts, CMA-h and CMA-m can robustly estimate the coefficients under each $\delta$ value. The BK, KKB and CMA-$\delta$ methods only work when $\delta = 0$. As $|\delta|$ approaches one, the biases of these three methods increase dramatically.

## F Additional analysis results on the fMRI experiment

### F.1 Imaging protocol

Ninety-six right-handed participants were recruited in the study at Yale University. The participants were asked to complete a STOP/GO task, where STOP and GO trials were randomly intermixed with probability 3/4 to be a GO trial ($Z = 0$) and with probability 1/4 to be a STOP trial ($Z = 1$). Each participant was scanned under a 3T scanner (Siemens Trio) for four ten-minute sessions of task. T1-weighted images were acquired for slice localization. In each of the four sessions,

Table E.3: Average variance estimates of CMA-ts, CMA-h and CMA-m with $K = 4$ sessions and $N = 50$ participants over 200 runs, as well as the estimates of KKB method with $\delta = 0$.

| *Method* | $\sigma^2_\alpha$ | $\sigma^2_\beta$ | $\sigma^2_\gamma$ | $\lambda^2_\alpha$ | $\lambda^2_\beta$ | $\lambda^2_\gamma$ |
|---|---|---|---|---|---|---|
| True value | 0.5 | 0.5 | 0.5 | 0.5 | 0.5 | 0.5 |
| CMA-ts | 0.509 | 0.504 | 0.484 | 0.516 | 0.539 | 0.636 |
| CMA-h | 0.382 | 0.363 | 0.298 | 0.412 | 0.431 | 0.511 |
| CMA-m | 0.509 | 0.504 | 0.485 | 0.473 | 0.460 | 0.523 |
| KKB ($\delta = 0$) | 0.509 | 0.503 | 1.004 | 0.516 | 0.530 | 1.107 |
| True value | 0.5 | 0.5 | 0.5 | 0.5 | 0.5 | 0.5 |
| CMA-ts | 0.487 | 0.491 | 0.492 | 0.517 | 0.539 | 0.608 |
| CMA-h | 0.360 | 0.352 | 0.311 | 0.413 | 0.430 | 0.488 |
| CMA-m | 0.487 | 0.492 | 0.491 | 0.474 | 0.460 | 0.497 |
| KKB ($\delta = 0$) | 0.487 | 0.492 | 0.993 | 0.517 | 0.529 | 1.066 |
| True value | 0.5 | 0.5 | 0.5 | 0.5 | 0.5 | 0.5 |
| CMA-ts | 0.515 | 0.505 | 0.503 | 0.523 | 0.536 | 0.626 |
| CMA-h | 0.386 | 0.365 | 0.317 | 0.418 | 0.429 | 0.501 |
| CMA-m | 0.514 | 0.505 | 0.504 | 0.480 | 0.458 | 0.514 |
| KKB ($\delta = 0$) | 0.515 | 0.505 | 1.026 | 0.523 | 0.527 | 1.095 |
| True value | 0.5 | 0.5 | 0.5 | 0.5 | 0.5 | 0.5 |
| CMA-ts | 0.480 | 0.497 | 0.498 | 0.523 | 0.540 | 0.616 |
| CMA-h | 0.353 | 0.357 | 0.316 | 0.418 | 0.432 | 0.493 |
| CMA-m | 0.480 | 0.498 | 0.501 | 0.479 | 0.462 | 0.503 |
| KKB ($\delta = 0$) | 0.480 | 0.499 | 0.981 | 0.523 | 0.531 | 1.098 |
| True value | 0.5 | 0.5 | 0.5 | 0.5 | 0.5 | 0.5 |
| CMA-ts | 0.502 | 0.523 | 0.486 | 0.517 | 0.529 | 0.623 |
| CMA-h | 0.377 | 0.386 | 0.309 | 0.413 | 0.422 | 0.497 |
| CMA-m | 0.502 | 0.523 | 0.490 | 0.476 | 0.452 | 0.503 |
| KKB ($\delta = 0$) | 0.502 | 0.523 | 0.489 | 0.517 | 0.529 | 0.626 |

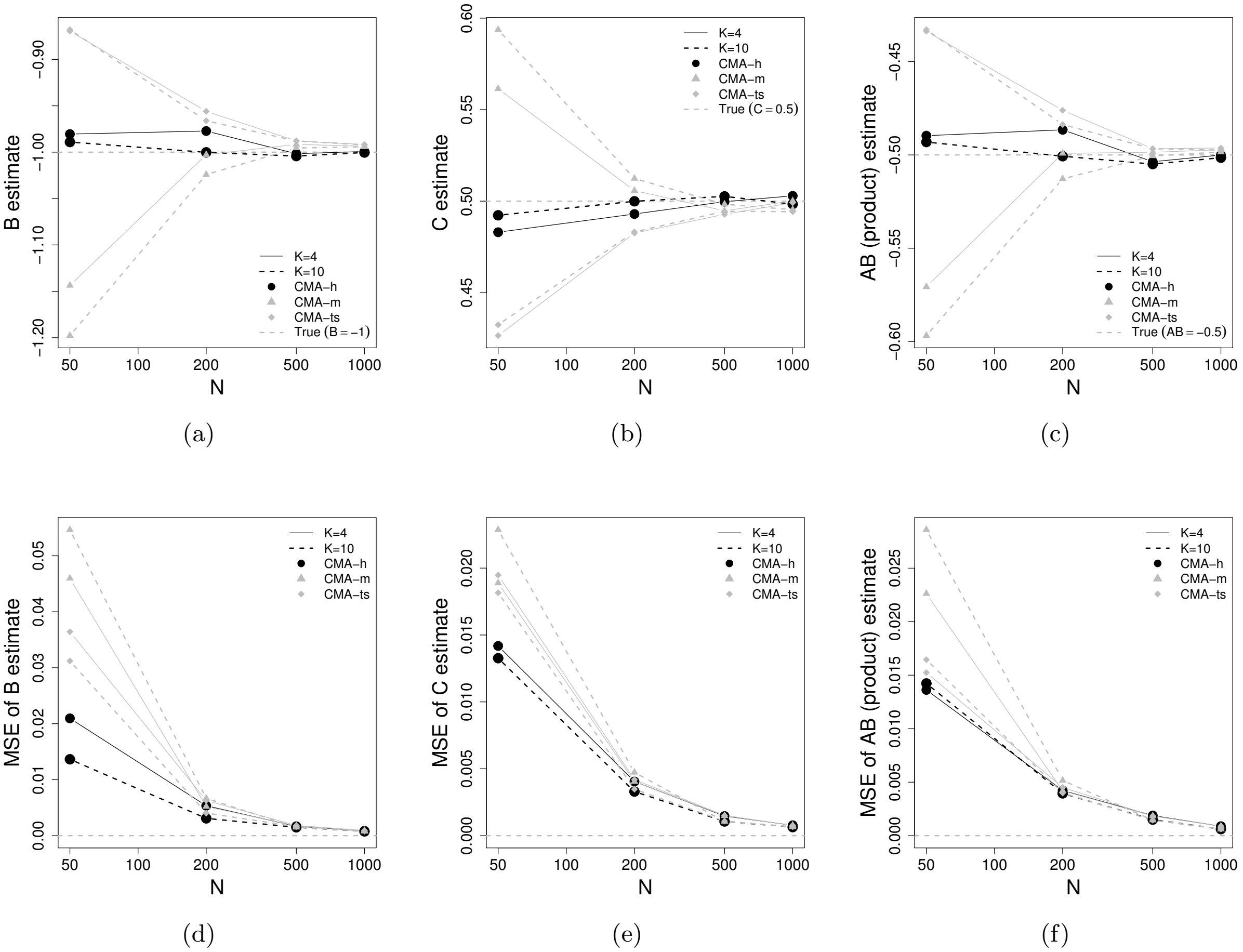

Figure E.5: Average point estimates of (a) $B$, (b) $C$, and (c) $AB$ (product), and the mean squared errors of (d) $\hat{B}$, (e) $\hat{C}$, and (f) $\widehat{AB}_p$. The solid circles are from CMA-h, the solid triangles are from CMA-m, and the solid diamonds are from CMA-ts. The true parameter values is shown by the dashed lines.

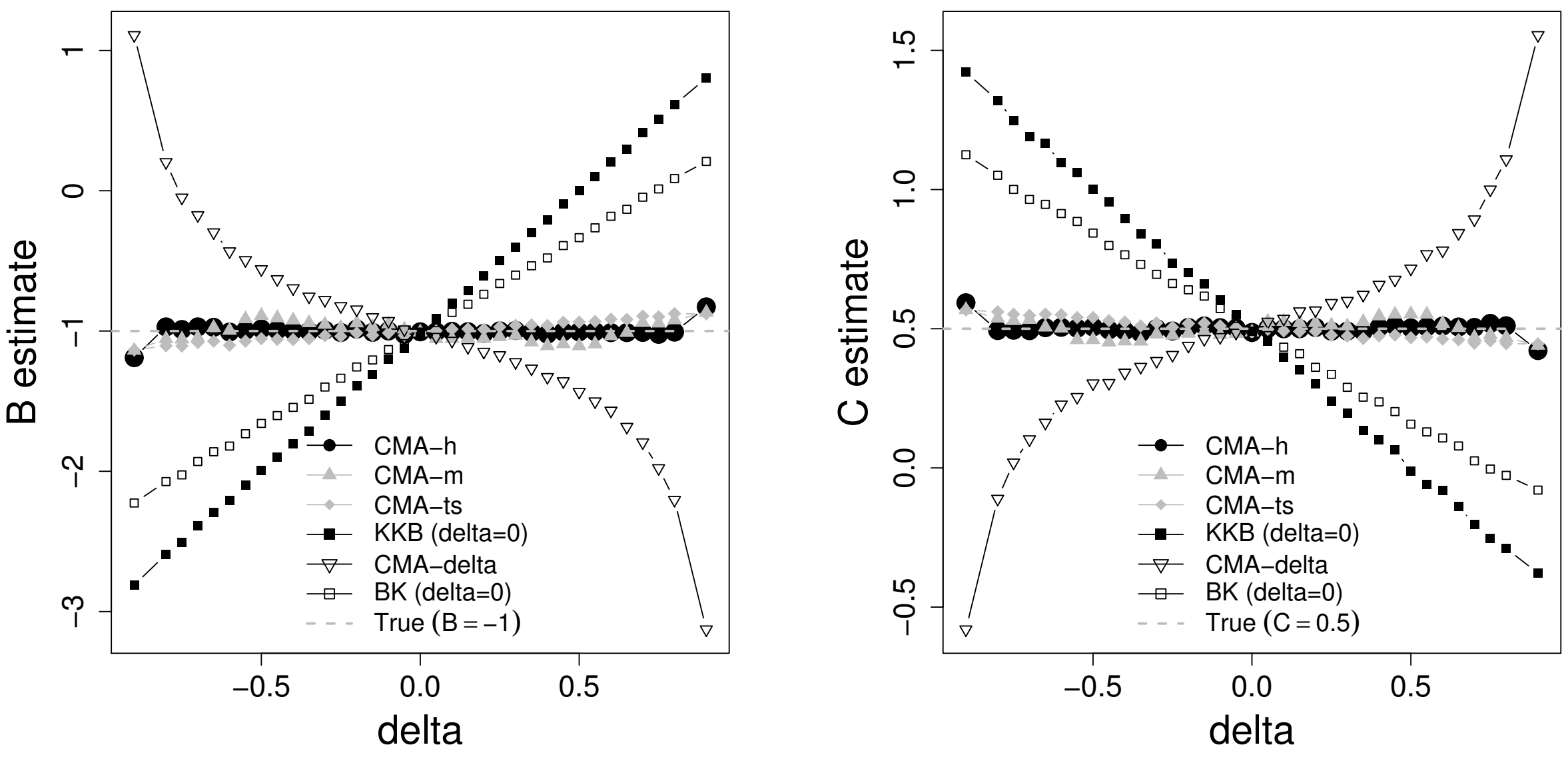

(a) (b)

Figure E.6: Point estimate of (a) $B$ and (b) $C$. The simulations are repeated 200 times with $K = 4$ sessions and $N = 50$ participants. The solid circles are from CMA-h, the solid triangles are from CMA-m, the solid diamonds are from CMA-ts, the solid squares are from KKB with $\delta = 0$, the triangles are from CMA-$\delta$ with true $\delta$ value and the squares are from BK with $\delta = 0$. The dashed lines are the true parameter value

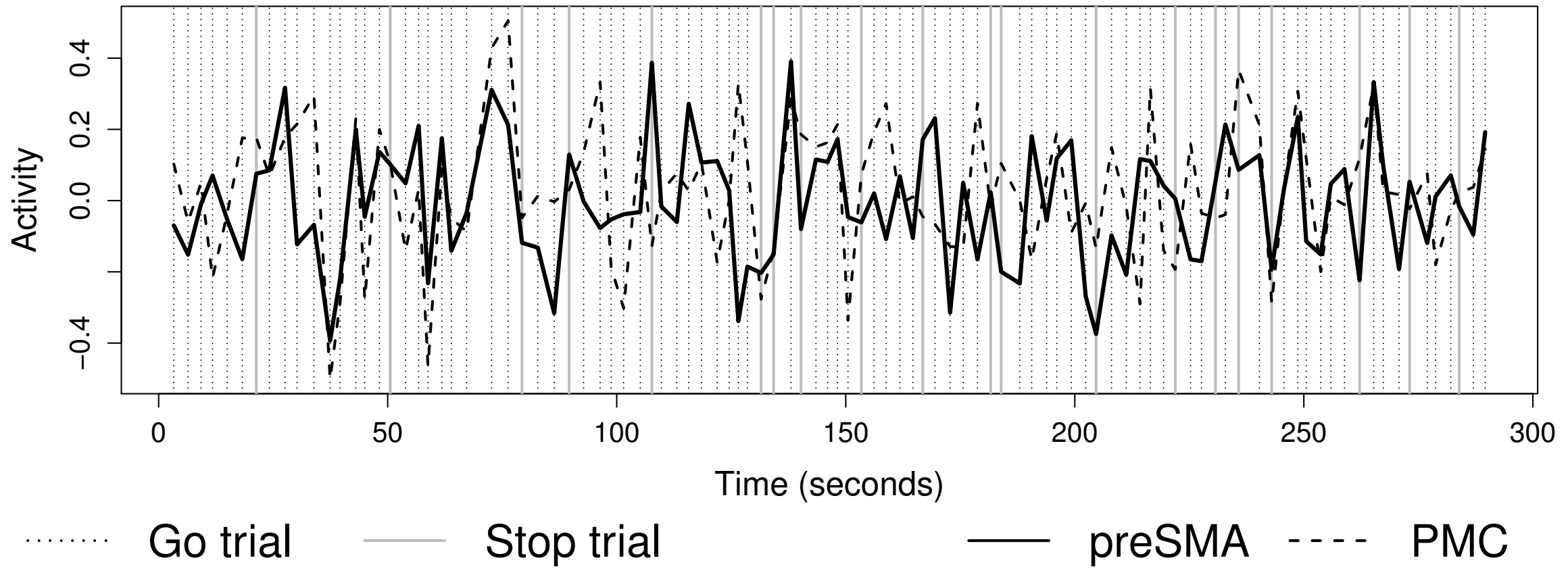

Figure F.7: Single trial brain activity of preSMA (black solid line) and PMC (black dashed line) for a representative session of representative participant.

300 images were acquired with 32 axial slices (spin echo parallel to the anterior commissure-posterior commissure, TR = 2000 ms, TE = 25 ms, bandwidth = 300 Hz/pixel, flip angle = 85°; field of view, 220 × 220 mm, matrix 64 × 64, 32 slices with slice thickness 4 mm and no gap). Anatomical and functional MRI images were preprocessed first with Statistical Parametric Mapping version 5 (SPM5) (Wellcome Department of Imaging Neuroscience, University College London, Londonj, UK), using standard processing steps, including slice timing correction, realignment, coregistration, normalization, and smoothing.

## F.2 Extraction of single trial activations or beta values

In the fMRI experiment, 96 participants are recruited and four experimental sessions are conducted for each participant. Each session is about ten minutes in length with a median of 90 randomized trials. In each session, the randomized stimuli are given at different time points. Brian activities for each stimulus are extracted using a GLM approach. In the model, the dependent variable is the BOLD signal at each time point that an event occurs; the design matrices are the hemodynamic response function (HRF) and its first derivative, which is upsampled first and then downsampled to attain better time resolution. A small ridge regularization term ($\lambda = 0.01$) is added to GLM to ensure stability of the estimates. Figure F.7 shows the extracted brain activity under each trail in both preSMA and PMC regions for a representative session of representative participant.

## F.3 Checking the normality assumption in the first-level model

Figure F.8 shows the Quantile-Quantile (Q-Q) plot of the mediator and outcome measurement for a representative session of representative participant, indicating that the normality assumption in the first-level model is valid.

## F.4 Sensitivity analysis under the first-level model

Figure F.9 presents the sensitivity analysis of the indirect effect estimates from the concatenated data across participants and sessions under the first-level model. The figure shows that the estimates depend heavily on $\delta$ and their confidence intervals may not overlap as $\delta$ varies.

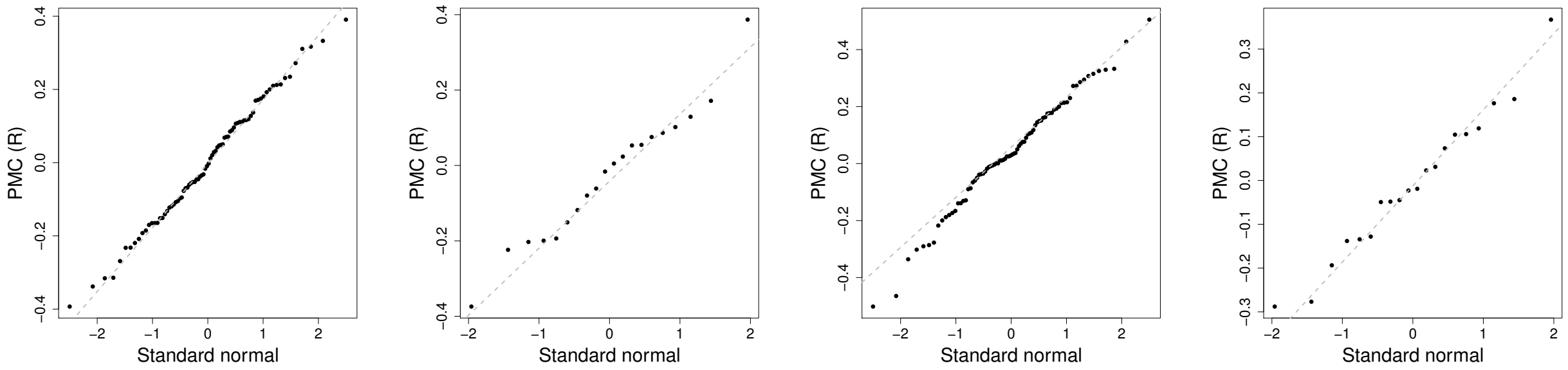

(a) (b) (c) (d)

Figure F.8: Q-Q plot of preSMA measure of (a) GO trial and (b) STOP trial, and PMC measure of (c) GO trial and (d) STOP trial in a representative session of representative participant.

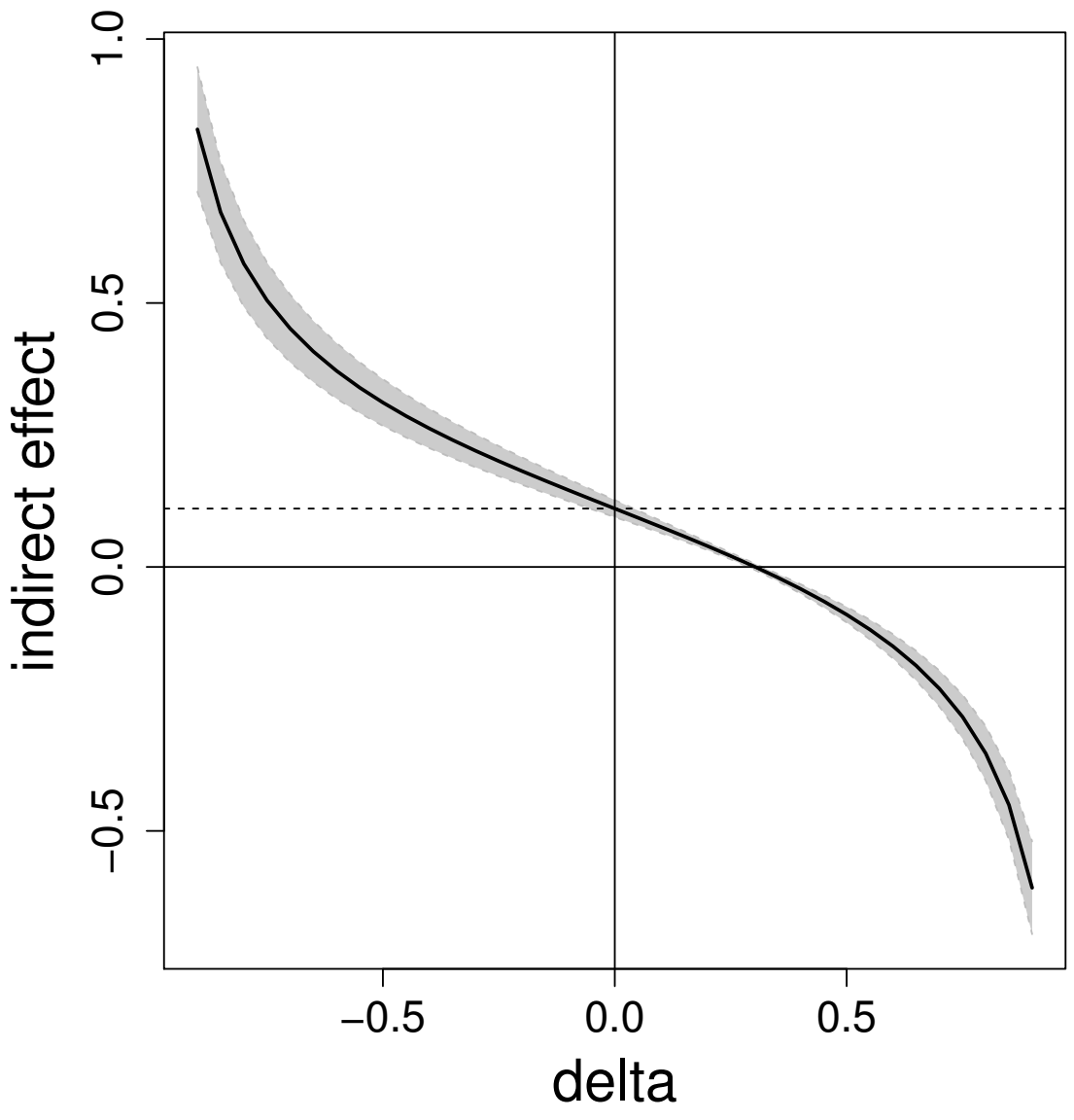

Figure F.9: Sensitivity analysis of the indirect effect estimates under the first-level model by concatenating trials from all participants and all sessions.

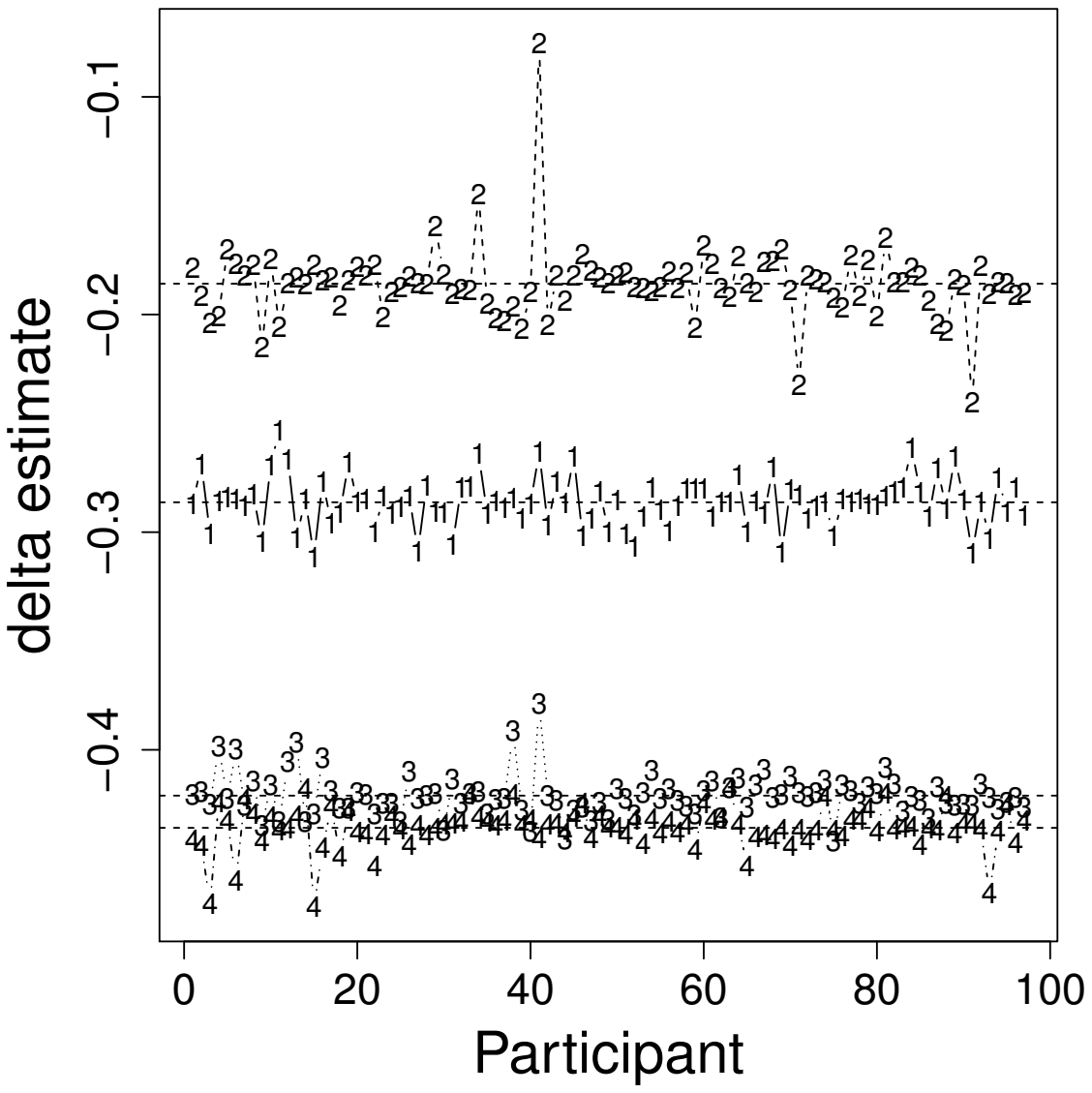

Figure F.10: Leave-one-out estimates of $\delta$ for each session under two-level model introduced in Section 2.2 using the coordinate-descent algorithm. The numbers in the figure denote the corresponding session number.

### F.5 Checking the assumption that $\delta_{ik} = \delta$

We have shown that $\delta$ is not identifiable under our first-level model. It is impossible to check the assumption that $\delta$ is a constant across both sessions and participants, but we can test the assumption that for each session $\delta$ is a constant across participants (assumption (A5$''$)) and obtain the estimate of $\delta_k$ $(k = 1, \ldots, K)$ for each session using the two-level model introduced in Section 2.2. Figure F.10 presents the leave-one-out estimates of $\delta_k$ for each session. From the figure, $\delta$ estimates of sessions 3 and 4 are very close, and estimates of these four sessions are all similar with negative values. These negative estimates are consistent with our estimates using the three-level model under assumption (A5$'$) in the main text. Therefore, we assume $\delta$ to be a constant across sessions as well. We leave the relaxation of this assumption to future study. In the future, we will also consider a random effects model for $\delta$, if it is identifiable, to account for within/between participant variations.

### F.6 Comparison of coefficient estimates from the three-level model

Figure F.11 shows the scatter plot of the estimated coefficients using our CMA methods under the proposed three-level model compared to the estimates from KKB method. From the figure, we can see the difference between our CMA methods and KKB method, especially in estimating $B$. The figure also explains that the normality assumption for the coefficients in the mixed effects model is appropriate.

### F.7 Average estimates and 95% confidence intervals of the rest parameters

Table F.4 compares the inference of the rest parameters using the first-level and three-level methods.

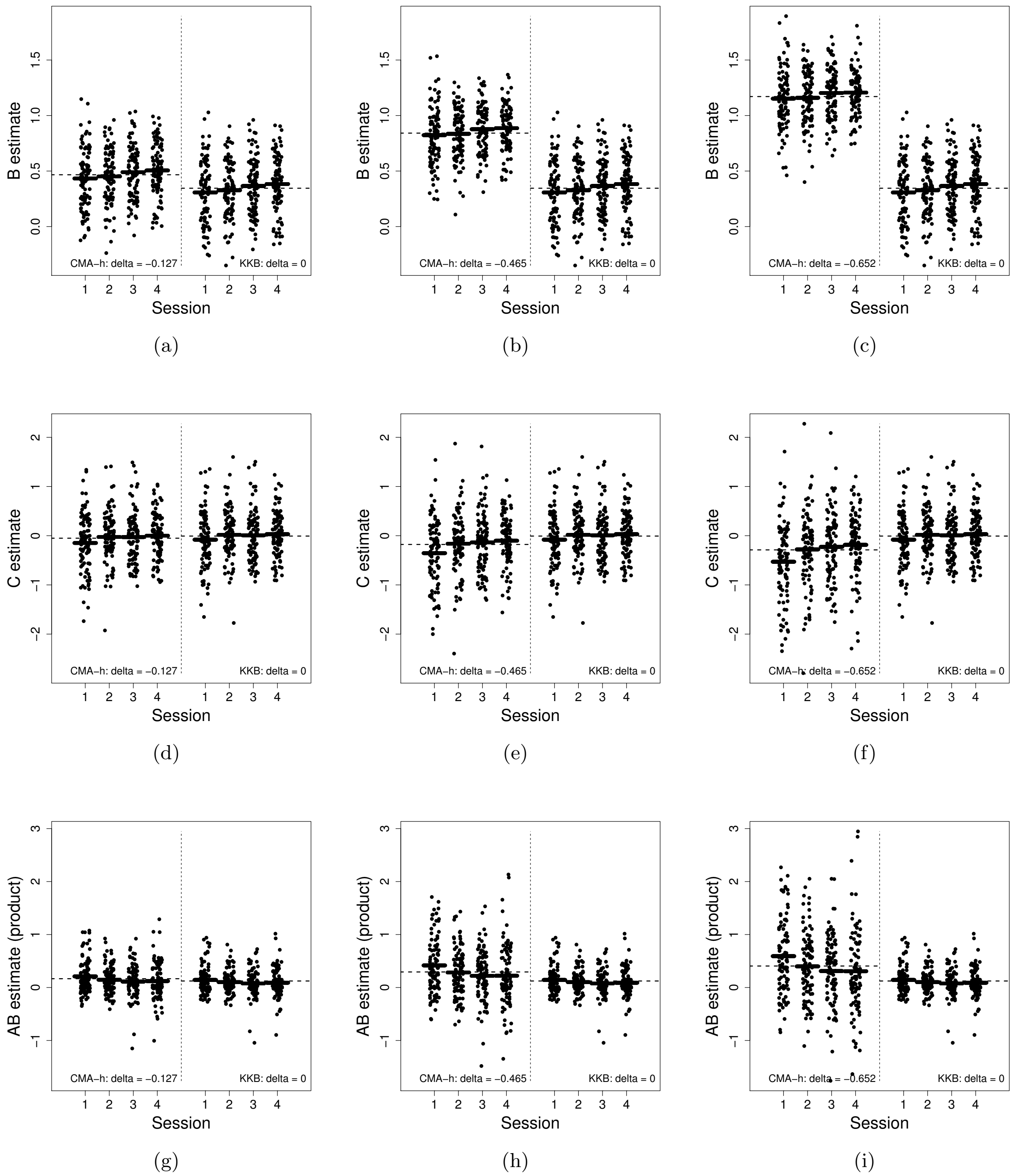

Figure F.11: Comparing our CMA methods (CMA-ts, CMA-h, CMA-m) versus the KKB method in estimating (a)-(c) $B$, (d)-(f) $C$ and (g)-(i) $AB$ (product) for each participant in each session.

Table F.4: Average estimates and 95% confidence intervals from our three-level CMA methods, CMA-$\delta$ method with $\delta$ estimated from CMA-h, KKB and BK on the fMRI datasets with and without motion correction (MC), using 500 bootstrap samples.

| *Data* | *Method* | $A$ | $B$ | $C'$ | $AB_d$ |
|---|---|---|---|---|---|
| Without MC | CMA-ts | 0.357 (0.275, 0.438) | 0.466 (0.398, 0.534) | 0.096 (0.053, 0.138) | 0.146 (0.121, 0.171) |
| | CMA-h | 0.349 (0.280, 0.417) | 0.841 (0.677, 1.006) | 0.096 (0.053, 0.138) | 0.273 (0.215, 0.331) |
| | CMA-m | 0.347 (0.278, 0.415) | 1.171 (0.825, 1.516) | 0.096 (0.053, 0.138) | 0.384 (0.268, 0.500) |
| | CMA-$\delta$ | 0.338 (0.293, 0.384) | 0.872 (0.683, 1.061) | 0.081 (0.040, 0.122) | 0.295 (0.231, 0.359) |
| | KKB | 0.357 (0.275, 0.438) | 0.345 (0.333, 0.358) | 0.096 (0.053, 0.138) | 0.103 (0.094, 0.112) |
| | BK | 0.338 (0.293, 0.384) | 0.327 (0.315, 0.339) | 0.081 (0.040, 0.122) | 0.111 (0.107, 0.115) |
| With MC | CMA-ts | 0.339 (0.260, 0.418) | 0.457 (0.396, 0.518) | 0.056 (0.011, 0.101) | 0.130 (0.108, 0.152) |
| | CMA-h | 0.334 (0.268, 0.400) | 0.779 (0.641, 0.917) | 0.056 (0.011, 0.101) | 0.234 (0.188, 0.281) |
| | CMA-m | 0.332 (0.266, 0.399) | 1.013 (0.718, 1.307) | 0.056 (0.011, 0.101) | 0.310 (0.216, 0.404) |
| | CMA-$\delta$ | 0.323 (0.277, 0.368) | 0.809 (0.650, 0.968) | 0.036 (-0.008, 0.080) | 0.261 (0.210, 0.312) |
| | KKB | 0.339 (0.260, 0.418) | 0.312 (0.300, 0.325) | 0.056 (0.011, 0.101) | 0.080 (0.071, 0.088) |
| | BK | 0.323 (0.277, 0.368) | 0.300 (0.287, 0.312) | 0.036 (-0.008, 0.080) | 0.097 (0.093, 0.101) |